\documentclass[10pt]{fcs}

\PassOptionsToPackage{hyphens}{url}
\usepackage[hang, flushmargin]{footmisc}
\usepackage{tikz}
\usepackage{amsmath,amsfonts}
\usepackage{multirow}
\usepackage{makecell}
\usepackage{verbatim}
\usepackage{url}
\usepackage[flushleft]{threeparttable}
\usepackage{ulem}
\usepackage{pifont}
\usepackage{graphicx}
\usepackage{color, colortbl}
\usepackage{epsfig}
\usepackage{amsmath}
\usepackage{float}
\usepackage{caption}
\usepackage{tabularx}
\usepackage{boldline}
\usepackage{multirow}
\usepackage{array}
\usepackage{threeparttable}
\usepackage{tabularx}
\usepackage{booktabs}
\usepackage{makecell}
\usepackage{multirow}
\usepackage{cite}
\usepackage{ulem} 
\usepackage{amssymb}
\usepackage[inkscapelatex=false]{svg}
\usepackage{pdfpages}
\usepackage{longtable}
\usepackage{tcolorbox}
\usepackage{setspace}
\usepackage{hyperref}
\usepackage{ulem}
\usepackage{graphicx} % 导入graphicx包以使用rotatebox

\usepackage[protrusion=false, expansion=false]{microtype}

\tcbuselibrary{breakable}
\tcbset{width=0.5\textwidth,boxrule=0pt,colback=red,arc=0pt,auto outer arc,left=0pt,right=0pt,boxsep=5pt}

\hypersetup{
    colorlinks=true,
    linkcolor=blue,
    filecolor=blue,      
    urlcolor=blue,
    citecolor=blue,
}

\definecolor{Raisins}{RGB}{160,32,240}

\definecolor{Revise}{RGB}{0,0,0}

\newcommand{\revise}[1][\textcolor{Revise}]{#1}
\newcommand{\ignore}[1]{}

\title{Analyzing Consumer IoT Traffic from Security and Privacy Perspectives: a Comprehensive Survey}

\author[1]{Yan Jia}
\author[1]{Yuxin Song}
\author[2]{Zihou Liu}
\author[1]{Qingyin Tan}
\author[3]{Yang Song}
\author*[1]{Yu Zhang}
\author[1]{Zheli Liu}

\address[1]{Key Laboratory of Data and Intelligent System Security, Ministry of Education, China (DISSec), Tianjin Key Laboratory of Network and Data Security Technology (NDST), College of Cryptology and Cyber Science, Nankai University, Tianjin 300350, China. }
\address[2]{DISSec, NDST, College of Computer Science, Nankai
University, Tianjin 300350, China. }
\address[3]{School of Computer Science and Technology, Hangzhou Dianzi University, HangZhou 310018, China. }
\fcssetup{
  received       = {March 14, 2025},
  accepted       = {May 6, 2025},
  corr-email     = {zhangyu1981@nankai.edu.cn},
}
\begin{abstract}
The Consumer Internet of Things (CIoT), a notable segment within the IoT domain, involves the integration of IoT technology into consumer electronics and devices, such as smart homes and smart wearables. Compared to traditional IoT fields, CIoT differs notably in target users, product types, and design approaches. While offering convenience to users, it also raises new security and privacy concerns. Network traffic analysis, a widely used technique in the security community, has been extensively applied to investigate these concerns about CIoT. \revise{Compared to traditional network traffic analysis in fields like mobile apps and websites, CIoT introduces unique characteristics that pose new challenges and research opportunities.} Researchers have made significant contributions in this area. To aid researchers in understanding the application of traffic analysis tools for assessing CIoT security and privacy risks, this survey reviews 310 publications on traffic analysis within the CIoT security and privacy domain from January 2018 to June 2024, focusing on three research questions. Our work: 1) outlines the CIoT traffic analysis process and highlights its differences from general network traffic analysis. 2) summarizes and classifies existing research into four categories according to its application objectives: device fingerprinting, user activity inference, malicious traffic detection, and measurement. 3) explores emerging challenges and potential future research directions based on each step of the CIoT traffic analysis process. This will provide new insights to the community and guide the industry towards safer product designs.

\end{abstract}


\begin{document}
\section{Introduction}

In recent years, numerous sectors related to the Internet of Things (IoT) have become part of everyday life, such as smart cities, industrial automation, smart homes, and smart healthcare~\cite{electronics12122590, 10053525, AIIoT58121202310174484, ICOEI56765202310125936}. A report by IoT Analytics\footnote{IoT market size by IoT Analytics, visit \url{https://iot-analytics.com/iot-market-size/}} predicts that the global IoT market will grow by 15\% in 2025, reaching \$347 billion. 

Compared with Industrial IoT (IIoT) and Medical IoT (MIoT), CIoT exhibits distinct differences in target users, device types, and product design objectives. Emerging security and privacy concerns in CIoT are increasingly prominent, with users expressing growing apprehension~\cite{205174,10.5555/3235924.3235931,289536}. 
Firstly, CIoT targets general consumers and frequently collects much personal information, such as location, health status, and daily routines, which may not be adequately protected. Secondly, the market hosts a wide variety of CIoT devices from numerous brands, resulting in low standardization. Thirdly, to improve user experience, the security measures in these devices are often simplified, that is, the short production cycles and limited capabilities of CIoT devices hinder effective defense against security threats~\cite{tawalbeh2020iot}.
Researchers have explored CIoT security issues and identified many risks~\cite{10.1145/3460120.3484592,9152619,274709}.

Network traffic analysis is an essential tool for security and privacy research.
Given the aforementioned risks associated with CIoT, new research scenarios and works have emerged in the field of CIoT traffic, distinct from traditional network traffic. 
By conducting an in-depth analysis of network traffic generated by CIoT devices, researchers can better understand device behavior patterns and promptly detect and prevent potential security threats. Furthermore, traffic analysis assists in assessing and improving data privacy protection within the CIoT environment.
Although many researchers have concentrated on examining CIoT traffic to bolster its security and privacy, CIoT, as a relatively new application compared to traditional ones, has special characteristics and may present new challenges. 

\begin{figure*}
\centering
\includegraphics[width=0.9\linewidth]{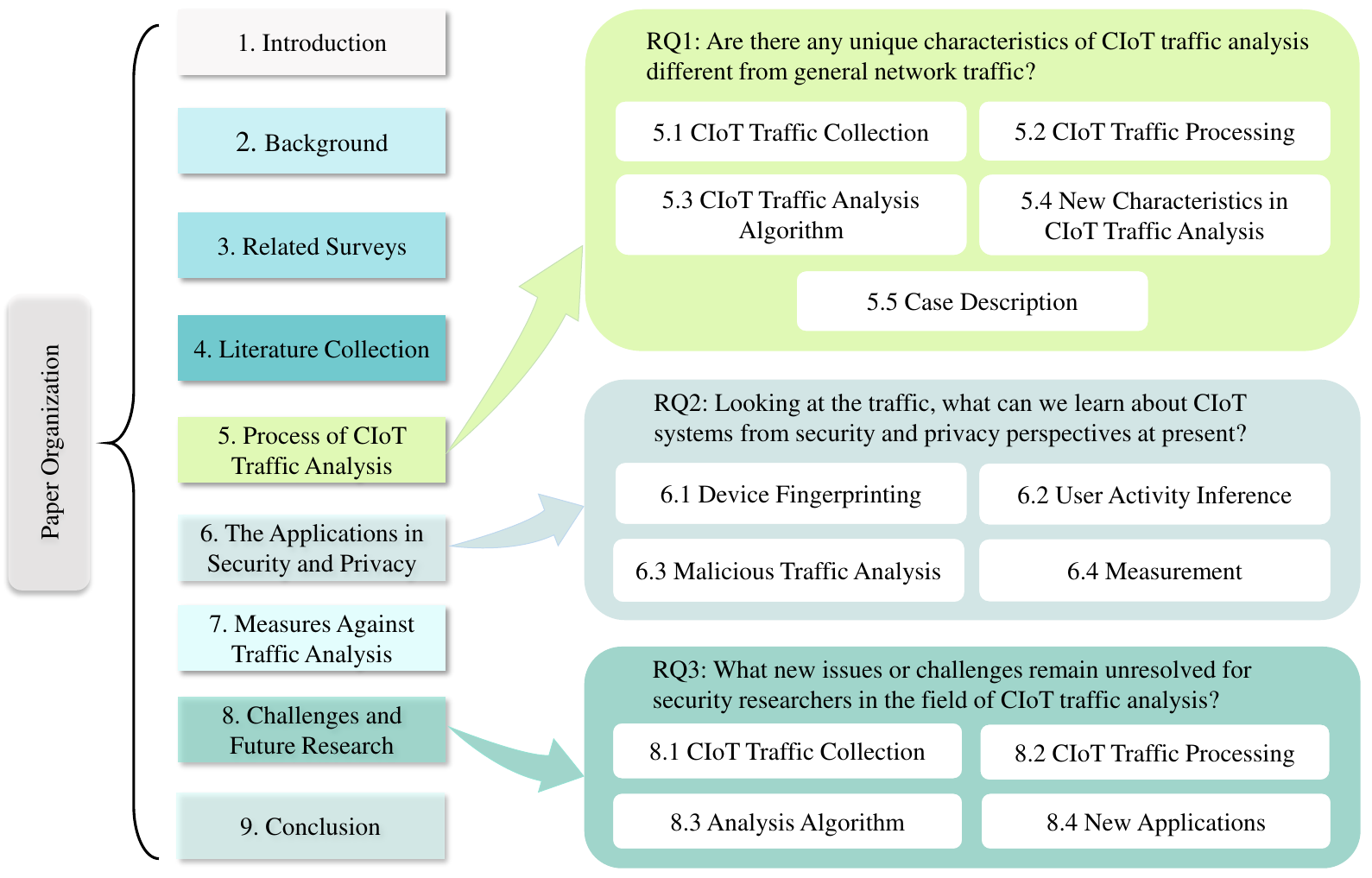}%
\caption{The organization of survey}
\label{figure0_paperorganization}
\vspace{-2.0em}
\end{figure*}

\revise{Emerging technologies such as machine learning (ML) and deep learning (DL) have significantly enhanced the effectiveness of traffic analysis within the CIoT domain. These technologies provide improved capabilities for processing and analyzing large-scale traffic data. AI-based methods, particularly in the areas of feature extraction and pattern recognition, play a crucial role in addressing the complexities of CIoT traffic, offering advanced solutions for security and privacy challenges.}

Considering that no other researchers in the community have explored how traffic analysis can guide CIoT security and privacy practices and that there is still an upward trend in this area, this paper systematically reviews the literature on CIoT traffic analysis over the past six years from the perspective of security and privacy. It conducts a detailed analysis around the following three research questions (RQs): 
\begin{itemize}
    \item \textbf{RQ1:} Are there any unique characteristics of CIoT traffic analysis different from general network traffic?
    
    \item \textbf{RQ2:} Looking at the traffic, what can we learn about CIoT systems from security and privacy perspectives at present?
    
    \item \textbf{RQ3:} What new issues or challenges remain unresolved for security researchers in the field of CIoT traffic analysis?
\end{itemize}

We systematically reviewed 310 papers from top-tier conferences and journals from January 2018 to June 2024 using the literature retrieval method described in Section~\ref{sec_literatureretrieval}. Based on these studies, we first summarize the process of CIoT traffic analysis, which comprises three key steps: CIoT traffic collection, CIoT traffic processing, and analysis. We point out the differences between CIoT traffic analysis and general network traffic analysis across these three steps (\textbf{RQ1}). 
Next, We categorize and summarize existing research according to their application goals, including device fingerprinting, user activity inference, malicious traffic analysis, and measurement, while also examining the latest advancements in each area (\textbf{RQ2}). 
\revise{Our research shows that AI-based technologies play an important role in this field. }
Finally, based on the traffic analysis process and its various application scenarios, we identify key challenges and outline future research directions and opportunities (\textbf{RQ3}). 
This article aims to provide researchers in the community with a deeper understanding of how traffic analysis can be leveraged to assess security and privacy practices within the CIoT ecosystem. \revise{In addition to} previous studies that primarily emphasize machine learning technology in traffic analysis, our work provides a more integrated and comprehensive perspective grounded in security and privacy, offering practical and targeted guidance for strengthening protection in the CIoT domain.

The contributions of this paper are summarized as follows: 
\begin{itemize}
    \item This is the first survey that focuses on CIoT traffic from security and privacy perspectives. 
    We employed a standardized literature retrieval methodology and conducted an in-depth review and classification of existing works, enabling researchers to efficiently grasp the current landscape of the field.

    \item We extracted the CIoT traffic analysis process from the literature and conducted a detailed comparison with general network traffic analysis, emphasizing the unique characteristics of CIoT.

    \item Based on this process, we provide new insights into the challenges associated with each step of CIoT traffic analysis and propose promising directions for future research.
\end{itemize}

\vspace{3pt}\noindent\textbf{Paper Organization.} 
The paper is organized as follows: 
Section~\ref{sec_background} covers the basics of CIoT and traffic analysis. 
Section~\ref{sec_relatedsurvey} reviews relevant surveys. 
In Section~\ref{sec_literatureretrieval}, we introduce the methodology for collecting the literature. 
Section~\ref{sec_AnalysisProcedure} presents the CIoT traffic analysis process and its unique characteristics. Section~\ref{sec_applications} discusses current works categorized by application goals. 
We summarize the measures against traffic analysis in Section~\ref{sec_masking}.
The challenges and future research directions are discussed in Section~\ref{sec_challenges}. We conclude the survey in Section~\ref{sec_conclusion}. 
The organization of our survey is shown in Figure~\ref{figure0_paperorganization}. 

\section{Background}
\label{sec_background}

% 由于我们的调查主要关注流量分析在消费物联网安全和隐私的应用，因此在本节中，我们首先介绍消费物联网的基本架构、生命周期和控制方式。随后，我们会简要介绍general流量分析的基本流程，如特征提取方式和常用的AI分析算法。具体CIoT流量分析框架将会在第五章详细展开，并将其与general网络流量分析进行对比
\revise{This survey focuses on the application of traffic analysis to the security and privacy of the CIoT. In this section, we first introduce the fundamental architecture, lifecycle, and control methods of CIoT. Next, we provide an overview of the general traffic analysis process. The specific process of CIoT traffic analysis will be detailed in Section~\ref{sec_AnalysisProcedure}, where it will also be compared to general network traffic analysis.}

\subsection{Consumer Internet of Things}

\label{sec_background_consumeriot}

CIoT devices are typically monitored via mobile apps or software API interfaces, enabling users to manage connected devices remotely or locally. For example, users can remotely open the lights or thermostats in a home, or use wearable devices such as smartwatches to monitor users' health data and provide corresponding feedback and suggestions. Given the highly sensitive nature of the data collected by these devices~\cite{10.1145/3355369.3355577}, ensuring security and privacy for CIoT systems is essential to protecting users' personal information. 

\begin{figure}[htbp]
\centering
\includegraphics[width=\linewidth]{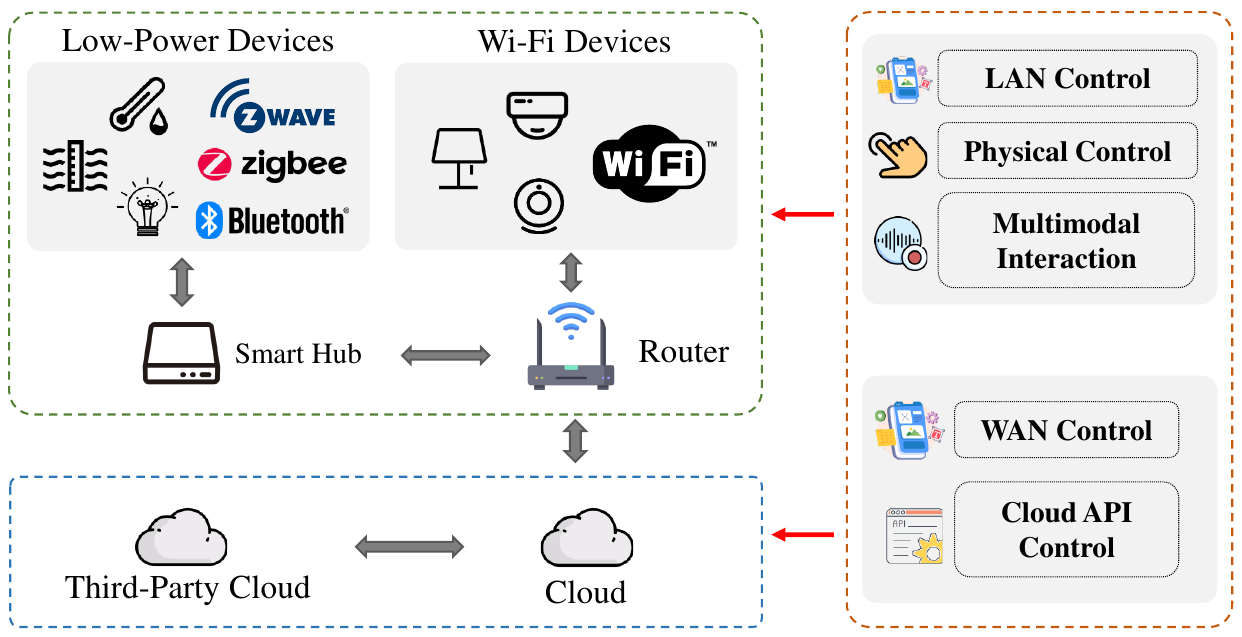}%
\caption{The architecture of CIoT}
\label{Fig:CIoTModel}
\end{figure}

Representative CIoT scenarios are shown in Figure~\ref{Fig:CIoTModel}. 
There are multiple control methods for the devices. 
Wi-Fi-enabled devices can communicate with remote users through cloud services. 
The transmitted information includes device command, status, and heartbeat packets which maintain the connection. 
It is worth noting that in addition to the first-party cloud communicated directly by the device, third-party clouds, such as advertisers, may also obtain device information~\cite{10.1145/3618257.3624803, girish-imc23}.
Devices using low-power protocols typically connect to a smart gateway, which acts as an intermediary to the Internet. Furthermore, third-party platforms can be authorized to use device control APIs. 
As shown in the figure, we summarize the following five control methods: 
\begin{itemize}
\setlength{\itemsep}{0pt}
\setlength{\parsep}{0pt}
\setlength{\parskip}{0pt}
\item \noindent\textbf{Physical Control}. Users can physically interact with devices. 
\item \noindent\textbf{Multimodal Interaction}. CIoT devices, equipped with various sensors, support multimodal interactions. For example, motion sensors detect activity to control smart lights, and smart speakers like Amazon Alexa and Xiaomi XiaoAi support voice commands. 
\item \noindent\textbf{Local Area Network (LAN) Control}. When the device and smartphone are on the same network, they communicate via Bluetooth or Wi-Fi for basic functions, firmware updates, and settings.
\item \noindent\textbf{Wide Area Network (WAN) Control}. When the smartphone and device are not on the same LAN, commands and status updates are relayed through the cloud.
\item \noindent\textbf{Cloud API Control}. Besides companion apps, some platforms offer cloud APIs for third-party access via authorization, enabling automation control like IFTTT \footnote{If This Then That, a service that lets users create simple conditional statements, called ``applets'', to automate tasks across different web services and devices.}.
\end{itemize}

Besides multiple control methods, 
the workflow of CIoT devices typically follows four \textit{lifecycle phases}: setup, interaction, idle, and deletion~\cite{10.1145/3355369.3355577}. 
Initially, the user setups the device, including configuring the network and account binding before using it. After setup, the user interacts with the device during the interaction phase. When there is no interaction, the device enters an idle state. Finally, a user can remove the device from the account, marking the deletion phase. 

In summary, the CIoT ecosystem supports diverse control methods and features a lifecycle distinct from PC and mobile apps. These characteristics potentially affect the analysis of CIoT traffic. 

\subsection{Traffic Analysis}

\label{Sec_background_traffic}

Traffic analysis is an important tool for network security and privacy, which extracts valuable insights from network data~\cite{tan2013system}. Its process includes four steps: traffic collection, traffic representation, analysis, and evaluation. 

The purpose of traffic collection is to capture traffic data packets at key network nodes, such as the output of the intranet or the public network server. Common tools include wireshark or tcpdump.

\begin{table*}[htbp]
  \centering
  \renewcommand\arraystretch{1}
    \caption{Related surveys}
    \resizebox{\textwidth}{!}{
    \begin{tabular}{p{4.125em}ccccccccc}
    \toprule
    \specialrule{0.05em}{1.5pt}{2pt}
    \multirow{2}[4]{*}{Literature} & \multicolumn{4}{c}{Application}       & \multicolumn{2}{c}{Technologies} & \multirow{2}[4]{*}{Amount} & \multirow{2}[4]{*}{Years covered\textsuperscript{1}} \\
\cmidrule(r){2-5}  \cmidrule(r){6-7}  \multicolumn{1}{c}{} & \multicolumn{1}{p{5.19em}}{Device \newline{}Fingerprinting} & \multicolumn{1}{p{6.19em}}{User Activity\newline{}Inference} & \multicolumn{1}{p{7.19em}}{Malicious Traffic\newline{} Analysis} &  \multirow{2}{*}{Measurement}  & \multirow{2}{*}{non-ML} & \multirow{2}{*}{ML}    &       &  \\
    \midrule
    \multicolumn{1}{l}{Seliem et al. \cite{seliem2018towards}} & $\times$ & $\surd$ & $\times$ & $\times$ & $\surd$ & $\surd$ & 105   & -2018 \\
    \multicolumn{1}{l}{Gupta et al. \cite{gupta2022privacy}} & $\times$ & $\surd$ & $\times$ & $\times$ & $\surd$ & $\surd$ & 153   & 2010-2021 \\
    \multicolumn{1}{l}{ Zavalyshyn et al. \cite{zavalyshyn2022sok}} & $\times$ & $\surd$ & $\times$ & $\times$ & $\surd$ & $\surd$ & 137 & 2010-2021 \\
    \multicolumn{1}{l}{Alrawi et al. \cite{alrawi2019sok}} & $\times$ & $\times$ & $\times$ & $\times$ & $\surd$ & $\surd$ & 108   & -2018 \\
    \multicolumn{1}{l}{Abosata et al. \cite{abosata2021internet}} & $\times$ & $\times$ & $\surd$ & $\surd$ & $\surd$ & $\times$ & 114   & -2021 \\
    \multicolumn{1}{l}{Wang et al. \cite{wang2022survey}} & $\times$ & $\times$ & $\surd$ & $\times$ & $\surd$ & $\surd$ & 222   & -2022 \\
    \multicolumn{1}{l}{Papadogiannaki et al.\cite{papadogiannaki2021survey}} & $\times$ & $\times$ & $\surd$ & $\times$ & $\surd$ & $\surd$ & 176   & -2021 \\
    \multicolumn{1}{l}{Shen et al. \cite{shen2022machine}} & $\surd$ & $\times$ & $\surd$ & $\times$ & $\times$ & $\surd$ & 108   & 2007-2021 \\
    \multicolumn{1}{l}{Shahraki et al. \cite{shahraki2022comparative}} & $\times$ & $\times$ & $\surd$ & $\times$ & $\times$ & $\surd$ & 118   & -2022 \\
    \multicolumn{1}{l}{Mathews et al. \cite{mathews2023sok}} & $\surd$ & $\times$ & $\times$ & $\times$ & $\times$ & $\surd$ & 10    & -2021 \\
    \multicolumn{1}{l}{Bhatiaa et al. \cite{bhatiaa2020survey}} & $\times$ & $\times$ & $\surd$ & $\times$ & $\times$ & $\surd$ & 97    & -2022 \\
    \multicolumn{1}{l}{Sanchez et al. \cite{sanchez2021survey}} & $\surd$ & $\times$ & $\times$ & $\times$ & $\surd$ & $\surd$ & 212   & -2021 \\
    \multicolumn{1}{l}{Jmila et al. \cite{jmila2022survey}} & $\surd$ & $\times$ & $\times$ & $\times$ & $\times$ & $\surd$ & 58    & 2018-2022 \\
    \multicolumn{1}{l}{Tahaei et al. \cite{tahaei2020rise}} & $\surd$ & $\times$ & $\times$ & $\times$ & $\times$ & $\surd$ & 159   & -2020 \\
    \specialrule{0.05em}{2pt}{2pt}
    \multicolumn{1}{l}{\textbf{Ours}}& \textbf{$\surd$} & \textbf{$\surd$} & \textbf{$\surd$} & \textbf{$\surd$} & \textbf{$\surd$} &\textbf{$\surd$} & \textbf{303}   & \textbf{2018-2024} \\
    \specialrule{0.05em}{2pt}{1.5pt}
    \bottomrule
    \end{tabular}
    }
    \label{tab:survey}
    
  \begin{footnotesize}
    \begin{itemize}
      \item[1] Some surveys do not specify the quantity or time of the literature they encompass, so we summarize this information according to references.
    \end{itemize}
    \end{footnotesize}  
    \vspace{-1.5em}
\end{table*}

Secondly, features are extracted from the analysis unit and represented in a format suitable for the subsequent analysis.
For non-encrypted traffic, Deep Packet Inspection (DPI) can be used to analyze the payloads of packets.
However, with the rise in packet encryption in data transmission, encrypted traffic analysis has become mainstream, which requires researchers to extract traffic features independent of payload.
In the beginning, the focus was primarily on packet-level features. By examining the values in packet headers, researchers could gather extensive information like packet length or number of bytes grouped. However, the rapid increase in traffic volume necessitates the development of flow-level features. \revise{A flow is a collection of packets with the same five-tuple (source IP, destination IP, source port, destination port, and protocol).}
Flow analysis redirects attention from individual packets to the interactions between the source and destination.
\texttt{NetFlow} and \texttt{sFlow} are popular tools for flow-level analysis.
Statistical analysis methods were integrated into traffic analysis to combine features from multiple flows~\cite{10.20533/ijisr.2042.4639.2020.0107}. 
The mean and median are combined with packet or flow-level features to generate statistical features. 
Advancements in deep learning have made it possible to convert raw packets into images or sequences that can then be processed by deep learning models, eliminating manual feature extraction. 

Next, we explore the diverse algorithms used in traffic analysis, categorized into \textit{machine learning (ML)} and \textit{non-machine learning (non-ML)} approaches. Corresponding to the extracted features, these methods are mainly applied to encrypted traffic analysis. 
ML algorithms are further divided into \textit{traditional machine learning (TML)}, \textit{deep learning (DL)}, and \textit{reinforcement learning (RL)}.
TML includes algorithms that learn patterns and make predictions or decisions based on data~\cite{ML, ML1959}. 
Feature selection in TML algorithms is crucial yet challenging.  
In contrast, DL models excel at automatically learning hierarchical features from data through various abstraction layers~\cite{DL}.
RL is a unique subset of ML that equips agents to observe the environment, select actions, and adjust behavior based on rewards or punishments~\cite{RL}, leading to optimal decision-making strategies. 
Despite their advantages, AI algorithms sometimes cannot explain false positives or negatives and are limited by data size~\cite{duan2023iota}.
To address these challenges, researchers have explored non-ML algorithms as alternatives, including model-based methods~\cite{duan2023iota, 10154338}, locality-sensitive hashing (LSH)~\cite{9246572,charyyev2020iotlocality}, and other techniques.

Finally, studies evaluate the performance of algorithms using various metrics. In Section~\ref{sec_AnalysisProcedure}, we will summarize the CIoT traffic analysis process and explain the framework in detail, considering the unique characteristics of CIoT devices.

\section{Related Surveys}
\label{sec_relatedsurvey}

Earlier surveys have investigated IoT security and privacy, general network traffic analysis, and IoT device fingerprinting, which overlaps with Section~\ref{sec_applications_devicesFingerprinting}. In contrast, our paper offers the first exhaustive review of traffic analysis focusing on CIoT security and privacy aspects, seeking to identify the information contained in CIoT traffic and the challenges encountered in its analysis. 
Table \ref{tab:survey} highlights the distinctions between our work and the existing surveys. 

\subsection{IoT Privacy and Security} 
As IoT devices become increasingly prevalent, safeguarding user privacy has emerged as a key issue. Several researchers have explored various aspects of IoT privacy protection. Seliem et al. 
\cite{seliem2018towards} reviewed existing research and solutions to privacy issues. Gupta and Ghanavati
\cite{gupta2022privacy} conducted a systematic literature review on IoT privacy practices and technologies, providing a comprehensive summary of several issues related to privacy protection; Zavalyshyn et al.
\cite{zavalyshyn2022sok} focused on privacy-enhancing technologies of the smart gateway. 
Several surveys focus on IoT security.  Alrawi et al. \cite{alrawi2019sok} summarized the literature on IoT device security and organized a systematic evaluation method for assessing device security attributes;  Abosata et al. 
\cite{abosata2021internet} discusses the security risks caused by the implementation of industrial IoT in smart cities and intelligent manufacturing and then categorizes attacks and potential security solutions;  
Due to Home Automation (HA) systems being vulnerable, Wang et al. \cite{wang2022survey} studied the security of HA from the perspectives of attacks and defense and summarized relevant literature.
By contrast, our research investigates user privacy and security through the perspective of network traffic, highlighting differences in applying traffic analysis in the field of CIoT.

\subsection{Traffic Analysis}
Some researchers review the advancements in traffic analysis research, with a particular emphasis on encrypted network traffic and the use of machine learning. Papadogiannaki and
Ioannidis
\cite{papadogiannaki2021survey} investigated the techniques, applications, and countermeasures related to encrypted network traffic analysis. They summarized relevant literature from four aspects: network analysis, network security, user privacy, and middleware network functionality.  Shen et al. 
\cite{shen2022machine} focused on the application of ML techniques in encrypted traffic analysis. This work organizes the existing literature in four directions: network asset identification, network characterization, privacy leak detection, and attack detection. Bhatiaa et al. 
\cite{bhatiaa2020survey} specifically discussed the encrypted traffic of smartphones. 
Furthermore, Shahraki et al. \cite{shahraki2022comparative} highlighted the benefits of employing online machine learning for traffic analysis. Mathews et al. \cite{mathews2023sok} examined methods to defend against website fingerprinting. Overall, these studies focus on analyzing network traffic from personal computers and smartphones, rather than concentrating on the traffic generated by CIoT devices. Our work specifically focuses on the traffic generated by CIoT devices, including both encrypted and unencrypted traffic.

\subsection{IoT Fingerprinting}
Some surveys focus on traffic analysis on CIoT devices, but only cover device fingerprinting.
Sanchez et al. \cite{sanchez2021survey} reviewed device behavior fingerprints, covering not only smart home devices but also non-IoT devices such as PCs and personal smartphones.  
Jmila et al. \cite{jmila2022survey} summarized the application of ML in the field of device classification and highlighted key issues to consider in device classification, such as feature costs and learning quality. Tahaei et al. 
% \cite{tahaei2020rise} investigated the application of network traffic classification in different fields of IoT\revise{\sout{, including common IoT devices, smart cities, and healthcare systems}}. 
\cite{tahaei2020rise} investigated the application of network traffic classification in different fields of IoT\revise{, including common IoT devices, smart cities, and healthcare systems}. 
Different from other surveys, this survey starts from the perspective of security and privacy, covers more application targets, and discusses future research directions based on the uniqueness of CIoT.

\section{Literature Collection}
\label{sec_literatureretrieval}

\begin{figure}[h]
\centering
\includegraphics[width=0.8\linewidth]{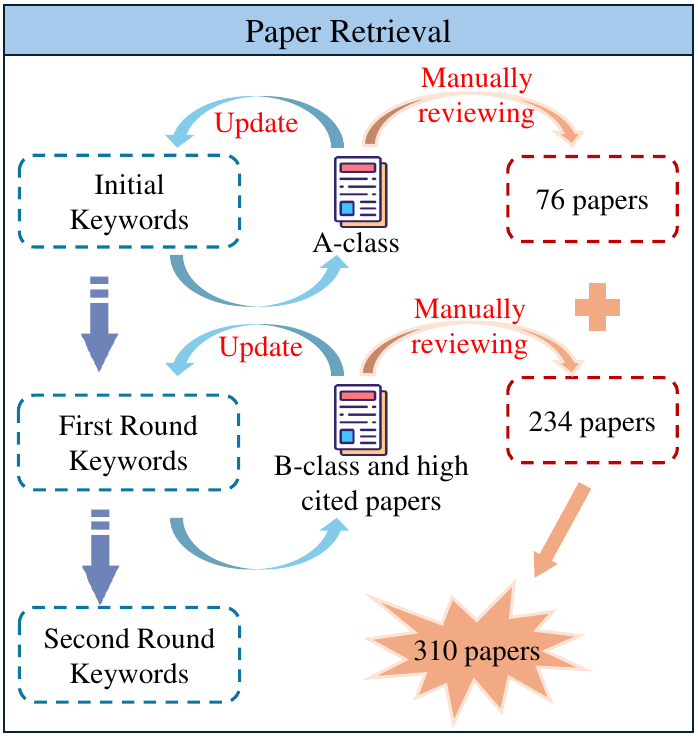}%
\caption{The process of paper retrieval}
\label{Fig:paper_retrieve}
\end{figure}

% To improve the quality of literature retrieval, we refer to well-known public lists of academic conferences and journals about network security, including the recommendations of Professor Guofei Gu from Texas A\&M University~\cite{conferencerankingbyguguofei}, Professor Jianying Zhou from Singapore University of Technology and Design~\cite{conferencerankingbyzhoujianying}, and Tsinghua University's Computer Science Discipline Group (TH-CPL) and the China Computer Federation (CCF)~\cite{conferencerankingbyTsinghua, conferencerankingbyccf}. 

To improve the quality of literature retrieval, we refer to well-known public lists of academic conferences and journals about network security, including the recommendations of Professor Guofei Gu from Texas A\&M University, Professor Jianying Zhou from Singapore University of Technology and Design, and Tsinghua University's Computer Science Discipline Group (TH-CPL) and the China Computer Federation (CCF). 
We particularly emphasize conferences and journals in the A and B categories related to network and information security, computer networks, high-performance computing, and systems software and software engineering. 
Based on these sources, we categorize the journals and conferences into classes A and B by comprehensively evaluating their rankings across the above lists.

The process of paper retrieval is shown in Figure~\ref{Fig:paper_retrieve}. We adopted the snowball generations approach to expand the search keywords dynamically, ensuring both accuracy and breadth in our literature search. We first identified the initial keywords that are closely related to our topic, as shown in Table~\ref{keywords} under ``Initial''. ``IoT Traffic'' is a core keyword that must appear in the retrieved articles. ``security'' and ``privacy'' are auxiliary keywords, meaning at least one or more of them are included in the search results. We conducted searches using these keywords in all 17 A-class conferences and journals (such as USENIX, S\&P, NDSS, and CCS), resulting in 2,966 papers over the past years (2018-2024). 

By manually reviewing each paper, we carefully selected a subset of 76 closely relevant papers about CIoT from A-class conferences and journals as the core literature for our survey. We dynamically expand initial keywords while reviewing.
Based on the keywords and abstracts of these core papers, we gained the auxiliary keywords (referred to as ``First Round'' in Table~\ref{keywords}). Subsequently, we searched B-class literature and extra papers that are highly cited but not within the scope using the first-round keywords. In this process, we used the same method to dynamically expand the keywords. Finally, the second round of keywords (referred to as ``Second Round'' in Table~\ref{keywords}) and 234 related articles were obtained. 

In total, we identified 310 relevant papers. The change in the number of papers over time is shown in Figure 4. Considering the length limitation, we have selected 156 of the most classic and core articles for a detailed introduction.
The complete list of journals and papers is available online.
\footnote{Complete list of journals and papers, visit \href{https://github.com/NKUHack4FGroup/CIoT-traffic-survey}{https://github.com/NKUHack4FGroup/CIoT-traffic-survey}}

\begin{table}
\small
\centering
\renewcommand\arraystretch{1.5}
\caption{The keywords used in literature search.}
\label{keywords}
% \usepackage{multirow}
%\begin{threeparttable}
\newcolumntype{M}[1]{>{\centering\arraybackslash}m{#1}}
\resizebox{0.49\textwidth}{!}{
\begin{tabular}{M{0.05\textwidth}|M{0.48\textwidth}}
\toprule
\specialrule{0.05em}{1.5pt}{0pt}
% \textbf{Keyword categories} & \textbf{Keywords}  \\
% \hline
% security & privacy, vulnerable, vulnerabilities, attacker, 
%  attack, attacks, attacking, security, secure, secures  \\
% \hline
% traffic & IoT traffic, smart home traffic, device traffic, IoT network, smart home network  \\
% \hline
% detection/identification & Detection, detecting, identification, classification, fingerprint, fingerprinting  \\
% \hline
% others & Monitoring, anomalous, anomalies, botnet, malicious  \\
% \hline
\textbf{Rounds} & \textbf{Keywords}  \\
\hline
Initial & IoT traffic, security, privacy  \\
\hline
First Round & IoT traffic, security, privacy, detection, fingerprint, vulnerable, attack, malicious, botnet, measurement  \\
\hline
Second Round & IoT traffic, security, privacy, detection, fingerprint, identification, classification, vulnerable, attack, hack, malicious, anomalous, botnet, DDoS, measurement, smart home, smart wearable, intrusion \\
\specialrule{0.05em}{0pt}{1.5pt}
\bottomrule

\end{tabular}
}
    \vspace{-1.5em}
\end{table}

\begin{figure}[htbp]
\centering
\includegraphics[width=\linewidth]{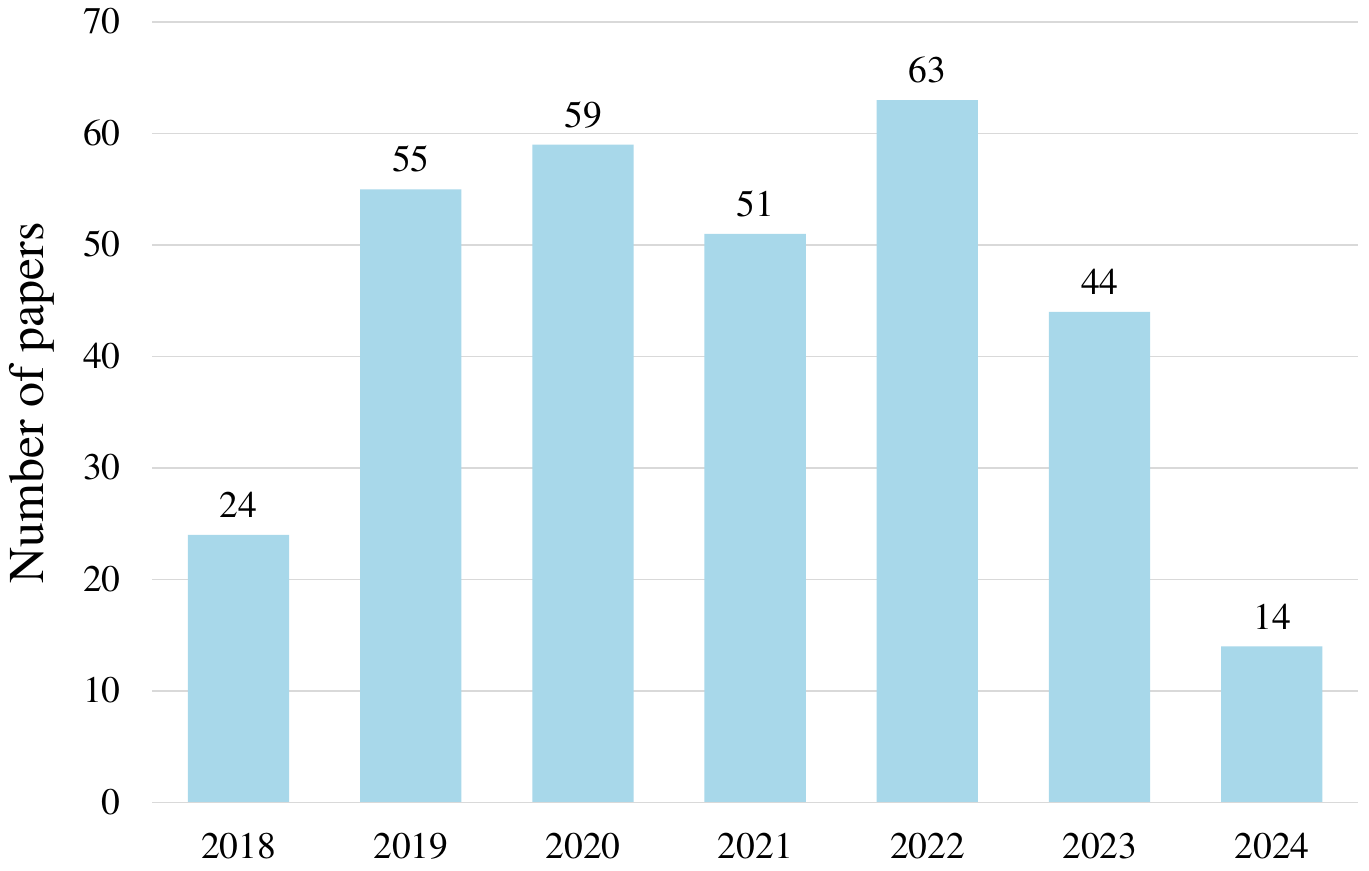}%
\caption{The number of papers published from 2018 to 2024}
\label{Fig:totaltrend}
    \vspace{-1.5em}
\end{figure}

\section{Process of CIoT Traffic Analysis}

\label{sec_AnalysisProcedure}
\begin{figure*}
\centering
\includegraphics[width=\linewidth]{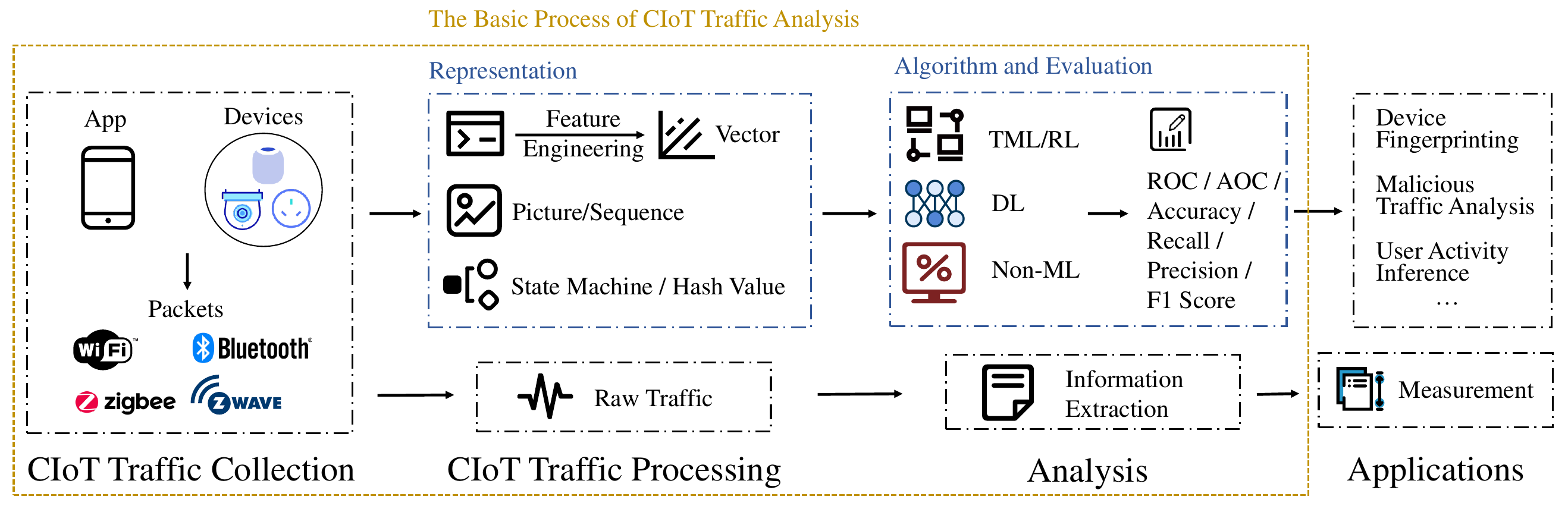}
\caption{The basic process of CIoT traffic analysis}
\label{Fig:trafficanalysis}
\end{figure*}

% 在第二章我们已经简短介绍过流量分析的基本流程，而在本节，我们将目光聚焦到本文的调查重点——消费物联网
\revise{In Section~\ref{Sec_background_traffic}, we've briefly covered the basic process of traffic analysis. In this section, we summarized the process of CIoT traffic analysis and its unique characteristics to answer RQ1.} 
By reviewing the existing literature, we outline the basic process of CIoT traffic analysis in 5.1, 5.2, and 5.3, which includes \textit{CIoT traffic collection}, \textit{CIoT traffic processing}, \textit{analysis}, which is depicted in Figure~\ref{Fig:trafficanalysis}. The \textit{application} will be discussed in detail in Section~\ref{sec_applications}. Based on this process, we summarize the unique characteristics of CIoT compared to the general network traffic analysis in 5.4.

\subsection{CIoT Traffic Collection}

\label{sec_systematic_trafficcollection}

\begin{figure}[htbp]
\centering
\includegraphics[width=\linewidth]{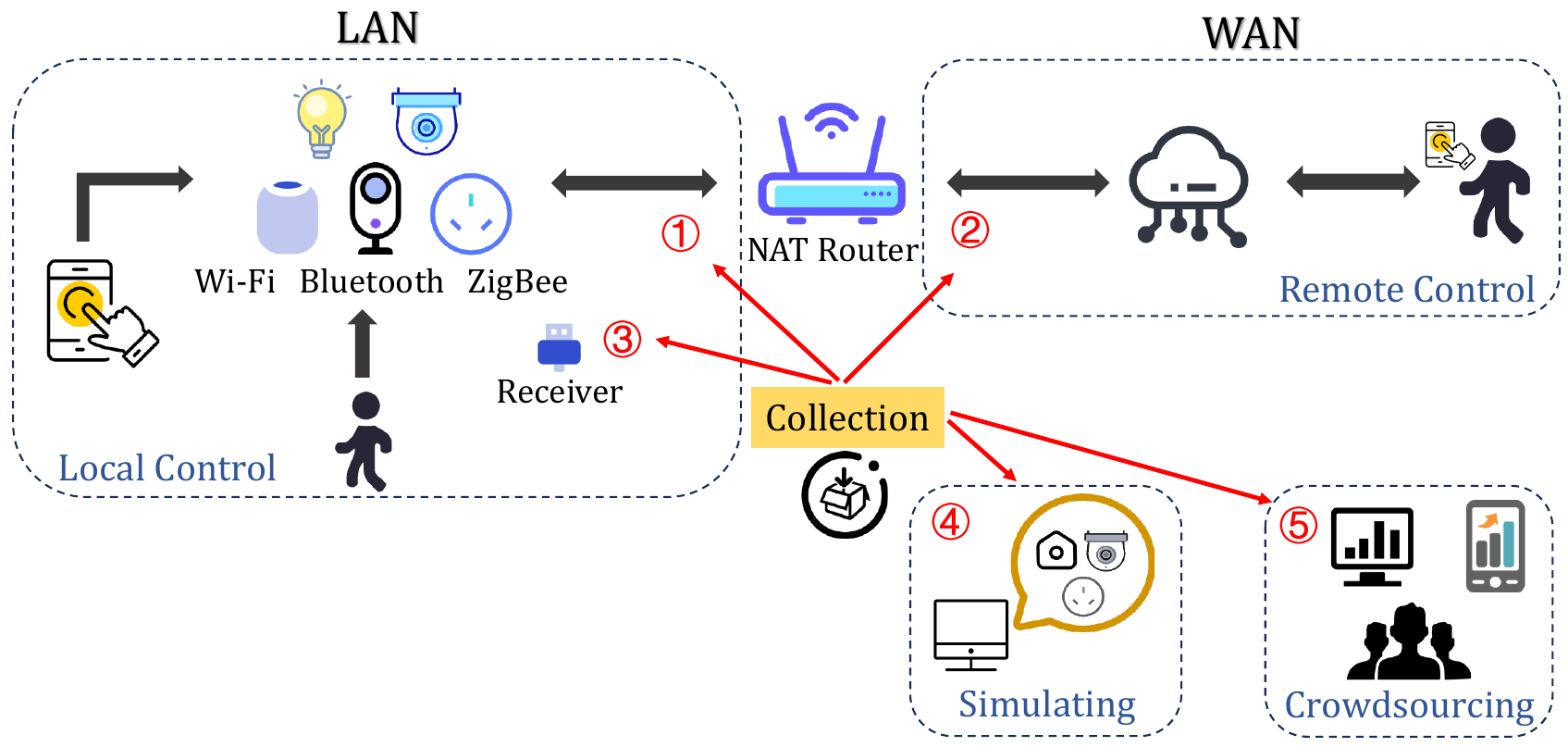}
\caption{Traffic collection setups}
\label{Fig:traffic_collection}
    \vspace{-1.5em}
\end{figure}

\subsubsection{Collection Process}
The traffic collection process is the first step in CIoT traffic analysis and differs significantly from general network traffic analysis. Firstly, CIoT devices exhibit a wider variety, with traffic patterns varying substantially among different types of devices. Secondly, CIoT devices demonstrate diverse interaction patterns, necessitating tailored traffic collection setups for various interaction scenarios. Thirdly, different communication techniques devices use, such as Wi-Fi and Bluetooth, require different traffic collection methods, posing challenges in constructing comprehensive CIoT traffic datasets.  

Figure~\ref{Fig:traffic_collection} illustrates five methods for acquiring traffic:
\begin{itemize}
\setlength{\itemsep}{0pt}
\setlength{\parsep}{0pt}
\setlength{\parskip}{0pt}
\item \noindent\textbf{Collection from the Router (Methods 1 ($M_1$) and 2 ($M_2$))}: Traffic can be collected at the inside or outside interface of the Network Address Translation (NAT) router. The inside interface of the NAT router connects to the private network, allowing local IP addresses to distinguish the traffic of each device. Conversely, capturing traffic at the interface after NAT mixes the traffic of all devices within the LAN, which is also the traffic aggregation point for regular ISPs.
\item \textbf{Collection through Receivers ($M_3$)}: CIoT devices use diverse communication protocols such as Zigbee and Bluetooth, in addition to Wi-Fi. Specialized receivers can capture the link-layer packets of these protocols.
\item \textbf{Generation through Simulators ($M_4$)}: Due to the challenges of collecting malicious CIoT traffic in the wild, researchers sometimes use simulators to generate special traffic patterns. For instance, Koroniotis et al. \cite{koroniotis2019towards} utilized a tool called Node-RED to simulate devices in a virtual network.
\item \textbf{Crowdsourced Collection ($M_5$)}: ``Crowdsourcing'' refers to the practice of gathering information or data about network traffic through a large number of individuals, typically users or volunteers. Since acquiring devices from various brands and categories can be costly, and simulating realistic user interaction traffic is challenging, some researchers leverage crowdsourcing to gather data~\cite{zhang2017crowdsourcing, gan2019crowdsourcing}. 
\end{itemize}

\subsubsection{Available Datasets}
\label{sec_systematic_dataset}
Considering that De Keersmaeker et al.\cite{10160090} have conducted a comprehensive review of public datasets in the IoT field, we only summarize the most classic and frequently cited datasets applicable to the CIoT field in Table~\ref{dataset}.
Among all self-collected CIoT datasets (data collected by researchers in their lab), the Mon(IoT)r dataset is the most frequently cited and contains the highest number of devices, followed by UNSW, YT, Ours, and PingPong. 

By analyzing the related dataset papers and Table~\ref{dataset}, we identify several shortcomings in the current CIoT datasets.
According to the survey by De Keersmaeker
et al.~\cite{10160090}, researchers have created nearly 70 datasets. However, we observe that the number and types of CIoT devices included in these datasets are very limited, and most of the data collection occurred before 2021, failing to reflect the current trends in CIoT devices. 
Furthermore, echoing the insights from De Keersmaeker
et al.~\cite{10160090}, future datasets should encompass a broader range of protocols (such as LoRa, Sigfox, etc.) and place greater emphasis on link-layer traffic analysis. 

Although De Keersmaeker et al. have comprehensively classified datasets, the article lacks insights into the geographical location of traffic collection. Devices in different regions adhere to distinct laws and regulations, leading to variations in transmitted data content. we observed that most datasets appear to be self-collected in laboratory settings, likely due to the convenience of traffic labeling. Additionally, the majority of devices in these datasets are from America and Europe, with a notable absence of CIoT traffic data from Asia. 
Thirdly, current datasets do not provide fine-grained labels regarding the lifecycle of devices. 
In conclusion, there is a need for a more state-of-the-art and comprehensive dataset that considers the unique characteristics and rapid development of CIoT. 

\subsection{CIoT Traffic Processing}
\label{sec_AnalysisProcedure_processing}
Following the general traffic analysis methodology, the second step involves extracting features or information and processing the data for specific application purposes. 

\subsubsection{Features Extraction} 
In this part, we introduce common features extracted from CIoT traffic. 
\begin{itemize}
\setlength{\itemsep}{0pt}
\setlength{\parsep}{0pt}
\setlength{\parskip}{0pt}
\item \textbf{Packet-level Feature.} Packet-level features primarily involve fields from packet headers, such as IP address, port number, TTL value, payload length, and TCP initial window size. 
\item \textbf{Flow-level Feature.} Flow features capture the overall characteristics of a flow, including the total input and output bytes, transmission byte rate, and flow duration. In 2005, Moore et al. \cite{moore2013discriminators} summarized 249 flow-level features. 
\item \textbf{Statistical Feature.} Based on packet-level and flow-level features, statistical features of the traffic can be computed, including measures such as maximum, minimum, mean, variance, and standard deviation. 
\item \textbf{Deep Learning Feature.} DL algorithms can automatically encode raw packets into sequences or images for advanced analysis. 
\end{itemize}

Community members have developed several tools to facilitate the extraction of network features from raw \texttt{pcap} files, including CICFlowmeter, Zeek, and Joy. \texttt{CICFlowmeter}, an open-source Java tool, can extract over 80 dimensions of features. 
\texttt{Zeek}, a network traffic analysis tool, enables custom feature extraction through its own Domain Specific Language (DSL). \texttt{Joy}, which focuses on the application layer, outputs data in JSON format, thereby complementing the feature sets provided by the other tools.
It is noteworthy that some researchers~\cite{8761559, 9269044}  extract real-time features by setting time windows to evaluate the model in real-time scenarios. That is, flow-level or packet-level features are extracted within a fixed time window. Choosing the suitable window length is challenging: Longer time windows increase the delay of model classification, while shorter ones cannot accurately reflect device characteristics. Pinheiro et al.\cite{pinheiro20198} use the number of bytes transmitted over a one-second window to identify devices and events.  Bai et al.\cite{8638232} implement continuous overlapping windows (5 minutes) to eliminate errors, finding that larger windows enhance classification performance. However, when the time window is greater than 8 minutes, it will not significantly affect the performance of the model. Due to the significant differences in traffic volume caused by the diversity of CIoT devices, it is necessary to set different time windows for each device type. 

\subsubsection{Traffic Representation}
For the TML algorithm, traffic is represented as a vector containing various features. Different neural networks (NNs) require distinct traffic representations for DL frameworks. 
For example, traffic can be viewed as time-series data and input into recurrent neural networks (RNNs). For graph neural networks (GNNs), the input can be a subgraph of network traffic, such as a communication graph. 
For non-ML algorithms, the traffic is typically represented as hash value~\cite{9246572} or a state transition graph~\cite{duan2023iota}.

% Table generated by Excel2LaTeX from sheet '数据集整理'
\begin{table*}
  \centering
  \caption{The summary of existing datasets}
  \label{dataset}
  \resizebox{\textwidth}{!}{
    %\begin{tabular}{ccccccccccccccc}
    \begin{tabular}{@{\hspace{10pt}}c@{\hspace{10pt}}c@{\hspace{10pt}}c@{\hspace{10pt}}c@{\hspace{10pt}}c@{\hspace{10pt}}c@{\hspace{10pt}}c@{\hspace{10pt}}c@{\hspace{10pt}}c@{\hspace{10pt}}c@{\hspace{10pt}}c@{\hspace{10pt}}c@{\hspace{10pt}}c@{\hspace{10pt}}c@{\hspace{10pt}}c@{\hspace{10pt}}}
    \toprule
    \specialrule{0.05em}{1.5pt}{3pt}
    % \hline
    \multirow{2}[1]{*}{\textbf{Name}} & \multirow{2}[1]{*}{\textbf{Area\textsuperscript{1}}} & \multirow{2}[1]{*}{\textbf{Source\textsuperscript{2}}} & \multirow{2}[1]{*}{\textbf{Categories\textsuperscript{3}}}& \multicolumn{2}{c}{\textbf{Number}} & \multicolumn{2}{c}{\textbf{Communication\textsuperscript{4}}} & \multirow{2}[1]{*}{\textbf{Period}} & \multirow{2}[1]{*}{\textbf{Size}} & \multirow{2}[1]{*}{\textbf{Time}} & \multicolumn{4}{c}{\textbf{Lifecycle\textsuperscript{5}}}  \\
\cmidrule(lr){5-8}\cmidrule(lr){12-15}          &       &   &    & \multirow{1}{*}{\textbf{IoT}} & \multirow{1}{*}{\textbf{N-IoT}} & \multirow{1}{*}{\textbf{Wi-Fi}} &  \multirow{1}{*}{\textbf{Low-energy}} &       &       &       & \multirow{1}{*}{\textbf{SU}} & \multirow{1}{*}{\textbf{ID}} & \multirow{1}{*}{\textbf{IR}} & \multirow{1}{*}{\textbf{DE}} \\
    % \hline
    \specialrule{0.05em}{1.5pt}{3pt}
    \multicolumn{1}{c}{Ours~\cite{osti_10314043}}  & US & SC  & 10 &    8   &    3   &   $\checkmark$   &   $\times$    &   2020.3    &    11.5GB    & 11 days &    $\times$   &    $\checkmark$   &    $\checkmark$   &    $\times$   \\
    \specialrule{0em}{2pt}{2pt}
    \multicolumn{1}{c}{YourThings~\cite{alrawi2019sok}}   &    US   & SC   & 15  &    46   &    0   &    $\checkmark$   &    $\checkmark$   &    2018.3   &      233GB   & 13 days &    $\times$   &    $\times$   &    $\checkmark$   &    $\times$   \\
    \specialrule{0em}{2pt}{2pt}
    \multicolumn{1}{c}{IoTDNS~\cite{9230403}}      &    US   & SC  &  28 &  53      &    12   &    $\checkmark$   &    $\checkmark$   &    2019.8   &   366MB    & 2 months &   $\times$    &    $\times$   &  $\checkmark$     &    $\times$   \\
    \specialrule{0em}{2pt}{2pt}
    \multicolumn{1}{c}{UNSW~\cite{8440758}}      &    AUS   &   SC  &  17  &   28    &    3   &    $\checkmark$   &    $\checkmark$   &   2016.10   &   9.72GB   &  6 months  &   $\times$   &   $\checkmark$   &    $\checkmark$   &   $\times$   \\
    \specialrule{0em}{2pt}{2pt}
    \multicolumn{1}{c}{BoT-IoT~\cite{koroniotis2019towards}}      &    AUS   &  SL  & 5 &    5    &    0   &   -  &    -   &   2018.4    &    69.3GB   &   2 months   &   $\times$   &   $\checkmark$   &    $\checkmark$   &   $\times$   \\
    \specialrule{0em}{2pt}{2pt}
    \multicolumn{1}{c}{Mon(IoT)r~\cite{10.1145/3355369.3355577}}      &    US\&UK   &   SC  &  15 &  81  &   0    &    $\checkmark$   &   $\checkmark$    &   2018.9   &   12.9GB   &  -  &   $\times$   &   $\checkmark$   &     $\checkmark$  &   $\times$   \\
    \specialrule{0em}{2pt}{2pt}
    \multicolumn{1}{c}{PingPong~\cite{trimananda2020packet}}      &    US   &   SC  & 12 & 19  &    3    &    $\checkmark$   &   $\checkmark$    &   2019    &   40.3GB   &   51 days   &  $\times$  &   $\times$   &   $\checkmark$   &  $\times$   \\
    \specialrule{0em}{2pt}{2pt}
    \multicolumn{1}{c}{HomeSnitch~\cite{campos2021towards}}      &   US    &  SC  &  13  &    57    &    0   &   $\checkmark$    &   $\checkmark$    &   2021.3   &   595MB   &  8 days  &   $\times$   &   $\times$   &    $\checkmark$   &   $\times$   \\
    \specialrule{0em}{2pt}{2pt}
    \multicolumn{1}{c}{IoT\_Sentinel~\cite{7980220}}      &   FI    &  SC  &  6  &    31    &    0   &    $\checkmark$   &    $\checkmark$   &   2016   &   61.4MB   &  -  &   $\checkmark$   &   $\times$   &  $\times$   &  $\times$  \\
    \specialrule{0em}{2pt}{2pt}
    \multicolumn{1}{c}{IoT\-23~\cite{bhandari2023distributed}}      &    CZ   &   SC  & 3 &   3  &  0    &    $\checkmark$   &    $\times$   &   2018    &   21GB   &   1 year   &  $\times$  &    $\times$  &   $\checkmark$   &   $\times$     \\
    \specialrule{0em}{2pt}{2pt}
    \multicolumn{1}{c}{N-BaIoT~\cite{meidan2018n}}  &   IL    &  SC  & 3 &    9     &    0   &   -    &    -   &   2018.3   &   240GB   &  -  &   $\times$   &   $\times$   &   $\checkmark$    &   $\times$   \\
    \specialrule{0em}{2pt}{2pt}
    \multicolumn{1}{c}{IoT Inspector~\cite{huang2020iotinspector}}      &    -   &  CR  &  -  &    65000+    &   -    &  -     &    -   &   2019.4   &   -   &  -  &   -   &   -   &   -    &   -   \\
    \specialrule{0em}{2pt}{2pt}
    \multicolumn{1}{c}{NSL-KDD~\cite{tavallaee2009detailed}}      &    US   &  SL  &  -  &    -    &   -    &  -     &    -   &   1998.5   &   4.06MB   &  7 weeks  &   $\times$   &   $\checkmark$   &    $\checkmark$   &   $\times$  \\
    
    \specialrule{0.05em}{2pt}{1.5pt}
    \bottomrule
    
    \end{tabular}}

    \begin{footnotesize}
    \begin{itemize}
      \item[1] ``US'' is the United States, ``UK'' is the United Kingdom, ``AUS'' is Australia, ``FI'' is Finland, ``CZ'' is Czech Republic, ``IL'' is Israel. 
      \item[2] ``SC'' stands for self-collection, ``CR'' is crowdsourcing, ``SL'' is devices simulation.
      \item[3] These datasets consist of a total of 58 types of devices, including IoT devices (smart speakers, TVs, doorbells, various sensors, etc.) and non-IoT devices (mobile phones, laptops, and game consoles, etc.).
      \item[4] ``Wi-Fi'' means the devices using WiFi protocol, ``Low-energy'' refers to the devices using the low-energy protocol like Bluetooth, ZigBee, and Z-Wave.
      \item[5] ``SU'' is setup, ``ID'' means idle, ``IR'' means interaction, ``DE'' is deletion.
    \end{itemize}
    \end{footnotesize}    
\vspace{-1.0em}
    
\end{table*}

\subsection{CIoT Traffic Analysis Algorithm}

\label{sec_systematic_algorithms}

This section summarizes the algorithms used for analyzing CIoT traffic. We found that the algorithms employed in CIoT traffic analysis are quite similar to those used in general traffic analysis. We summarize them as follows. 

\subsubsection{Machine Learning Algorithms}
ML algorithms analyze input data to identify relationships and dependencies within datasets~\cite{ML2021}. ML algorithms can be classified into TML, DL, and RL (briefly introduced in Section~\ref{Sec_background_traffic}). 
Additionally, the Federated Learning (FL) algorithm is used in scenarios involving multiple computational nodes. FL algorithm ensures user privacy by training models locally at each node and sharing only model updates instead of raw data. 

TML algorithms are advantageous due to their robustness and interpretability~\cite{carvalho2019machine}, making them valuable tools. 
Commonly used TML algorithms for CIoT traffic classification include decision trees (DT)~\cite{97458}, support vector machines (SVM)~\cite{SVM}, random forests (RF)~\cite{RF}, and $k$-nearest neighbor ($k$-NN)~\cite{KNN}.

DL algorithms can directly learn complex feature representations from raw data, making them particularly effective for processing large datasets and extracting valuable features from traffic. 
Key DL algorithms include convolutional neural networks (CNNs), which are crafted for the analysis of visual data \cite{CNN}; GNNs, designed to handle graph-structured data~\cite{GNN}; RNNs, which model sequential data~\cite{lipton2015critical}; Notably, long short-term memory (LSTM) Networks are variants of RNNs that address the vanishing gradient problem and allow for the modeling of long-term dependencies in sequential data~\cite{LSTM}.

RL~\cite{RL} is a type of ML that enables an agent to learn and perform tasks by interacting with its environment, receiving feedback through rewards or penalties based on its actions. 
However, traditional RL is not widely used for CIoT traffic analysis due to the time-consuming process of finding optimal solutions while exploring large state-action space and the challenges of the exploration-exploitation tradeoff~\cite{RLprob}.
To address these issues, deep reinforcement learning (DRL)~\cite{DRL} algorithms utilize advanced techniques, such as artificial neural networks, to handle high-dimensional and continuous state and action spaces. 
For example, Deep Q-Networks (DQNs)~\cite{mnih2013playing} use deep neural networks as function approximators to estimate value or policy functions.

\subsubsection{Non-Machine Learning Algorithms}
\revise{Although ML algorithms are inherently adaptive, they are prone to overfitting when the training data is limited in size. Furthermore, since model training often requires considerable time, changes in the environment necessitate model retraining, which may lead to delays in detection ~\cite{charyyev2020iotlocality}.} 
Consequently, some researchers opt for non-ML algorithms. Initially, basic rule-based methods~\cite{roesch1999snort} and signature-based methods~\cite{kumar2012signature} played important roles. 

Subsequently, more advanced non-ML analysis methods were developed. Locality-sensitive hashing (LSH) is a technique for quickly finding similar items in a large dataset. It maps each item to a hash value and uses a family of hash functions to group items with similar hash values. 
This approach is particularly relevant in identifying CIoT devices, as demonstrated by  Charyyev and Gunes~\cite{9246572} and Charyyev
and Gunes~\cite{charyyev2020iotlocality}.
Additionally, traffic can be modeled as a state machine for analysis, which has proven efficient in network intrusion detection systems (NIDS). For example,  Duan et al.
\cite{duan2023iota} constructed CIoT packet-level automaton to profile traffic patterns.  

Finally, the algorithm should be evaluated for its performance and effectiveness in solving a particular problem or task. Several commonly used metrics are employed to measure performance, as introduced in Table \ref{table:Eva}. Some of the most commonly used metrics are accuracy Precision \& Recall, which intuitively show the improvement of researchers' work.

\begin{table}
% \footnotesize
\centering
\renewcommand{\arraystretch}{1.2}
\caption{Evaluation metrics}
\label{table:Eva}
\newcolumntype{M}[1]{>{\centering\arraybackslash}m{#1}}

\begin{tabular}{c|M{20.5em}}
% \hline
\toprule
\specialrule{0.05em}{1.5pt}{0pt}
\textbf{Metrics} & \textbf{Detail} \\
\hline
\textbf{Accuracy} & $(P_{t}  + N_{t} ) / (P_{t}  + N_{t}  + P_{f}  + N_{f} )$  \\
\hline
\textbf{Precision \& Recall} & Precision=$P_{t} / (P_{t} + P_{f})$

Recall=$P_{t} / (P_{t} + N_{f})$  \\
\hline
\textbf{F1 score} & $2 *$ Precision $*$ Recall $/$ (Precision $+$ Recall)  \\
\hline
\textbf{ROC \& AUC} & ROC curve visually plots the true positive rate against the false positive rate at various classification thresholds. AUC represents the overall performance of the model by calculating the area under the ROC curve  \\
% \hline

\specialrule{0.05em}{0pt}{1.5pt}
\bottomrule
\end{tabular}
\begin{footnotesize}

\centering
    
      $P_{t}$: True positive example;
$P_{f}$: False positive example;

$N_{t}$: True negative example;
$N_{f}$: False negative example.

    \end{footnotesize} 
\vspace{-2.0em}

\end{table}

\subsection{New Characteristics in CIoT Traffic Analysis}

\label{sec_systematic_compare}

The traffic analysis process of CIoT and other fields (e.g., PC website traffic) share moderate similarities. 
However, the unique features of CIoT have led to numerous attempts to customize and improve the analysis process at every step to achieve specific application goals. In this subsection, we summarize the challenges faced by CIoT traffic analysis and its unique characteristics compared to network traffic, as shown in Table~\ref{table_IoTandNIoT}). 

\subsubsection{Traffic Collection}
\label{sec_systematic_newCharacteristics_trafficcollection}
The collection of CIoT traffic data is more complex than that of PC or mobile apps, as summarized in the following aspects.
First, there are many types of CIoT devices, each differing in hardware and software design, making collecting substantial training data a significant challenge~\cite{feng2018acquisitional}.
Second, CIoT devices typically have diverse interaction modes, complicating the automation of traffic collection. 
Third, traffic analysis must consider the features of various communication technologies (such as Wi-Fi, BLE, Zigbee, Z-Wave, LoRa, and NB-IoT) and use the appropriate receivers to capture the packets in different layers. 
Finally, different lifecycle phases of devices exhibit distinct traffic patterns. For example, during the setup phase, a device may engage in numerous TLS key negotiations and domain name requests, whereas, in the idle state, it typically sends only heartbeat packets to maintain connections. Each phase requires specific user configurations that are challenging to automate. 

In summary, the complexity of collecting CIoT traffic leads some researchers to prefer using public datasets rather than creating their own. 

% Table generated by Excel2LaTeX from sheet 'IoT和N-IoT'
% \begin{table}
%   \centering
%   \caption{The comparison of IoT traffic and network traffic}
%     \begin{tabular*}{0.3\textwidth}{|c|c|c|}
%     \hline
%     \multicolumn{1}{|c|}{Items} & \multicolumn{1}{c|}{\textbf{IoT}} & \multicolumn{1}{c|}{\textbf{N-IoT}} \\
%     \hline
%     \multicolumn{1}{|c|}{\textbf{Data scale}} &       &  \\
%     \hline
%           &       &  \\
%     \hline
%           &       &  \\
%     \hline
%     \end{tabular*}%
%   \label{IoTandNIoT}%
% \end{table}%

% \begin{table}
%   \centering
%   \caption{The comparison of IoT traffic and network traffic}
%   \begin{tabular}{|c|c|c|} % 直接使用 tabular 环境
%   \hline
%   \multicolumn{1}{|c|}{Items} & \multicolumn{1}{c|}{\textbf{IoT}} & \multicolumn{1}{c|}{\textbf{N-IoT}} \\
%   \hline
%   \textbf{Data scale} &      &  \\
%   \hline
%         &       &  \\
%   \hline
%         &       &  \\
%   \hline
%   \end{tabular}%
%   \label{IoTandNIoT}%
% \end{table}%

\begin{table*}
  \centering
  \renewcommand{\arraystretch}{1}
  % \vspace{-20pt} % 减少表格与正文之间的间距
  \caption{Comparison of CIoT traffic and general network traffic}
  \begin{tabular}{p{4cm} p{3cm} p{4cm}} % 去掉垂直线
  \toprule
  \specialrule{0.05em}{1.5pt}{3pt}
  Items & \textbf{CIoT Traffic} & \textbf{Non-IoT Taffic} \\
  \midrule
  %设备种类：多、少；环境搭建：复杂、相对简单；交互方式：多、少；通信技术：多，少；
  \textbf{Device Type} &  diverse   & simple \\
  \textbf{Protocol} & diverse and customized   & relatively diverse and standard \\
  \textbf{Interaction Mode}  &    complex   & easy \\
  \textbf{Communication Technology}  &   various   & mostly IP-based \\
 \textbf{Traffic Volume} &    small   & large \\
 \textbf{Update Frequency} &    low   & high \\
 \textbf{Available Datasets} &    relatively few   & numerous \\
 \specialrule{0.05em}{1pt}{1.5pt}
 \bottomrule
\end{tabular}
\label{table_IoTandNIoT}
\vspace{-2.0em}
  
\end{table*}

\subsubsection{Traffic Processing}

For traffic processing, the low power requirements and the diverse range of communication protocols, device types, and lifecycle phases of CIoT bring new traffic characteristics for feature extraction.
Firstly, CIoT devices generally have simpler hardware configurations than traditional network computing devices, often resulting in smaller TCP buffer sizes~\cite{9110451}. This limitation directly affects packet transmission and network congestion control mechanisms. 
Additionally, traditional metrics used in network traffic classification, such as the user-agent field in HTTP, may not always be effective for CIoT devices due to their limited use of online web services.
Furthermore, our survey indicates that the set of DNS domains or remote IPs contacted by CIoT devices is a popular feature used by researchers~\cite{8440758, le2019policy, 9230403}. This is because CIoT devices communicate with a limited number of endpoints. 
Significantly, many CIoT device manufacturers develop proprietary application layer protocols and implement encryption based on these protocols. Considering common traffic features that may not behave well, it is essential to precisely capture these unique characteristics. 

Secondly, traffic from CIoT devices varies significantly depending on the type and lifecycle phase. For example, devices like plugs or lamps exchange a few packets, whereas cameras generate large volumes of video and audio data. This variability poses a challenge for statistical feature extraction. 
In contrast, traffic analysis in other fields, like website fingerprinting, generates many packets during short-term visits. That is, the volume of CIoT traffic is caused by a specific event or device type. 
As a result, it is necessary to pay more attention to packet-level features rather than relying solely on flow-level or statistical features commonly used in general network traffic.  

\subsubsection{Algorithm}
The deployment characteristics of CIoT bring some differences in general traffic analysis algorithms.
First, network traffic characteristics change rapidly, necessitating frequent retraining of models to adapt to updates. However, this dynamic is less pronounced in CIoT devices due to their longer firmware update cycles~\cite{10.1145/3560835.3564551}. As a result, CIoT traffic classification models often retain their effectiveness over extended periods. This was confirmed by a study by  Ahmed et al.~\cite{osti_10314043}, who used a 2020 dataset to train fingerprints and tested them using a 2021 dataset. The experiment showed that temporality does not significantly affect the accuracy of device fingerprints. 
Second, the dispersed location, large number, and limited bandwidth of compromised CIoT devices present additional challenges. Specifically, the low bandwidth DDoS attack from CIoT botnet may pose additional difficulty for detection algorithms. 
Third, the low power requirements of CIoT necessitate the use of distributed algorithm. While traditional network traffic classification models are typically executed on personal computers or servers with substantial computational resources, CIoT scenarios often require deployment on gateways or servers. This increases the demand for distributed designs such as FL. 

\subsection{Case Description}
\revise{
Device identification is a typical application scenario in this field. Therefore, to facilitate reader understanding, this subsection outlines the core process of CIoT traffic analysis through a hypothetical case study of device fingerprinting.
}

\revise{
The first step in the analysis process is traffic collection. We assume that we have control over the router, which enables us to capture the traffic between devices and the cloud using port mirroring tools on the internal interface of the NAT router. It is important to note that devices generate different traffic patterns at various stages of their lifecycle. For example, during the setup phase, devices may perform a significant number of TLS handshakes, while in the idle state, they primarily generate periodic heartbeat packets. Firmware updates often involve more intensive data transfers. By collecting traffic from different stages of the device lifecycle, we can build more accurate models of device behavior for subsequent analysis
}
\revise{
Once traffic collection is complete, the next step is traffic processing. We extract packet-level features, flow-level features, and statistical features from the traffic captured at different lifecycle stages, and combine them into feature vectors. Multiple feature vectors are then aggregated to form a complete dataset.
}
\revise{
In the analysis phase, we opt for a machine learning approach for device classification, using algorithms such as Random Forest. The dataset obtained in the previous step serves as the input for constructing device fingerprints. We randomly split the dataset into two groups: one containing 70\% of the ``training'' instances and the other containing 30\% of the ``testing'' instances. After training the model on the training set, we evaluate its performance on the test set using metrics such as accuracy, precision, and recall. Additionally, cross-validation is performed to ensure the robustness of the model, resulting in a classification model with optimal performance.
}

\revise{
Through the process outlined above, it becomes evident that the core of CIoT traffic analysis lies in its ability to flexibly address the challenges posed by device diversity, protocol heterogeneity, and the dynamic nature of application scenarios. 
}

\begin{tcolorbox}[colback=black!5!white, colframe=black!30!white, breakable]

\textbf{Takeaways: }
   This section outlines the process of CIoT traffic analysis and addresses RQ1. 
   We introduced in detail the three basic processes: traffic collection, traffic processing, and analysis, and discussed separately how CIoT traffic differs from general network traffic analysis in these three steps. CIoT devices have various models and communication protocols, involve multiple interactions, and have intricate life cycles. Therefore, different CIoT devices exhibit unique behavior patterns, necessitating targeted traffic processing methods and algorithms for application in real-world scenarios. 
   
\end{tcolorbox}

\section{The Applications in Security and Privacy}
\label{sec_applications}
\begin{figure}[h]
\centering
\includegraphics[width=\linewidth]{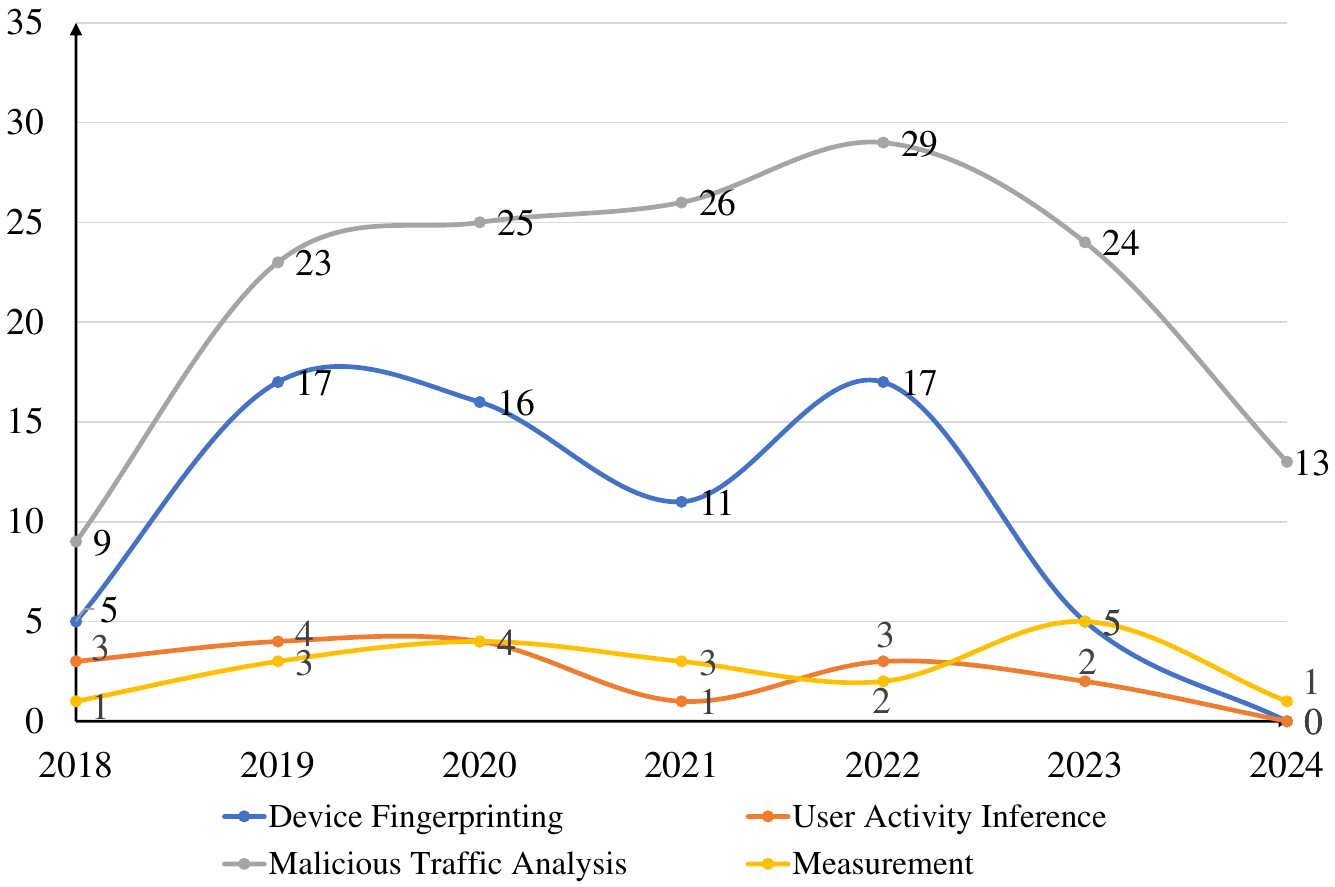}%
\caption{Publication trends of different application purposes}
\label{Fig:apptrend}
\end{figure}

Within this section, \revise{as the focus of our research work, we list the cutting-edge and representative literature in CIoT traffic security and privacy.} We group the current research into four primary categories according to application scenarios or objectives, addressing RQ2. These four categories include device fingerprinting, user activity inference, malicious traffic analysis, and measurement. 
Figure~\ref{Fig:apptrend} displays the publication trends for these application purposes.
\revise{At the end of this section, we summarize our findings by presenting the advantages, limitations, and application scenarios for each application goal and its subcategories.}

\subsection{Device Fingerprinting}

\label{sec_applications_devicesFingerprinting}

\begin{figure}[htbp]
\centering
\includegraphics[width=\linewidth]{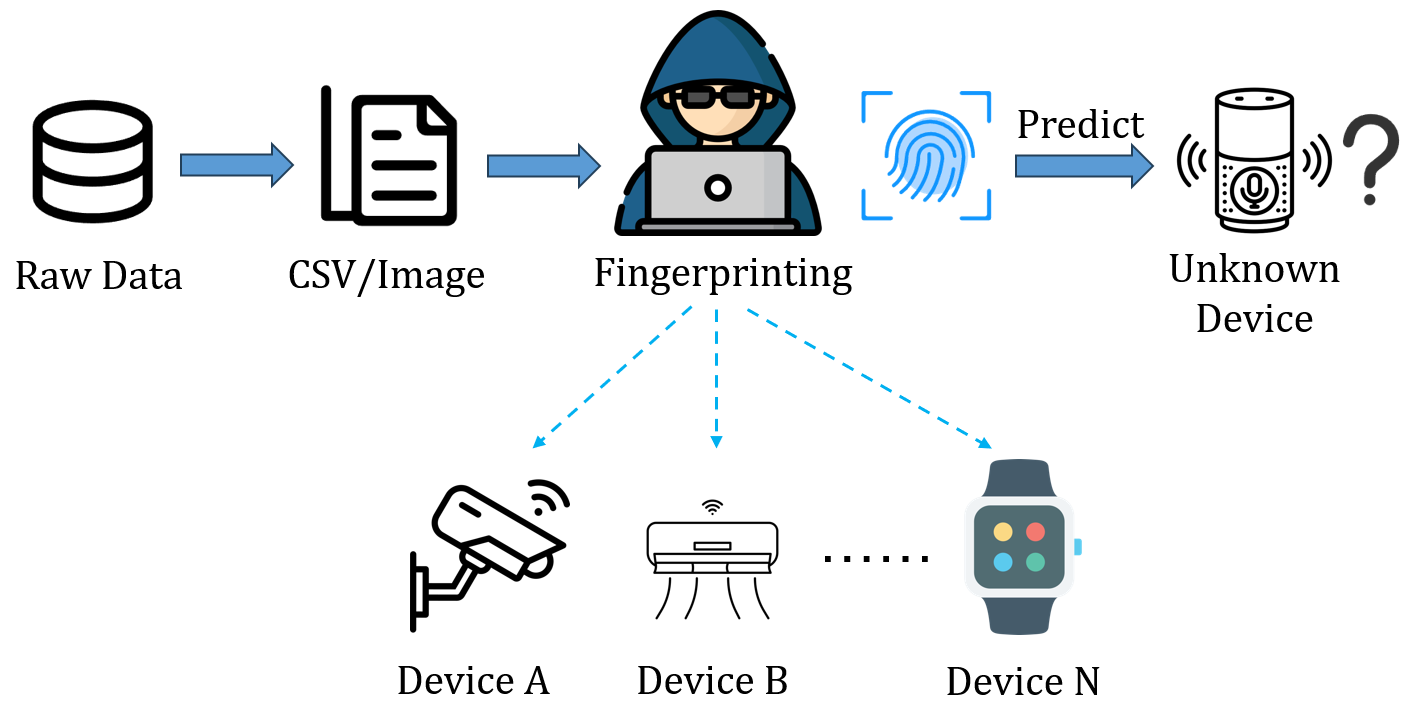}
\caption{The basic process of device fingerprinting}
\label{Fig:devicesfingerprinting}
\vspace{-2.0em}
\end{figure}

Different types, vendors, and behaviors of CIoT devices generate traffic with unique characteristics, which can be used to uniquely identify devices and their behaviors, just like a fingerprint.
The general process for constructing device fingerprints is shown in Figure~\ref{Fig:devicesfingerprinting}.
% As depicted, attackers can extract fingerprints from raw packets to predict device models and behaviors, thereby identifying hidden devices. 
\revise{As shown in the figure, traffic from devices is processed and fed into the model to train a fingerprint of the different devices, which can then be used to identify whether a segment of an unknown traffic pattern belongs to a device or not. }
% 如图所示，攻击者可以从原始数据包中提取指纹以预测设备模型和行为，从而识别隐藏的设备。
% 如图所示，将设备的流量处理后输入到模型中，以训练出不同设备的指纹，该指纹随后可被用于识别一段未知的流量模式是否属于某设备
Device fingerprinting can be further classified into three categories: device identification, device behavior identification, and hidden device detection.

\subsubsection{Device Identification}
\label{Sec:Sub-DI}
Device identification involves using unique traffic patterns to identify devices, such as their vendors and types or categories. This information facilitates fingerprinting attacks: attackers use the information to discover vulnerable targets. On the other hand, it also can help regulators recognize these vulnerable devices.
Considering that there are a large number of works in this field and the contributions of researchers are clearly distinguished at different stages in the CIoT traffic analysis framework (described in Section~\ref{sec_AnalysisProcedure}), in this part we classify the literature based on their contribution.

\vspace{3pt}\noindent\textbf{\textcircled{1}Traffic Collection.} 
For the traffic collection step, some work in the device identification field enriches the dataset. Specifically, they consider the devices that use low-energy protocols, certain kinds of devices (smart TVs or smartwatches), and the number of devices in the dataset. 

Certain research~\cite{9149285, 9548663, 9832419} endeavors to tackle these issues and integrate devices that use low-energy protocols into their studies. Babun et al.\cite{9149285} asserted themselves as the pioneering work that investigates the Zigbee and Z-Wave device fingerprinting framework. They constructed a density distribution based on inter-arrival time (IAT) by capturing packets at the link layer, which is divided into 300 equal intervals to build the signature. 
They conducted tests on 39 popular Zigbee and Z-Wave devices, resulting in accuracy rates of 91.2\% for Zigbee and 93.6\% for Z-Wave, respectively.
Considering the asymmetry of learning and testing by ISPs during device identification, Ma et al.\cite{9548663} monitored inbound and outbound packets and extracted Spatial-Temporal features to identify these devices that share a common IP (behind a NAT) from the ISP's perspective. The protocols used by the devices include Bluetooth, Zigbee, or LoRa. Kostas et al. 
\cite{9832419} used the entropy of the payload as the feature, which also allows them to identify devices with non-IP and low-power protocols. 

Some researchers~\cite{8299447, varmarken2022fingerprintv} focused on edge IoT devices such as smart wearables and smart TVs. Aksu et al.
\cite{8299447} focused on fingerprints of wearable devices using the Bluetooth protocol based on Bluetooth packet characteristics. The algorithm utilizes the inter-arrival time of packets as a feature and can automatically select the optimal solution from over 20 classifiers. 
In light of the advertising tracking and data leakage issues associated with smart TVs, Varmarken et al.\cite{varmarken2022fingerprintv} extracted application fingerprints based on domains, data packets, and TLS information. The process relies only on a few packets, making their method lightweight and applicable to encrypted traffic.

Ahmed et al. \cite{osti_10314043} unprecedentedly considered a remarkable number of 188 devices. The experiment integrated six public datasets along with a self-collected dataset (the ``Ours'' dataset). They innovatively considered five different fingerprinting granularities: device instances; devices have unique make and model; devices have the same manufacturer and type; devices have the same manufacturer; devices have the same device category. Employing RF as the classifier, its accuracy surpasses 97\% in all five cases.

Moreover,  Bremler-Barr et al. \cite{9110451} expanded the dataset to non-IoT devices. They designed a multi-stage classifier to distinguish IoT and non-IoT devices. This offers insights for extracting background traffic (i.e., traffic from non-IoT devices like smartphones and PCs) in CIoT traffic analysis. 

\vspace{3pt}\noindent\textbf{\textcircled{2}Traffic Processing. } 
%*****Unique feature processing method*****
Part of the works~\cite{8116438, 8664655} adopted personalized feature processing methods. Sivanathan et al. 
\cite{8116438} made the first systematic study on smart device identification. They collected three-week traffic from 20 CIoT devices and extracted 11 distinct features by observing device activity patterns. They divided the values of each feature into 5 ranges called cluster bins to distinguish different devices. This method achieved a 95\% accuracy. 
However, as the number of devices increases, only 11 features with 5 gradations become insufficient. Marchal et al. 
\cite{8664655} divided network traffic into multiple time-series ``flows'', which are defined as a collection of packets using a given MAC address and protocol. They then computed 33 periodic features obtained from the Discrete Fourier Transform (DFT) of traffic and employed the $k$-NN algorithm for device classification. 

%*****backend infrastructure feature*****
Some researchers~\cite{10.1145/3229565.3229572, 10.1145/3419394.3423650, 9230403} believe that the backend infrastructure for device connections has unique information. Guo and Heidemann
\cite{10.1145/3229565.3229572} utilized unique communication server domain names to label CIoT devices; it can detect devices behind Network Address Translation (NAT) from aggregated traffic. However, distinguishing devices of the same type from the same manufacturer remains challenging.  
Similarly,  Saidi et al. \cite{10.1145/3419394.3423650} identified devices by analyzing the domains and the backend infrastructure IPs and ports they communicate with. 
Likewise, Perdisci et al. \cite{9230403} found that DNS domain names and their corresponding frequencies show significant discrepancies across various devices, providing a basis for using the DNS feature. However, DNS depends on device services; for instance, a TV of one brand equipped with voice services from another can lead to confusion and reduce accuracy.

%*****unique network layer feature*****
In addition, researchers~\cite{7980220, 255244, 9548663} emphasize the importance of considering device lifecycle, protocol, and time information should also be considered.  Miettinen et al. 
\cite{7980220} innovatively considered the setup traffic. They selected 23 features from the first 12 packets during the setup stage to identify the types of new devices and further restricted their communication capabilities based on security levels. Analysis in fewer packets makes their method more lightweight. Yu et al.
\cite{255244} employed BC/MC (Broadcast/Multicast) packet features to identify devices. The features primarily fall into three categories: identifiers that uniquely identify the device model, main protocol fields from BC/MC packets, and auxiliary features acquired by active detection. 
To enhance the distinctiveness of features, Ma et al.\cite{9548663} devised an efficient and scalable system using spatial-temporal traffic fingerprinting. They integrated both the temporal sequence of packets and their spatial correlations across the network, which provides a more comprehensive and accurate depiction of traffic.

%*****link or physical layer feature*****
In addition to the feature of the network layer and above, some works~\cite{10.1145/3320269.3384732, 10.1007/978-3-319-66399-9_14} considered link or physical layer features. Dong et al.
\cite{10.1145/3320269.3384732} incorporated frame length and epoch time in the physical layer as features. Maiti et al.\cite{10.1007/978-3-319-66399-9_14} categorized the devices used into 10 classes, with features including but not limited to frame type, size, arrival time, and rate. The findings revealed instances of confusion between cameras and non-IoT devices like PCs. This indicates that there still are great challenges in utilizing link layer frames as the optimal distinguishing features.

\begin{table*}[h!]
  \centering
  \renewcommand\arraystretch{1}
  \caption{Summary of device identification literature}
  \resizebox{\textwidth}{!}{
  \begin{tabular}{|c|c|c|c|c|c|c|c|c|c|c|c|c|c|}
    \hline
    \multirow{2}[3]{*}{\textbf{Literature}} & \multirow{2}[3]{*}{\textbf{Year}} & \multicolumn{3}{c|}{\textbf{Contributions}} & \multirow{2}[3]{*}{\textbf{Feature}} & \multicolumn{2}{c|}{\textbf{Algorithm\textsuperscript{2}}} & \multicolumn{2}{c|}{\textbf{Datasets Source}} & \multicolumn{2}{c|}{\textbf{Communication}} & \multirow{2}[3]{*}{\textbf{\makecell{Collection \\ Location\textsuperscript{4}}}} \\
\cline{3-5}\cline{7-8}\cline{9-12}          &       & \textbf{Algorithm\textsuperscript{1}} & \textbf{Feature} & \textbf{Dataset} &       &    \textbf{Type\textsuperscript{3}}   &    \textbf{Name}   & \multicolumn{1}{c|}{\textbf{\makecell{Public \\ Datasets}}} & \multicolumn{1}{c|}{\textbf{\makecell{Self- \\ collection}}} & \textbf{Wi-Fi} & \textbf{Low-energy} & \\
    \hline
    
        \cite{10.1145/3019612.3019878}   & 2017  & $\surd$     & - & - & Flow & TML    & \makecell{GBM, RF, \\XGBoost} & - &    $\surd$   &  $\surd$   & - & $M_1$ \\  \hline
        \cite{10.1007/978-3-319-66399-9_14} & 2017 & - & $\surd$ & - & Packet & TML & \makecell{DT, RF, \\SVM} & - & $\surd$ & $\surd$ & - & $M_3$  \\ \hline
        \cite{7980220} & 2017 & - & $\surd$ & $\surd$ & Packet & TML & RF & - & $\surd$ & $\surd$ & $\surd$ & $M_1$  \\ \hline
        \cite{8116438} & 2017 & - & $\surd$ & - & Statistics & TML & RF & - & $\surd$ & $\surd$ & - & $M_1$  \\ \hline
        \cite{10.1145/3229565.3229572} & 2018 & - & $\surd$ & - & Packet & NML & - & - & $\surd$ & $\surd$ & $\surd$ & $M_4$  \\ \hline
        \cite{8638232} & 2018 & $\surd$ & - & - & DL & DL & LSTM & \makecell{Arunan et.al.'s \\IoT campus dataset} & - & $\surd$ & - & $M_1$  \\ \hline
        \cite{8538630} & 2018 & - & $\surd$ & - & Statistics & TML & RF & \makecell{Arunan et.al.'s \\IoT campus dataset} & - & $\surd$ & - & $M_1$  \\ \hline
        \cite{feng2018acquisitional} & 2018 & $\surd$ & - & - & - & NML & Apriori &  \makecell{dataset from the National\\Vulnerability Database} & $\surd$ & - & - & $M_1$  \\ \hline
        \cite{10.1145/3302505.3310073} & 2019 & $\surd$ & - & - & DL & DL & LSTM &  \makecell{UNSW,  a North America \\private lab's dataset}  & - & $\surd$ & - & $M_1$  \\ \hline
        \cite{8885429} & 2019 & $\surd$ & - & - & \makecell{Packet, \\Statistics} & TML & \makecell{RF, \\Extra-Trees, \\AdaBoost} & \makecell{UNSW,\\ Arunan et.al.'s \\IoT campus dataset} & - & $\surd$ & - & $M_1$  \\ \hline
        \cite{pinheiro20198} & 2019 & $\surd$ & - & - & Statistics & TML & \makecell{$k$-NN, RF,\\DT, SVM,\\ Majority Voting} & \makecell{Arunan et.al.'s \\IoT campus dataset} & $\surd$ & $\surd$ & $\surd$ & $M_1$  \\ \hline
        \cite{8437128} & 2019 & $\surd$ & - & - & \makecell{Packet, \\ Flow, \\Statistics}  & TML & $k$-means & - & $\surd$ & $\surd$ & $\surd$ & $M_2$  \\ \hline
        \cite{8440758} & 2019 & $\surd$ & - & $\surd$ & \makecell{Packet, \\Flow} & TML & \makecell{RF, \\ Naive Bayes} & - & $\surd$ & $\surd$ & $\surd$ & $M_1$  \\ \hline
        \cite{8761559} & 2019 & - & $\surd$ & - & Packet & TML & \makecell{J48 DT, \\OneR, PART} & M. Miettinen et.al.'s dataset & - & $\surd$ & $\surd$ & $M_1$  \\ \hline
        \cite{8664655} & 2019 & - & $\surd$ & - & \makecell{Flow, \\Statistics} & TML & $k$-NN & - & $\surd$ & $\surd$ & $\surd$ & $M_1$  \\ \hline
        \cite{charyyev2020iotlocality} & 2020 & $\surd$ & - & - & - & NML & LSH & - & $\surd$ & $\surd$ & - & $M_1$  \\ \hline
        \cite{10.1145/3320269.3384732} & 2020 & $\surd$ & $\surd$ & - & DL & DL & LSTM & - & $\surd$ & $\surd$ & - & $M_1$  \\ \hline
        \cite{9149285} & 2020 & - & - & $\surd$ & Packet & TML & Bayes Net & - & $\surd$ & - & $\surd$ & $M_3$  \\ \hline
        \cite{255244} & 2020 & $\surd$ & $\surd$ & - & DL & DL & Self-designed & - & $\surd$ & $\surd$ & $\surd$ & $M_1$$M_4$  \\ \hline
        \cite{9110451} & 2020 & - & - & $\surd$ & Statistics & TML & \makecell{DT, Logistic\\ Regression} & \makecell{Arunan et.al.'s \\IoT campus dataset} & $\surd$ & $\surd$ & $\surd$ & $M_1$  \\ \hline
        \cite{guo2020detecting} & 2020 & - & - & $\surd$ & Packet & NML & - & \makecell{ZMap’s 443-https\\-ssl\_3-full\_ipv4 \\TLS certificate dataset} & $\surd$ & $\surd$ & $\surd$ & $M_1$$M_4$  \\ \hline
        \cite{9148821} & 2020 & $\surd$ & - & - & \makecell{Packet, \\ Flow} & TML & RF & - & $\surd$ & $\surd$ & - & $M_1$  \\ \hline
        \cite{10.1145/3419394.3423650} & 2020 & - & $\surd$ & - & Packet & NML & - & - & $\surd$ & $\surd$ & - & $M_2$$M_4$  \\ \hline
        \cite{9230403} & 2020 & $\surd$ & $\surd$ & - & Packet & NML & TF-IDF & - & $\surd$ & $\surd$ & - & $M_1$$M_4$  \\ \hline
        \cite{chakraborty2021cost} & 2021 & - & $\surd$ & - & \makecell{Packet, \\ Flow, \\Statistics} & TML & DT & - & $\surd$ & $\surd$ & $\surd$ & $M_2$  \\ \hline
        \cite{9246572} & 2021 & $\surd$ & - & - & Flow & NML & LSH & \makecell{M. Miettinen et.al.'s Setup,\\ J. Ren et.al.'s Idle} & $\surd$ & $\surd$ & $\surd$ & $M_1$  \\ \hline
        \cite{8299447} & 2021 & - & - & $\surd$ & Packet & TML & RF & - & $\surd$ & - & $\surd$ & $M_1$  \\ \hline
        \cite{varmarken2022fingerprintv} & 2022 & - & - & $\surd$ & Packet & ML & \makecell{Agglomerative \\clustering} & - & $\surd$ & $\surd$ & - & $M_1$  \\ \hline
        \cite{du2022lightweight} & 2022 & - & $\surd$ & - & \makecell{Packet, \\Statistics} & TML & \makecell{RF, \\Extra-Trees} & UNSW & - & $\surd$ & $\surd$ & $M_1$  \\ \hline
        \cite{osti_10314043} & 2022 & - & - & $\surd$ & Statistics & TML & RF & - & $\surd$ & $\surd$ & $\surd$ & $M_1$  \\ \hline
        \cite{9548663} & 2022 & - & $\surd$ & $\surd$ & Packet & DL & CNN & UNSW & $\surd$ & $\surd$ & $\surd$ & $M_1$$M_4$  \\ \hline
        \cite{9832419} & 2022 & - & $\surd$ & $\surd$ & Packet & TML & DT & \makecell{Aalto University's dataset, \\UNSW} & - & $\surd$ & $\surd$ & $M_1$  \\ \hline
        \cite{wanode2022optimal} & 2022 & - & $\surd$ & - & \makecell{Packet, \\Statistics} & TML & RF & - & $\surd$ & $\surd$ & $\surd$ & $M_2$  \\ \hline
    \end{tabular}}
  \label{deviceidentification}%
  \begin{footnotesize}
    \begin{itemize}
      \item[1] ``Algorithm" denotes computational methods with provable improvements in either complexity (time/space) or task performance metrics.
      \item[2] The detail of algorithm of typical device identification paper is shown in section~\ref{Sec:Sub-DI}
      \item[3] ``TML" means traditional machine learning, ``DL" means deep learning, ``NML" means traditional analysis.
      \item[4] ``$M_1-M_5$" corresponds to the five methods to acquire traffic in section~\ref{sec_systematic_trafficcollection}.
    \end{itemize}
  \end{footnotesize}  
\vspace{-1.5em}
\end{table*}%

%*****feature reduction***** 
To save classification costs, some researchers have added feature reduction technologies. We found that Genetic Algorithms (GA) are the most common~\cite{8761559, 9832419}. In the work by  Aksoy and Gunes~\cite{8761559}, 30 chromosomes correspond to the number of feature subsets. Each chromosome comprises a string of 0/1 representing feature selection. The chromosomes are initially filled with 0/1 and then run a fitness function to ascertain the robustness of these features, resulting in an optimal feature subset. Kostas et al. \cite{9832419} first employed the feature-importance-based voting method to eliminate unnecessary features from the initial set. Then, GA is applied to select the most suitable feature subset from the remaining 52 features. 
In addition to GA, Santos et al. \cite{8538630} incorporated the CfsSubsetEval algorithm, which selects subsets from the original feature set with a high correlation with the target variable but a low correlation between features to reduce computational complexity. Wanode et al.
\cite{wanode2022optimal} compared three distinct feature reduction techniques: SVD, PCA, and MI. In the case of classifying 16 CIoT devices, MI performs significantly better than SCD and PCA.
Another part of works designed their own feature reduction algorithm~\cite{chakraborty2021cost, du2022lightweight}. Chakraborty et al.
\cite{chakraborty2021cost} emphasize the varying costs associated with different features during the extraction process. As a result, they devised a cross-entropy-based algorithm to tackle this concern. Similarly, Du et al.\cite{du2022lightweight} built upon NSGA-\uppercase\expandafter{\romannumeral3}, introduced concepts like symmetric uncertainty and correlation coefficient. They propose multiple objective functions that reduce feature dimensions and filter effective features.

\vspace{3pt}\noindent\textbf{\textcircled{3}Algorithm Analysis.} 
Early studies tended to favor TML algorithms. Gradually, researchers began considering constructing more complex identification frameworks based on basic classification models (e.g., SVM, $k$-NN, RF) or even neural networks. 
Meanwhile, with the development of edge computing, researchers began to consider distributed models. 
Apart from these approaches, a few studies employ non-ML methods that effectively shorten the calculation time and are very suitable for scenarios with high real-time requirements.

%*****traditional machine learning*****
Early studies tended to favor TML algorithms~\cite{10.1145/3019612.3019878, pinheiro20198, 9148821, 7980220, 8538630, 8116438, 8440758}. 
In 2017, Meidan et al.\cite{10.1145/3019612.3019878} trained a multi-stage meta classifier. The first stage differentiates IoT from non-IoT devices, and the second stage identifies specific device categories. However, the granularity of this approach only reaches device types (such as TVs, printers, motion sensors, etc.). 
Similarly, Pinheiro et al.\cite{pinheiro20198} demonstrated that using only packet statistics features, the RF algorithm outperforms $k$-NN, DT, SVM, and Majority Voting, achieving an accuracy of 96\% in device identification. 
To handle the frequent addition of new devices,  Ammar et al. \cite{9148821} constructed a binary RF classifier for each device. This approach eliminates the need to retrain the entire model whenever new devices are added.

%Integrate classifiers 
Gradually, some studies have considered constructing more complex algorithms based on basic models~\cite{8885429, 8440758}.Msadek et al.
\cite{8885429} emphasized the reduction of training data and the elimination of manual tuning. This is achieved by introducing a novel sliding window technique that dynamically segments traffic. As the activity of relevant traffic varies, the window automatically expands; otherwise, it contracts to discard irrelevant packets. 
To achieve higher accuracy, Sivanathan et al\cite{8440758} collected traffic from 28 CIoT devices and constructed a multi-stage classifier. The first stage employed a Naive Bayes Multinomial classifier, taking the bag of remote port numbers, domain names, and cipher suites as input. The bag comprises values and their corresponding frequencies in a matrix format. Then, class and confidence for the bag, flow volume, and flow rate were used as inputs for the second stage. The third stage further leveraged an RF classifier to determine categories and confidence scores. This architecture achieved an impressive device recognition accuracy of 99\%. 

%*****deep learning*****
Subsequently, there were studies using DL algorithms\cite{8638232, 10.1145/3302505.3310073, 255244}.  Bai et al.
\cite{8638232} constructed an LSTM-CNN cascade model to classify 4 device categories (hubs, Electronics, Cameras, Switches \& Triggers). However, while the algorithm performed well in binary classification, its accuracy dropped to 74.8\% in the four-class problem. Ortiz et al. 
\cite{10.1145/3302505.3310073} employed autoencoders to automatically learn features from traffic and probabilistically model devices as distributions of traffic classes. Yu et al.
\cite{255244} innovatively developed a novel multi-view wide and deep learning (MvWDL) framework. The six views constructed in the experiments are derived from the devices' BC/MC protocols. 
Meanwhile, they devised a hybrid-fusion multi-view artificial NN to achieve view fusion.

Although the above approaches achieve high accuracy in their designated scenarios, deploying the aforementioned algorithms at one network node presents challenges in scalability. Thangavelu et al.\cite{8437128} developed a Distributed Device Fingerprinting Technique (DEFT) to tackle this challenge. By using SDN technology, the DEFT controller maintains information while gateways perform classification. While robust and manageable, DEFT is not lightweight, necessitating the collaboration of multiple distinct gateways to maximize system efficacy.

The above ML methods consume massive computing resources in practical implementation. Therefore, a novel approach based on LSH was proposed by Charyyev and Gune~\cite{9246572, charyyev2020iotlocality}. This approach eliminates the need for feature extraction and doesn't require frequent model updates. The algorithm employs LSH functions to compute hash values of traffic for device identification, which are stored in a database.  Perdisci et al.
\cite{9230403} analogized the Term Frequency-Inverse Document Frequency (TF-IDF) algorithm from document retrieval to device identification. 
When employing a set composed of DNS request frequencies as recognition features, devices, and the requested domain names are treated as documents and their entries to create TF-IDF vectors. Finally, target device recognition relies on cosine similarity between vectors.

In addition to passive traffic capture and analysis methods, researchers have also adopted active probing techniques. Feng et al.
\cite{feng2018acquisitional} have proposed an innovative method for automatic discovery and annotation of IoT devices, known as ARE. The ARE focuses on the response information of the application layer and establishes a mapping between IoT devices and their official description websites by extracting banner information (usually containing details like device type and model). Compared to traditional network scanning tools like Nmap, ARE has shown superior capabilities in searching for IoT devices. Especially when new devices are connected to the network, ARE can dynamically and quickly learn and update the fingerprint information of new devices. 

\vspace{3pt}\noindent\textbf{Summary:}
We have summarized relevant papers in Table~\ref{deviceidentification}. Most researchers have focused on improving their algorithms and feature extraction techniques, reaching a relatively high accuracy within their datasets. However, practical application scenarios still face a significant issue: traffic characteristics may be confused between devices of the same type but different models~\cite{osti_10314043}. 
Meanwhile, only a small part of the literature focuses on devices using low-energy protocols. Identifying devices that support various protocols (including Zigbee and Bluetooth) will be challenging. 
We also observed that most studies assume that attackers can infiltrate the home router. If traffic is obtained after NAT (traffic collection method $\textcircled{2}$), the effective classification of devices needs more attention from the perspective of ISPs. 

\subsubsection{Device Behavior Identification}
\label{Sec:Sub-DBI}
Different device behaviors could generate different traffic patterns. 
The triggering of device behaviors involves physical control, LAN/WAN control, multimodal interaction, and cloud API control (for more details, refer to Section~\ref{sec_background_consumeriot}). 

\begin{table*}
  \centering
  \renewcommand\arraystretch{1}
  \caption{Summary of device behavior identification literature}
  \resizebox{\textwidth}{!}{
  \begin{tabular}{|c|c|c|c|c|c|c|c|c|c|c|c|c|c|}
    \hline
    \multirow{2}[3]{*}{\textbf{Literature}} & \multirow{2}[3]{*}{\textbf{Year}} & \multicolumn{3}{c|}{\textbf{Contribution}} & \multirow{2}[3]{*}{\textbf{Feature}} & \multicolumn{2}{c|}{\textbf{Algorithm\textsuperscript{2}}} & \multicolumn{2}{c|}{\textbf{Data Source}} & \multicolumn{2}{c|}{\textbf{Communication}} & \multirow{2}[3]{*}{\textbf{\makecell{Collection \\ Location\textsuperscript{4}}}} \\
\cline{3-5}\cline{7-8}\cline{9-12}          &       & \textbf{Algorithm\textsuperscript{1}} & \textbf{Feature} & \textbf{Dataset} &       &    \textbf{Type\textsuperscript{3}}   &    \textbf{Name}   & \multicolumn{1}{c|}{\textbf{\makecell{Public \\ Datasets}}} & \multicolumn{1}{c|}{\textbf{\makecell{Self- \\ collection}}} & \textbf{Wi-Fi} & \multirow{1}{*}{\textbf{Low-energy}} & \\
    \hline
        \cite{9110451} & 2019 & $\surd$ & - & - & Packet & TML & \makecell{$k$-NN,RF, \\ DT,SVM, \\Majority Voting} & $\surd$ & $\surd$ & $\surd$ & $\surd$ & $M_1$  \\ \hline
        \cite{oconnor2019homesnitch} & 2019 & $\surd$ & - & - & \makecell{Flow, \\Statistics} & TML & RF & YT & $\surd$ & $\surd$ & $\surd$ & $M_1$$M_4$  \\ \hline
        \cite{acar2020peek} & 2020 & $\surd$ & $\surd$ & - &\makecell{Packet, \\Statistics} & TML & RF & - & $\surd$ & $\surd$ & $\surd$ & $M_1$$M_2$$M_3$  \\ \hline
        \cite{trimananda2020packet} & 2020 & $\surd$ & $\surd$ & - & Packet & TML & DBSCAN & \makecell{UNSW, \\ YT, UNB Simulated \\Office-Space Traffic dataset} & $\surd$ & $\surd$ & $\surd$ & $M_2$$M_3$  \\ \hline
        \cite{charyyev2020iot} & 2020 & $\surd$ & - & $\surd$ & Statistics & ML & - & J. Ren et.al.'s dataset & - & $\surd$ & $\surd$ & $M_1$  \\ \hline
        \cite{gu2020iotgaze} & 2020 & $\surd$ & - & - & Packet & TML & RF & - & $\surd$ & - & $\surd$ & $M_3$  \\ \hline
        \cite{duan2021pinball} & 2021 & $\surd$ & $\surd$ & - & Packet & NML & - & \makecell{PingPong, Mon(IoT)r, \\UNSW, YT, \\CICIDS2017 dataset} & - & $\surd$ & $\surd$ & $M_2$$M_3$  \\ \hline
        \cite{wan2021iotathena} & 2022 & $\surd$ & $\surd$ & - & Packet & NML & - &  J. Ren et.al.'s dataset & $\surd$ & $\surd$ & $\surd$ & $M_1$  \\ \hline
        \cite{shafqat2022zleaks} & 2022 & - & - & $\surd$ & Packet & NML & - & Zigator CRAWDAD dataset & $\surd$ & - & $\surd$ & $M_1$$M_3$  \\ \hline
        \cite{Dilawer2023Spying} & 2023 & - & $\surd$ & $\surd$ & Flow & TML & RF & - & $\surd$ & $\surd$ & - & $M_2$  \\ \hline
    
  \end{tabular}}
  \label{DevicesBehaviorFingerprinting}
  \begin{footnotesize}
    \begin{itemize}
      \item[1] ``Algorithm" denotes computational methods with provable improvements in either complexity (time/space) or task performance metrics.
      \item[2] The detail of algorithm of typical device behavior identification paper is shown in section~\ref{Sec:Sub-DBI}
      \item[3] ``TML" means traditional machine learning, ``NML" means traditional analysis.
      \item[4] ``$M_1-M_5$" corresponds to the five methods to acquire traffic in section~\ref{sec_systematic_trafficcollection}.
    \end{itemize}
    \end{footnotesize}  
  \vspace{-1.5em}
\end{table*}%

Early works used statistical features to identify device behavior~\cite{apthorpe2017spying, apthorpe2017smart, pinheiro20198, oconnor2019homesnitch, charyyev2020iot}. Apthorpe et al.
\cite{apthorpe2017spying, apthorpe2017smart} were among the early explorers who investigated the impact of diverse user behaviors on traffic patterns. They observed that user interactions can trigger abrupt changes in traffic behavior. Subsequently, Pinheiro et al. \cite{pinheiro20198} found that devices show different packet length patterns in response to external commands. However, their scope of tested event types remained limited and couldn't distinguish similar behaviors among devices of the same model, such as opening/closing a speaker. 
Therefore,  OConnor et al. \cite{oconnor2019homesnitch} embraced a broader spectrum of device behaviors. They employed 13 features at the transport layer to characterize triplets. 
Similarly, Charyyev and Gunes\cite{charyyev2020iot} also used statistical features. Their contribution lies in evaluating and comparing the performance of 10 ML algorithms in classifying 128 device events stemming from 39 distinct devices.

Compared to previous work,  Trimananda et al.\cite{trimananda2020packet} innovatively used packet-level features for the first time. They employ packet-pair sets between devices and servers to distinguish device behaviors. However, it only works with TCP protocol. Nonetheless, this method inspired subsequent investigations. 
Some researchers draw inspiration from PingPong and address its limitations~\cite{duan2021pinball, wan2021iotathena}. Duan et al. \cite{duan2021pinball} resolved the constraint of being limited to TCP. Devices employing the UDP protocol can successfully extract signatures. Another advantage is that their signatures encompass more encoded information, making the impact of lost packet pairs minimal. Wan et al. 
\cite{wan2021iotathena} introduced a novel time-sensitive subsequence matching technique that generates more comprehensive signatures.

In addition to research on the behaviors of Wi-Fi devices, there is a growing body of work on studying the behaviors of devices using low-energy protocols~\cite{acar2020peek, shafqat2022zleaks, gu2020iotgaze}. Acar et al.
\cite{acar2020peek} proposed a ``multi-stage privacy attack'' that encompasses the recognition of Wi-Fi, BLE, and Zigbee devices. The traffic is represented as a triplet, including timestamp, direction, and packet length, from which statistical features are extracted.  Gu et al.
\cite{gu2020iotgaze} built a vulnerability detection framework called IoTGaze. It constructed wireless context by extracting the packet-level features of the device and identifying dependencies between events. This context is then used to detect anomalies by comparing it with the expected context. IoTGaze has an 98\% accuracy in anomaly detection for 5 types of Zigbee devices. Shafqat et al.
\cite{shafqat2022zleaks} leveraged the low-power characteristics of the Zigbee protocol that message lengths are matched during encryption. It allows inference of application layer (APL) commands from encrypted traffic. Moreover, they found Zigbee devices periodically report attributes like battery levels and temperature. This enables attackers to infer device events from payload lengths and reporting patterns.  

\vspace{3pt}\noindent\textbf{Summary:}
We summarized the papers about device behavior identification in Table~\ref{DevicesBehaviorFingerprinting}. The table shows that packet-level features are evidently more suitable than flow-level features in device behavior identification.
Many researchers draw inspiration from the work by Trimananda et al.~\cite{trimananda2020packet}, employing patterns concealed within request-reply packet pairs to achieve this goal. Notably, DL techniques are rarely used in the context of device behavior identification, which may be related to the dimensions of the sample.

\subsubsection{Hidden Device Detection}
While CIoT devices bring convenience for users, unexpected deployment of the device poses a threat to personal privacy. 
Therefore, some researchers have begun investigating ways to detect hidden IoT devices in unfamiliar environments. Existing approaches relying on radio frequency receivers are not entirely dependable, as they are susceptible to interference from other legitimate RF devices such as smartphones and PCs~\cite{usenix272130}. This situation offers an opportunity for hidden device detection based on network traffic. Taking cameras as an example, visual scenes trigger differences between adjacent frames~\cite{li2016side}, which can be used to confirm the potential cameras that are operational effectively. 
This part presents relevant research that utilizes passive traffic to detect hidden devices in unfamiliar environments. 

Due to concerns regarding unauthorized video recording, some works~\cite{cheng2018dewicam, wu2019yousee} focus on hiding cameras. Cheng et al\cite{cheng2018dewicam} proposed DeWiCam. It automatically analyzes physical and MAC layer data within interested rooms. By exploiting camera compression and fragmentation techniques, DeWiCam can employ differences in data transmission during both transient and stable states as features. However, this method heavily relies on common transmission modes, which may not be generic across different manufacturers and may change with camera firmware updates. Wu and Lagesse\cite{wu2019yousee} have designed a solution for dynamically detecting the presence of uploading cameras. However, it relies on comparing the similarity between user videos and videos uploaded by hidden cameras. The difficulty of detection increases if a camera does not engage in uploading behavior.

Some studies~\cite{singh2021always, sharma2022lumos} have extended the scope of detectable devices.  Singh et al. \cite{singh2021always} leveraged the concept of ``human motion'' from Wu and Lagesse~\cite{wu2019yousee}, which involves activating trustworthy sensor values and observing whether there exist similar traffic patterns from other devices. 
Furthermore, they introduced an innovative sensor coverage technique to locate the detected sensors. 
Sharma et al. \cite{sharma2022lumos} addressed device diversity by extracting device-specific attributes using multiple time scales. They also improved upon previous spectrum sensing methods~\cite{shi2015beyond}, utilizing learned approximate transmission patterns over time to acquire device data transmission timing and channels. Their device fingerprint module computed features through an XGBoost classifier. The channel-aware module identified subsets of active channels through cyclic channel hopping. Lastly, a rough device positioning was achieved through RSSI-VIO. 

\vspace{3pt}\noindent\textbf{Summary:}
Compared to regular device traffic analysis, the existing literature primarily focuses on the link layer 802.11 protocol, resulting in a limited set of features from traffic. Meanwhile, these methods often have limited usability, requiring additional user assistance or being tailored to specific operating systems and usage scenarios. More importantly, the generalizability of the algorithm needs to be improved. 

\subsection{User Activity Inference}
The leakage of user privacy has remained a prominent topic in network traffic analysis, such as user web browsing history~\cite{felten2000timing} or user interactions in mobile Apps~\cite{conti2015analyzing, hasara2021user}. 
Attackers who understand user behavior can learn about the living habits and further commit crimes~\cite{9152619}. 

Some researchers infer user activity from CIoT traffic, who use fingerprints of devices and their behaviors to establish a mapping relationship between traffic patterns and user activities, including devices using Wi-Fi protocol~\cite{li2016side, acar2020peek, wan2022iotmosaic} and low-energy protocol~\cite{gu2020iotspy}. 
Li et al. \cite{li2016side} discovered that differential coding in surveillance cameras could inadvertently leak side-channel information. Distinct body movements by users can lead to significant inter-frame differences between packets. Consequently, they used frames and applied $k$-NN and DBSCAN for activity recognition. 
In addition to cameras,  Acar et al.\cite{acar2020peek} have considered scenarios with more device types. They modeled user activities through multiple stages. 
The first three stages identified the device type, whether it is active, and its specific state. Then, they modeled user activities using Hidden Markov Model (HMM) in the final stage. However, this model only achieved a coarse-grained user behavior representation, identifying aspects such as whether a user remotely controlled a device or moved between locations.
Gu et al. \cite{gu2020iotspy} innovatively focused on 5 Zigbee devices on the SmartThings platform. combined with the idea of dynamic programming, they revealed the user activity dependency, which can be used to infer the user's living habits and routines. However, they did not evaluate their methods on Wi-Fi devices where their communications are more complex. 
Based on the previous works, Wan et al. \cite{wan2022iotmosaic} considered the presence of missing or unordered device events and develops an approximate user activity signature matching algorithm. Additionally, they design a heuristic trimming step to address multiple matches involving overlapping CIoT device events.

Different from the above work,  some researchers~\cite {subahi2019detecting, chu2018security, schmidt2023iotflow, babun2019real} studied the traffic behavior and privacy leakage problems on the App side.  
Subahi and Theodorakopoulos~\cite{subahi2019detecting} studied the interactions between users and CIoT devices and the exposure of sensitive Personally Identifiable Information (PII) and its type. They employ three random forest classifiers to achieve this goal. 
However, this study doesn't provide a solution for companion Apps using fixed certificates. 
\revise{
Recent research~\cite{schmidt2023iotflow, babun2019real, chu2018security} combines static and traffic analysis techniques to uncover privacy risks and device behaviors through companion Apps.
IoTFlow~\cite{schmidt2023iotflow} utilizes Value Set Analysis (VSA) and Data-flow Analysis (DFA) to reconstruct network protocols and endpoints, and assess potential vulnerabilities. This approach focuses on understanding how these apps communicate with IoT devices and remote backends, identifying what data is shared and with whom.
Similarly, Babun et al.~\cite{babun2019real} introduce IOTWATCH, a dynamic analysis tool, which monitors and collects data traffic during app execution, using Natural Language Processing (NLP) techniques to classify sensitive information and detect unauthorized data leaks. The tool focuses on analyzing how IoT apps communicate and transmit data, ensuring that sensitive information is not shared with unauthorized parties. By leveraging traffic analysis, IOTWATCH enables real-time monitoring of IoT app behavior, allowing users to better understand and control their privacy by identifying potential leaks and unauthorized recipients of their data.
Chu et al.~\cite{chu2018security} have uncovered similar security flaws in smart toys, such as unencrypted data transmissions and the lack of authentication in toy-app communications, violating privacy regulations like COPPA. These studies demonstrate how mobile app analysis can reveal privacy risks in networked devices. 
}

\vspace{3pt}\noindent\textbf{Summary:}
Existing research shows that attackers can commit crimes by analyzing user behavior. Researchers try to infer user activities through device fingerprints and traffic patterns, but there are limitations in device types and protocols.
The core challenge of inferring user privacy from CIoT traffic lies in identifying dependencies between device events and user activities. 
\revise{Combining app behavior and network traffic analysis provides valuable insights into user activity and privacy risks. Tools like IoTFlow and IOTWATCH use static and dynamic analysis to identify privacy issues at scale.} 
In the future, there is an opportunity for a refined exploration of diverse device types and events, particularly in multi-user scenarios, where different household members trigger devices at distinct time intervals, just like  Wan et al.\cite{wan2022iotmosaic} did.  

\subsection{Malicious Traffic Analysis}

\label{sec_applications_MaliciousTrafficAnalysis}
Like device fingerprinting, malicious traffic analysis is another popular direction in CIoT traffic research.
In this subsection, we introduce the malicious traffic analysis papers from two perspectives: detecting attacks on IoT and CIoT botnet detection. 

\subsubsection{Detecting Attacks on CIoT}
\label{Sec:Sub-DAC}
\begin{table*}
  \centering
  \renewcommand\arraystretch{1}
  \caption{Summary of attacks on CIoT detection literature}
  \resizebox{\textwidth}{!}{
  \begin{tabular}{|c|c|c|c|c|c|c|c|c|c|c|c|c|c|}

    \hline
    \multirow{2}[3]{*}{\textbf{Literature}} & \multirow{2}[3]{*}{\textbf{Year}} & \multicolumn{3}{c|}{\textbf{Contribution}} & \multirow{2}[3]{*}{\textbf{Feature}} & \multicolumn{2}{c|}{\textbf{Algorithm\textsuperscript{2}}} & \multicolumn{2}{c|}{\textbf{Data Source}} & \multicolumn{2}{c|}{\textbf{Communication}} & \multirow{2}[3]{*}{\textbf{\makecell{Collection \\ Location\textsuperscript{4}}}} \\
\cline{3-5}\cline{7-8}\cline{9-12}          &       & \textbf{Algorithm\textsuperscript{1}} & \textbf{Feature} & \textbf{Dataset} &       &    \textbf{Type\textsuperscript{3}}   &    \textbf{Name}   & \multicolumn{1}{c|}{\textbf{\makecell{Public \\ Datasets}}} & \multicolumn{1}{c|}{\textbf{\makecell{Self- \\ collection}}} & \textbf{Wi-Fi} & \multirow{1}{*}{\textbf{Low-energy}} & \\
    \hline
    \cite{heartfield2020self} & 2021 & $\surd$ & - & $\surd$ & \makecell{Statistics} & TML+RL & \makecell{RF, \\iForest, \\MAB-RL} & - & $\surd$ & $\surd$ & $\surd$ & $M_3$  \\ \hline  
    \cite{anthi2019supervised} & 2019  & - & $\surd$ & $\surd$ & \makecell{Flow, \\Packet} & TML & DT & - & $\surd$ & $\surd$ & $\surd$ & $M_1$ \\   \hline
    \cite{9155424} & 2020 & $\surd$ & $\surd$ & - & \makecell{Flow, \\Packet, \\Statistics} & TML & \makecell{RF, \\PCA} & - & $\surd$ & $\surd$ & $\surd$ & $M_1$  \\ \hline
    \cite{9500825} & 2021 & $\surd$ & - & $\surd$ & \makecell{Packet, \\Statistics} & TML & \makecell{iForest, \\DT} & - & $\surd$ & - & $\surd$ & $M_1$  \\ \hline    
    \cite{tekiner2022lightweight} & 2022 & - & $\surd$ & $\surd$ & \makecell{Statistics} & TML & SVM & - & $\surd$ & $\surd$ & - & $M_2$  \\ \hline  
    \cite{zhang2018homonit} & 2018& $\surd$ & - & $\surd$ & - & NML & DFA & - & $\surd$ & - & $\surd$ & $M_1$  \\ \hline
    \cite{charyyev2020detecting} & 2020 & $\surd$ & - & - & - & NML & LSH & N-BaIoT & - & - & $\surd$ & $M_1$  \\ \hline
    \cite{259723} & 2020 & $\surd$ & $\surd$ & - & \makecell{Flow, \\Packet, \\Physical} & NML & - & - & $\surd$ & - & $\surd$ & $M_3$  \\ \hline
    
  \end{tabular}}
    
  \label{table:malicious-A}%
  \begin{footnotesize}
    \begin{itemize}
      \item[1] ``Algorithm" denotes computational methods with provable improvements in either complexity (time/space) or task performance metrics.
      \item[2] The detail of algorithm of typical CIoT attack detection paper is shown in section~\ref{Sec:Sub-DAC}
      \item[3] ``TML" means traditional machine learning, ``NML" means traditional analysis, ``RL" means reinforcement learning.
      \item[4] ``$M_1-M_5$" corresponds to the five methods to acquire traffic in section~\ref{sec_systematic_trafficcollection}.
    %\item[**] This is another footnote.
    \end{itemize}
  \end{footnotesize}  
  \vspace{-1.5em}
\end{table*}%

Due to low hardware configuration and long update cycles, CIoT devices are vulnerable to various attacks, including Scanning attacks, Brute Force attacks, DoS attacks, and Cryptojacking~\cite{prasad2020cyber}.
Therefore, many researchers devoted themselves to detecting attack traffic targeting CIoT devices~\cite{anthi2019supervised,charyyev2020detecting,259723,tekiner2022lightweight,9500825,zhang2018homonit,9155424,heartfield2020self}.
Most of these work provided the ability to detect several general attacks~\cite{anthi2019supervised, charyyev2020detecting, 9155424, tekiner2022lightweight}, as is represented by DDoS and Scanning attacks. In 2019, Anthi et al.\cite{anthi2019supervised} presented a NIDS with 3 layers to detect 12 attacks (e.g., DDoS, MITM, Scanning) in a CIoT network environment. 
In 2020, Charyyev and Gunes\cite{charyyev2020detecting} proposed LSAD, based on LSH, to detect various attacks such as ARP Spoofing and DDoS attacks. Unlike ML-based algorithms, their method does not need to extract features from data.
Similarly, aiming at specific attacks, in 2022, Tekiner et al.\cite{tekiner2022lightweight} presented a lightweight traffic-feature-based method to detect CIoT Cryptojacking.
They trained with an SVM classifier and proved that their algorithm can obtain 99\% accuracy with one hour’s training data.
Meanwhile, special attacks targeting IoT devices are also detected. To solve the IoT security sensor tampering issue, Pathak et al. \cite{9500825} developed two algorithms to detect sensor tampering attacks: an unsupervised learning algorithm using iForest and a supervised learning algorithm CART based on C4.5 DT. 

Additionally, aiming at the attacks on low-energy CIoT devices, researchers also came up with solutions~\cite{259723,zhang2018homonit}. 
%zigbee+zwave
\begin{figure}[htbp]
\centering
\includegraphics[width=0.9\linewidth]{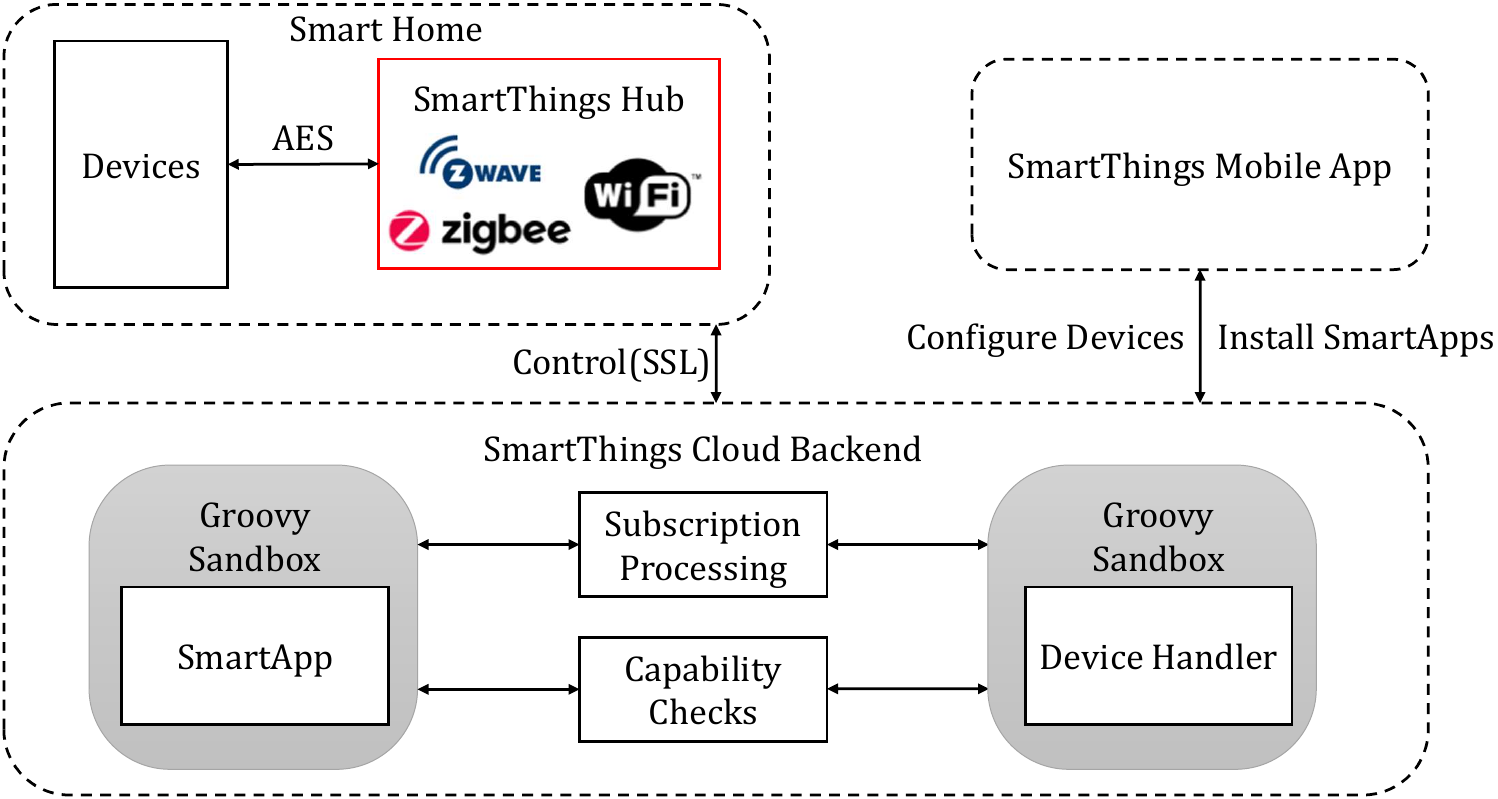}%
\caption{Architecture of SmartThings platform~\cite{zhang2018homonit}}
\label{Fig:SmartApp}
\end{figure}
SmartApp, proposed by SmartThings, is a type of program running on the cloud (as shown in Figure~\ref{Fig:SmartApp}).
Targeting the issues of over-privileged permissions and spoofing attacks within the application layer of SmartApp, in 2018,  Zhang et al.\cite{zhang2018homonit} conducted a notable study centered around identifying malicious SmartApps via traffic analysis. They first derived Deterministic Finite Automata (DFA) from textual descriptions and user interfaces, to model each App.
By monitoring encrypted traffic captured from wireless channels, they compare the observed state transitions associated with a behavior to the predefined DFA. If a match fails, it suggests the possibility of a malicious App. Notably, this research targeted Zigbee and Z-Wave devices on the SmartThings platform. However, the scalability of this approach when dealing with more complex functionalities and a greater number of states in Wi-Fi devices remains the problem for further consideration.

When facing unknown attacks, researchers put forward algorithms based on unsupervised learning~\cite{9155424,heartfield2020self} and RL~\cite{heartfield2020self} algorithms. In 2020, Wan et al.\cite{9155424}introduced a security monitoring system IoTArgos, which detects attacks such as Scanning and Brute-force at the system, network, and application layers of Smart Home IoT system by a supervised learning algorithm RF, and integrated an unsupervised learning algorithm principal component analysis (PCA) to detect zero-day or unknown attacks. Through the evaluation, IoTArgos can detect anomalous activities that target Smart Home IoT devices with high precision \& recall. 
In 2021, Heartfield et al.\cite{heartfield2020self} presented MAGPIE, which is able to autonomously adjust the function of its underlying anomaly classification models to a smart home's changing conditions (such as newly-added devices, new automation rules, and human interaction) to detect both known and unknown threat in Smart Home IoT network. Researchers applied a probabilistic cluster-based reward mechanism to RL and combined them with supervised learning classifier RF and unsupervised learning model iForest to classify traffic. Evaluation experiments in a real-home smart home environment containing Wi-Fi and Zigbee devices showed that MAGPIE provides high accuracy.

% At the same year, Peng et al.\cite{10118854} proposed a DDoS attack detection algorithm targeting vulnerabilities in 5G network security architecture by using a novel improved naive Bayes algorithm integrated with genetic algorithms. They extracts seven features from signaling changes in the authentication process, constructs a classifier, and achieves efficient detection through structural improvements and attribute weight optimization.

\vspace{3pt}\noindent\textbf{Summary: }
Researchers have designed specific algorithms to detect various attacks aiming at CIoT devices, as is shown in Table~\ref{table:malicious-A}. The intrusion detection research is mostly based on non-ML and TML algorithms. Meanwhile, researchers combine unsupervised learning algorithms into their research to detect unknown threats. Nevertheless, due to potential limitations in storage or data, DL-based detection methods have not yet been investigated in the CIoT domain.

\subsubsection{CIoT Botnet Detection}
\label{Sec:Sub-CBD}
\begin{table*}[h!]
  \centering
  \renewcommand\arraystretch{1}
  \caption{Summary of the CIoT botnet detection literature}
  \resizebox{\textwidth}{!}{
  \begin{tabular}{|c|c|c|c|c|c|c|c|c|c|c|c|c|c|}

    \hline
    \multirow{2}[3]{*}{\textbf{Literature}} & \multirow{2}[3]{*}{\textbf{Year}} & \multicolumn{3}{c|}{\textbf{Contribution}} & \multirow{2}[3]{*}{\textbf{Feature}} & \multicolumn{2}{c|}{\textbf{Algorithm\textsuperscript{2}}} & \multicolumn{2}{c|}{\textbf{Data Source}} & \multicolumn{2}{c|}{\textbf{Communication}} & \multirow{2}[3]{*}{\textbf{\makecell{Collection \\ Location\textsuperscript{4}}}} \\
\cline{3-5}\cline{7-8}\cline{9-12}          &       & \textbf{Algorithm\textsuperscript{1}} & \textbf{Feature} & \textbf{Dataset} &       &    \textbf{Type\textsuperscript{3}}   &    \textbf{Name}   & \multicolumn{1}{c|}{\textbf{\makecell{Public \\ Datasets}}} & \multicolumn{1}{c|}{\textbf{\makecell{Self- \\ collection}}} & \textbf{Wi-Fi} & \multirow{1}{*}{\textbf{Low-energy}} & \\
    \hline
    \cite{reed2021reliable} & 2021  & $\surd$ & - & - & \makecell{Packet} & NML & - & - & $\surd$ & $\surd$ & - & $M_1$ \\   \hline
    \cite{10278685} & 2023 & $\surd$ & - & - & \makecell{DL} & DL+RL & CNN & \makecell{NSL-KDD, \\IoT-23, \\N-BaIoT} & - & $\surd$ & $\surd$ & $M_4$  \\ \hline
    \cite{nguyen2019diot} & 2019 & $\surd$ & $\surd$ & - & \makecell{Flow, \\Packet, \\Statistics} & DL+FL & RNN & - & $\surd$ & $\surd$ & $\surd$ & $M_2$$M_5$  \\ \hline       
    \cite{9499122} & 2022 & $\surd$ & $\surd$ & - & \makecell{DL} & DL+FL & DNN & - & $\surd$ & $\surd$ & $\surd$ & $M_1$  \\ \hline
    \cite{10000633} & 2022 & $\surd$ & - & - & \makecell{Statistics} & DL+FL & AE & - & $\surd$ & $\surd$ & $\surd$ & $M_1$  \\ \hline     
    \cite{9838256} & 2022 & $\surd$ & - & - & \makecell{DL} & DL+FL & CNN & N-BaIoT & - & $\surd$ & $\surd$ & $M_1$  \\ \hline
    \cite{zhang2023towards} & 2023 & $\surd$ & - & - & \makecell{DL} & DL+FL & AE & N-BaIoT & - & $\surd$ & - & $M_1$  \\ \hline
    \cite{9024425} & 2019 & $\surd$ & - & - & \makecell{DL} & DL & \makecell{AE, \\CNN} & USTC-TFC 2016 & $\surd$ & $\surd$ & - & $M_1$  \\ \hline    
    \cite{9348169} & 2020 & $\surd$ & $\surd$ & - & \makecell{Flow, \\Statistics} & DL & \makecell{VAE, RNN} & CTU-13 & - & $\surd$ & $\surd$ & $M_5$   \\ \hline    
    \cite{bovenzi2020hierarchical} & 2020 & $\surd$ & - & - & \makecell{DL} & DL & AE & Bot-IoT& - & $\surd$ & $\surd$ & $M_1$$M_2$  \\ \hline    
    \cite{zixu2020generative} & 2020 & $\surd$ & - & - & \makecell{Statistics} & DL & \makecell{GAN, AE} & Bot-IoT & - & $\surd$ & - & $M_1$  \\ \hline    
    \cite{popoola2020hybrid} & 2020 & - & $\surd$ & - & \makecell{DL} & DL & \makecell{AE, \\LSTM} & BoT-IoT & - & $\surd$ & $\surd$ & $M_1$  \\ \hline
    \cite{9685306} & 2021 & $\surd$ & $\surd$ & - & \makecell{Statistics} & DL & RNN & Kitsune & - & $\surd$ & $\surd$ & $M_1$  \\ \hline    
    \cite{dinh2022twin} & 2022 & $\surd$ & - & - & \makecell{DL} & DL & AE & \makecell{NSL-KDD and \\five IoT botnet datasets} & - & $\surd$ & $\surd$ & $M_4$  \\ \hline      
    \cite{9838882} & 2022 & - & $\surd$ & - & \makecell{DL} & DL & Transfomer & N-BaIoT & - & $\surd$ & $\surd$ & $M_1$  \\ \hline    
    \cite{10118702} & 2023 & - & $\surd$ & - & \makecell{DL} & DL & CNN & \makecell{NSL-KDD, \\BoT-IoT} & - & $\surd$ & $\surd$ & $M_4$  \\ \hline
    %   NSL-KDD=4 N-BaIoT=1 MedBIoT=2 Bot-IoT=1
    \cite{9798307} & 2022 & $\surd$ & $\surd$ & - & \makecell{Statistics} & ML & ELM & \makecell{MedBIoT, \\ETF botnet dataset} & $\surd$ & $\surd$ & $\surd$ & $M_2$  \\ \hline
    \cite{illy2019securing} & 2019 & $\surd$ & - & - & \makecell{Flow, \\Packet, \\Statistics} & TML & \makecell{RF, Bagging, \\AdaBoost, Voting} & NSL-KDD & - & $\surd$ & $\surd$ & $M_4$  \\ \hline
    \cite{9014300} & 2019 & $\surd$ & $\surd$ & $\surd$ & \makecell{Packet, \\Statistics, \\Application} & TML & RF & - & $\surd$ & $\surd$ & $\surd$ & $M_1$  \\ \hline
    \cite{hafeez2020iot} & 2020 & $\surd$ & - & $\surd$ & \makecell{Statistics} & TML & FCM & - & $\surd$ & $\surd$ & $\surd$ & $M_1$  \\ \hline
    \cite{9307994} & 2020 & $\surd$ & $\surd$ & - & \makecell{Statistics} & TML & DT & N-BaIoT & - & $\surd$ & - & $M_1$  \\ \hline
    \cite{wang2021non} & 2021 & $\surd$ & - & - & \makecell{Statistics} & TML+FL & $k$-NN & \makecell{N-BaIoT, \\BoT-IoT} & - & $\surd$ & $\surd$ & $M_1$  \\ \hline
    \cite{9316792} & 2021 & $\surd$ & - & $\surd$ & \makecell{Packet} & TML & $k$-NN & N-BaIoT & - & $\surd$ & $\surd$ & $M_1$$M_4$  \\ \hline
    \cite{9116932} & 2021 & $\surd$ & $\surd$ & - & \makecell{Statistics} & TML & \makecell{DT, RF} & - & - & $\surd$ & $\surd$ & $M_1$  \\ \hline
    \cite{zhou2022metric} & 2022 & $\surd$ & - & - & \makecell{Flow} & TML & Metric Learning & N-BaIoT & - & $\surd$ & $\surd$ & $M_1$  \\ \hline
    \cite{9027816} & 2022 & $\surd$ & - & $\surd$ & \makecell{Flow} & TML & DBSCAN & - & $\surd$ & $\surd$ & $\surd$ & $M_5$  \\ \hline
    
  \end{tabular}}
    
  \label{table:malicious-B}%
  \begin{footnotesize}
    \begin{itemize}
      \item[1] ``Algorithm" denotes computational methods with provable improvements in either complexity (time/space) or task performance metrics.
      \item[2] The detail of algorithm of typical CIoT botnet detection paper is shown in section~\ref{Sec:Sub-CBD}
      \item[3] ``TML" means traditional machine learning, ``DL" means deep learning, ``RL" means reinforcement learning, ``ML" means machine learning, ``FL" means federated learning and ``NML" means non-machine learning.
      \item[4] ``$M_1$-$M_5$" corresponds to the five methods to acquire traffic, see section~\ref{sec_systematic_trafficcollection}.
    %\item[**] This is another footnote.
    \end{itemize}
  \end{footnotesize}  
\vspace{-1.5em}
\end{table*}%

DDoS attacks, primarily launched by botnets consisting of compromised CIoT devices like Mirai, Satori, and BASHLITE, have posed a significant threat and resulted in substantial damage to the network infrastructure. Among them, the infamous Mirai botnet caused widespread network disruptions~\cite{antonakakis2017understanding}. 
According to the research methodologies, we have categorized them into four types as outlined below.

\vspace{3pt}\noindent\textbf{\textcircled{1}Traditional ML-based Detection Methods}. 
Among our survey, many researchers use DT and RF algorithms to detect CIoT botnet traffic in the research of CIoT botnet attack detection based on TML~\cite{9307994,9014300,9116932}. In 2020, OKUR
and DENER\cite{9307994} compared two different ML algorithms in detecting normal traffic and the attack traffic from botnet. In their evaluation, the supervised learning algorithm (J48 DT) behaved better than the unsupervised learning algorithm (Expectation Maximization).

Furthermore, some researchers~\cite{9014300,9116932} concentrate on the feature selection of the CIoT botnet traffic. 
In 2019,  Dwyer et al.\cite{9014300} developed an analysis method based on DNS to detect CIoT botnet. They put forward a DNS feature set and evaluated a variety of TML classifiers, including RF, $k$-NN, and Naïve Bayes. RF classifier behaved the best among them (shows 99\% accuracy) and indicated that the feature-set based on DNS can significantly reduce the time of botnet detection. 

In addition to the centralized approach, researchers have also proposed distributed DDoS detection. 
In 2021, Doshi et al. \cite{9316792} proposed a novel NIDS based on a modified version of the Online Discrepancy Test (ODIT) to timely detect and mitigate Mongolian DDoS attacks characterized by widely distributed attack sources and small attack scales. The researchers used a $k$-NN-based algorithm to calculate the abnormal traffic conditions at each node. 
They then used a cooperative detector to aggregate the local statistical data of each node and obtain the global statistical data to determine whether an attack had occurred. 
This approach was validated using the N-BaIoT dataset, IoT testbeds, and simulations, proving its effectiveness against various DDoS scenarios. 

Based on the detection of CIoT botnet, researchers have made deeper discussion in some campaigns of CIoT botnet~\cite{9027816}. 
In 2022, Torabi et al. \cite{9027816} proposed a system to detect and analyze scanning campaigns of CIoT botnet. 
The author extracted the traffic from CIoT devices using the Shodan search engine and over 6TB network from the Dark web and detected compromised devices by examining whether they emitted unsolicited scanning. 
In their discussion, they pointed out that their work may be affected by the dataset, which was collected too early in August 2017. Some of these compromised devices may have already been removed from the Internet. Meanwhile, the researchers also detected and classified the scanning campaigns in compromised CIoT devices based on DBSCAN. Then, they grouped CIoT devices with similar scanning behavior and showed the campaign feature of CIoT botnet.

\vspace{3pt}\noindent\textbf{\textcircled{2}DL-based Detection Methods}. 
Till now, as one of the most popular types of ML algorithms, DL has plenty of applications in botnet detection~\cite{9024425,9685306,9348169,nguyen2019diot,10000633,9838256,10118702,zhang2023towards,9499122,9838882,bovenzi2020hierarchical,zixu2020generative,popoola2020hybrid,dinh2022twin}. 
Among them, most use NNs to detect the CIoT botnet. 
In 2019, Hwang et al.\cite{9024425} proposed a DL-based IoT malicious traffic detection mechanism. Researchers extracted flow features with the help of CNN and classified traffic with AE. The authors evaluated the mechanism with the traffic dataset collected from their Mirai network and USTC-TFC 2016 dataset and pointed out that the mechanism can achieve nearly 100\% accuracy. 
To solve the problem that only known botnets can be detected offline by existing technology, in 2020, Kim et al.\cite{9348169} proposed a new botnet detection method based on the Recurrent Variational Autoencoder (RVAE).
By testing in scenarios (including botnets not used for training), they demonstrated the robustness of the method in detecting unknown botnets.

Till now, most of researchers that use DL to detect CIoT botnets focused on the effect of their methods without conducting comparative tests or only comparing their methods with TML methods to show their advantages. Only a few researchers compared their methods with other DL methods in their evaluation; however, some of these methods did not train with datasets collected from CIoT devices.

In addition, the weak computing power and low storage of devices in the CIoT network challenge the deployment of DL models. To solve this problem, researchers tried to combine FL with DL~\cite{nguyen2019diot, 10000633, 9838256}. 
In 2022, Nishio et al.\cite{10000633} trained an anomaly detection FL model based on AE to detect botnet traffic to detect easily infected software. When assessed using their datasets collected from CIoT devices and simulating malware traffic, their method demonstrated enhanced efficiency in detecting malware under reasonable conditions. They got a more efficient detection model than AE and iForest models.

Based on the above algorithms, researchers conducted further research to solve the problems of privacy leakage and deployment difficulty. 
In 2022, Zhao et al.\cite{9838256} pointed out that FL-based NIDS may cause privacy breaches because the transmitted model data may be used to recover private data. Meanwhile, not independent and identically distributed (non-IID) private data can affect FL in training effect, especially the distil-based FL. The typically large models are difficult to deploy.
To solve these problems, they proposed a Semi-supervised FL (SSFL) NIDS scheme, based on knowledge distillation of unlabeled data and used CNN as a classifier and discriminator network to build the model. 
They evaluated SSFL with the N-BaIoT dataset and showed SSFL has the advantage in classifying performance and communication overhead compared with common algorithms such as FL-based algorithms and LSTM-based algorithms. 

\vspace{3pt}\noindent\textbf{\textcircled{3}RL-based Detection Method}. 
Some researchers have pro\\ved that RL-based algorithms are effective in general traffic analysis~\cite{10278685}. Baby et al. 
\cite{10278685} designed an RL-based NIDS. They adapt DRL algorithms to detect malicious DoS and DDoS traffic raised by CIoT botnets.
In their evaluation, researchers tested DRL models in different attacking and defending situations with three datasets, NSL-KDD, IoT23, and N-BaIoT, which are constructed mainly by botnets formed by compromised CIoT devices and pointed out that DRL algorithms are much more successful than TML and DL algorithms. 

\begin{figure}[htbp]
\centering
\includegraphics[width=0.75\linewidth]{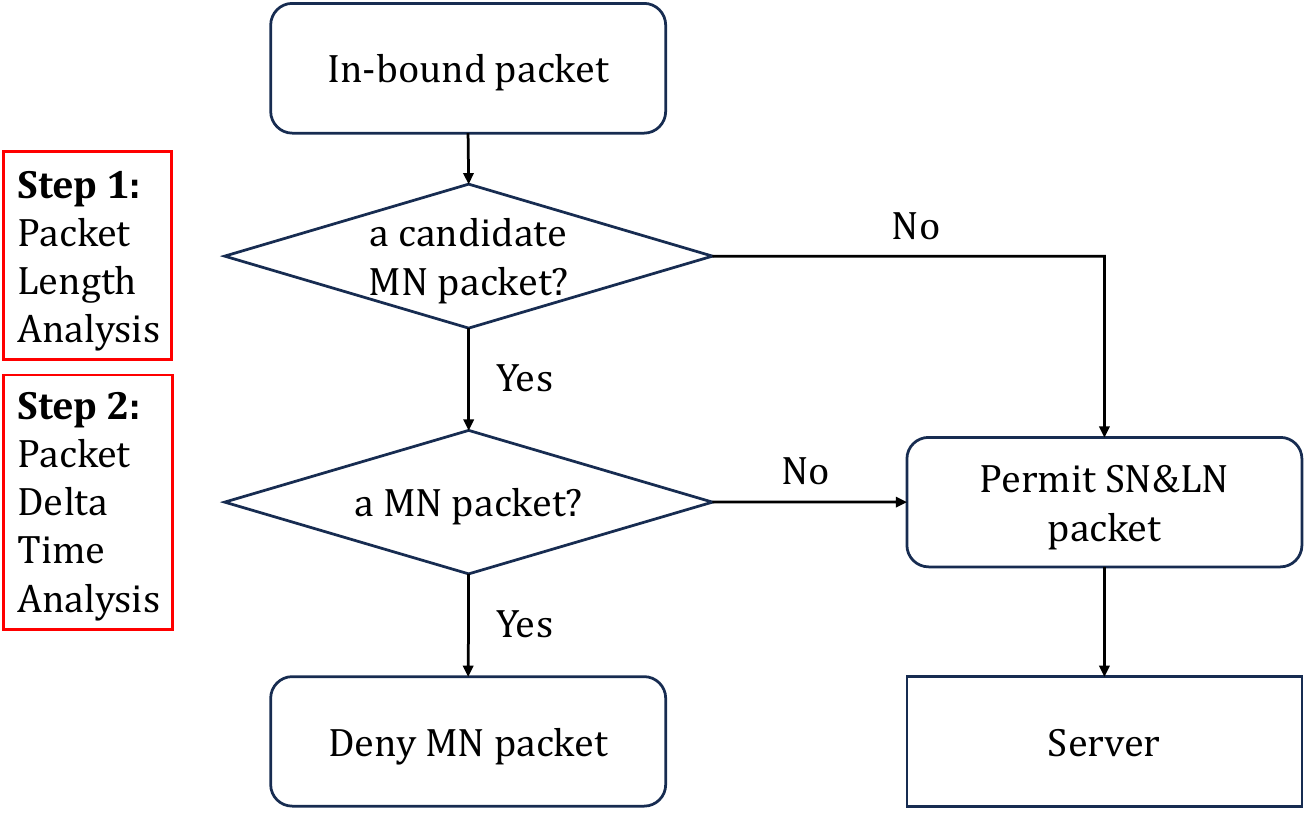}%
\caption{Real-Time slow DoS detection framework~\cite{reed2021reliable}}
\label{Fig:SlowDoS}
\vspace{-1em}
\end{figure}

\vspace{3pt}\noindent\textbf{\textcircled{4}Non-ML-based Detection Methods}. 
In recent years, ML-based botnet detection methods have become popular among researchers; however, nonML-based methods are an important complement of ML-based methods.
In 2021, Reed et al.\cite{reed2021reliable} proposed a lightweight framework that detects ``Slow DoS” attacks (attacks that can only provide low bandwidth and limited device resources) in resource-constrained IoT networks. 
Their method is shown in Figure~\ref{Fig:SlowDoS}. They analyze real-time IoT packets by two steps based on a set of only four attributes and classify them into three types: legitimate nodes (LN), genuine nodes with slow-to-intermittent connections (SN) and malicious nodes (MN).
This lightweight NIDS framework can classify genuine nodes experiencing slow or intermittent network connections and malicious nodes effectively.

\revise{In addition, for emerging threats such as DDoS attacks on 5G networks, in 2023, Pineda et al. \cite{10154440} proposed a defense method against internal DDoS attacks in 5G core networks by monitoring GPRS Tunneling Protocol User Plane (GTP-U) traffic and deploying traffic filtering mechanisms using Software Defined Networks (SDN). They deployed their approach in a 5G testbed to block malicious IoT traffic in real time. In their evaluation, they proved the performance and efficiency of their method in factual scenarios.}

\vspace{3pt}\noindent\textbf{Summary: } As shown in Table~\ref{table:malicious-B}, to date, researchers have proposed various ML algorithms to detect CIoT botnet. In the TML domain, DT and RF are predominantly employed, while NNs are utilized in the DL domain. In addition, researchers are adopting distribution strategies to address the issue of ML model deployment. However, much of the research lacks comparative experiments as their models are not evaluated on the same datasets.

\subsection{Measurement}
\label{sec_applications_measurement}
Researchers did measurement studies on CIoT traffic in order to gain insights into CIoT security and privacy status. IoT backends, vendors, communication protocols (especially TLS), IoT botnet, traffic destination, and private data exposure are considered in the studies, summarized in Table~\ref{table_Measurement}. We review current research in measurement from two perspectives: security and privacy. 

\subsubsection{Security Perspective}
\revise{Kumar et al.\cite{kumar2019all}, Paracha et al.
\cite{paracha2021iotls}, Saidi et al.\cite{saidi2022deep} and Tagliaro et al.~\cite{tagliaro2024large} discussed the security of CIoT from different perspectives, including secure deployment, TLS security, and backend.}
First, the general deployment and security status of CIoT devices is studied.  Kumar et al.\cite{kumar2019all} collaborated with Avast Software, an antivirus company, and conducted empirical analysis on traffic of 83 million devices across 16 million homes. This study reveals the significant regional variations in device types and manufacturers of CIoT devices. Open services, weak default credentials, and susceptibility to known attacks are also explored.
As TLS is a prominent security protocol used in CIoT, Paracha et al.\cite{paracha2021iotls} analyzed two years of TLS traffic and assessed the security of TLS connections established by IoT devices and how these connections changed over time. They revealed that TLS 1.2 was the most widely used version, while TLS 1.3 was less frequently adopted. Additionally, approximately 1/3 of devices were found to be vulnerable to interception attacks during TLS practices, potentially exposing sensitive data.
Similarly, Huang et al.\cite{huang2020iotinspector} expanded the dataset through crowdsourcing. They developed a tool called ``IoT Inspector'' to collect the traffic of 44,956 smart devices worldwide. By analyzing the data, researchers found that many device vendors used outdated TLS versions and that third-party advertising and tracking services on TV were prevalent.  Saidi et al.
\cite{saidi2022deep} emphasized that the security and functionality of IoT devices often rely on the IoT backend, the server on which the device downloads resources or the cloud-hosted for computing. 
By analyzing ISP's passive traffic data, they constructed a detailed map of IoT backend servers and revealed the relationships among these backend providers. 
\revise{Similarly to Saidi et al., Tagliaro et al.~\cite{tagliaro2024large} investigates the security of IoT backend servers, focusing on the MQTT, CoAP, and XMPP protocols. The study finds that 9.44\% of backends expose sensitive information, 30.38\% of CoAP backends are vulnerable to denial of service attacks, and 99.84\% of MQTT and XMPP backends do not use secure transport protocols. Through large-scale datasets and non-invasive measurements, the authors reveal significant security vulnerabilities and provide recommendations for improvement.} 

Some researchers~\cite{noroozian2021can, almazarqi2022macroscopic, herwig2019measurement} focused on the compromised ones, especially the IoT botnet.  Noroozian et al. 
\cite{noroozian2021can} evaluated the impact of two ISP security policies on Mirai.
By analyzing four years of dark web data, the research revealed that the strategy of closing ports to reduce the attack surface had no significant effect. In contrast, improving overall network health and remediation efforts significantly reduced the infection rate of Mirai. 
Almazarqi et al.\cite{almazarqi2022macroscopic} investigated the impact of AS structural properties on the spread of Mirai-like IoT botnets. They pointed out that commonly and widely used IP blacklist databases are incapable of tracking concentrated botnets. At the same time, they found that if the degree of an AS, that is, the number of direct connections between this AS and other ASes, is low, then the AS is more likely to become a host for malware downloaders.
Herwig et al. \cite{herwig2019measurement} investigated the Hajime botnet. 
Through active scanning and passive DNS backscatter traffic collection, the study reveals that there is a higher number of compromised IoT devices than previously reported. These devices use a variety of CPU architectures, and their popularity varies widely between countries. 

\begin{table*}
  \centering
  \renewcommand\arraystretch{1.2}
  \caption{Summary of measurement literature}
  %\resizebox{\textwidth}{!}{
    %\begin{tabular}{cccc}
    \begin{tabular}{p{0.06\textwidth}p{0.04\textwidth}<{\centering}p{0.06\textwidth}<{\centering}p{0.75\textwidth}<{\raggedright}}

    \toprule
    % \hline
    \specialrule{0.05em}{1.5pt}{2pt}
     \textbf{Topic} & \textbf{Paper} &\textbf{Year} & \multicolumn{1}{c}{\textbf{Measurement Description}} \\
    %\hline
    \specialrule{0.05em}{2pt}{2pt}
  
     \multirow{8}[1]{*}{Security} & \cite{kumar2019all} & 2019 & Evaluating the deployment of CIoT devices in different regions and security issues that include open services and weak default credentials.  \\ 
     & \cite{herwig2019measurement}  & 2019 & Investigating the Hajime botnet.  \\
     & \cite{huang2020iotinspector} & 2020 & Measuring insecure TLS implementations and the phenomena of third-party advertising and tracking services.  \\
     & \cite{paracha2021iotls} & 2021 & Assessing the security of TLS connections established by IoT devices and how these connections changed over time.  \\
     & \cite{noroozian2021can} & 2021 & Evaluating the impact of two ISP security policies on Mirai: closing propagation ports of malicious software and strengthening regulatory efforts.  \\
     & \cite{saidi2022deep} & 2022 & Constructing a detailed map of IoT backend servers and revealing the relationships among these backend providers.  \\    
     & \cite{almazarqi2022macroscopic} & 2022 & Investigating the impact of AS structural properties on the spread of Mirai-like IoT botnets.  \\
     % \hline
    \specialrule{0.05em}{2pt}{2pt}
     \multirow{8}[1]{*}{Privacy} & \cite{10.1145/3355369.3355577} & 2019 & The first group to study cross-regional data privacy on a large scale, which includes the destination of traffic, encryption status, distribution of plaintext and ciphertext content, as well as the possible exposure of device information.  \\
     & \cite{dubois-pets20} & 2020 & The privacy risk about speaker misactivations.  \\
     & \cite{li2020yourprivilege} & 2020 &  A large-scale empirical measurement focusing on Home Security Cameras (HSCs) in China and identifying three major behaviors that may leak user privacy: traffic surge, traffic regularity, and traffic rate change.  \\
     & \cite{mandalari-pets21} & 2021 & Extracting the non-essential destinations of the device.  \\
     & \cite{10.1145/3618257.3624803} & 2023 & Focusing on how the smart speaker ecosystem, especially Amazon Echo, collects, uses, and shares data.  \\    
     & \cite{girish-imc23} & 2023 &  Measuring the privacy leakage of local network interactions of IoT devices.  \\
     
    \specialrule{0.05em}{2pt}{1.5pt}
    \bottomrule
    \end{tabular}
    %}

    \begin{footnotesize}
    % \begin{itemize}
      
    % %\item[**] This is another footnote.
    % \end{itemize}
    \end{footnotesize}  
    
    \label{table_Measurement}
    \vspace{-1em}
\end{table*}

\vspace{3pt}\noindent\textbf{Summary:}
CIoT traffic security measurement provides valuable insights and guidance for building a more secure CIoT ecosystem, as summarized in Table~\ref{table_Measurement}. Firstly, the user data used for measurements should be legally authorized and thoroughly desensitized to ensure user data privacy. Secondly, current research (such as~\cite{noroozian2021can, almazarqi2022macroscopic, herwig2019measurement}) primarily focuses on a few malicious software families, such as the Mirai botnet. Future research should pay attention to various types of malicious software families. 

\subsubsection{Privacy Perspective}
In addition to security measurements, researchers are also working to measure the leakage of private information by devices from the traffic perspective.

``Mon(IoT)r Research Group'' from Northeastern University has done a series of work related to CIoT privacy measurement~\cite{10.1145/3355369.3355577,dubois-pets20,mandalari-pets21,gunawan-2023-chi,mandalari-sp23,girish-imc23,10.1145/3618257.3624803}.
Ren et al.\cite{10.1145/3355369.3355577} are the first to study cross-regional data privacy on a large scale. By capturing traffic from 81 CIoT devices distributed across laboratories in the UK and the US, they delved into aspects like the destination of traffic, encryption status, distribution of plaintext and ciphertext content, and the possible exposure of device information. 
Next year,  Dubois et al.\cite{dubois-pets20} focused on the privacy risk of speaker misactivations. By playing different TV shows on Netflix around seven speakers for 134 hours, they found that smart speakers have a 95\% possibility of misactivations with unintentional and listed the wake words that caused misactivations for the specific speaker. 
Similarly, Iqbal et al.\cite{10.1145/3618257.3624803} focused on how the smart speaker ecosystem, especially Amazon Echo. They exposed that Alexa Echo smart speakers collect user data and are used to target ads and track users' interests, which may raise concerns about privacy.
Mandalari et al.\cite{mandalari-pets21} extracted the non-essential destinations of the device. The study found that 52\% of devices communicated with non-essential destinations. Among them, smart TVs and cameras contacted numerous non-essential destinations. 
Different from the above studies, Girish et al.\cite{girish-imc23} measured possible privacy leakage of local network interactions of CIoT devices. The authors identified uncontrolled dissemination of sensitive information and revealed that the companion apps and third-party SDKs could potentially abuse user-space discovery protocols to access local network information, resulting in privacy infringements. 

Another group of researchers~\cite{li2020yourprivilege} conducted a large-scale empirical measurement focusing on Home Security Cameras (HSCs) in China. They identified three main behaviors that can leak user privacy: A sudden increase in traffic indicates that video uploading is in progress; the regularity change in traffic can be used to infer whether users are active and specific activities; and the traffic rate change can reflect changes in user activities. 

\vspace{3pt}\noindent\textbf{Summary:}
The results of privacy measurements indicate that users' private information may be exposed through CIoT devices. Various measurement studies have confirmed that devices frequently connect to third-party servers, which can lead to violations of local regulations such as GDPR. Therefore, regulators must ensure that the statements about third-party organizations in privacy policies are accurate and that devices properly enforce these statements. Furthermore, firmware updates on CIoT devices can alter existing behaviors, necessitating evaluations of the impact of time on measurement results. Additionally, many studies focus on devices in the EU and the US but neglect other regions, such as Asia.

\begin{table*}
\centering
\caption{Summary of the pros and cons of different application goals}
\label{table_application_comprision}

\resizebox{\textwidth}{!}{%
\begin{tabular}{|>{\centering\arraybackslash}m{3cm}|>{\centering\arraybackslash}m{3.5cm}|>{\centering\arraybackslash}m{4.5cm}|>{\centering\arraybackslash}m{4.5cm}|>{\centering\arraybackslash}m{4.5cm}|}
\hline
% \rowcolor{lightgray} % Title row background color
\multicolumn{1}{|>{\centering\arraybackslash}m{3cm}|}{\textbf{Application Goals}} & \multicolumn{1}{>{\centering\arraybackslash}m{3.5cm}|}{\textbf{Subcategories}} & \multicolumn{1}{>{\centering\arraybackslash}m{4.5cm}|}{\textbf{Advantages}} & \multicolumn{1}{>{\centering\arraybackslash}m{4.5cm}|}{\textbf{Limitations}} & \multicolumn{1}{>{\centering\arraybackslash}m{4.5cm}|}{\textbf{Applicable Scenarios}} \\ \hline
\multirow{3}{*}{\vbox{\vskip35pt\hbox{\hss \textbf{Device Fingerprinting}}}}  & \multirow{1}{*}{Device Identification} & High accuracy (\textgreater{}90\%), applicable to various device types; capable of identifying devices behind NAT; suitable for large-scale networks. & Limited ability to differentiate devices of the same type, limited support for non-IP protocols, and challenges in feature extraction. & Device and asset management; ISP network monitoring; large-scale device management. \\ \cline{2-5} 
& \multirow{1}{*}{Device Behavior Identification} & High precision; capable of detecting changes in device status through behavioral characteristics; can identify abnormal device behaviors. & Existing methods have poor scalability, limited types of device behaviors, sensitivity to packet loss, and difficulty distinguishing similar behaviors. & Safety monitoring (e.g., device health checks); user behavior analysis. \\ \cline{2-5} 
& \multirow{1}{*}{Hidden Device Detection} & Real-time detection of hidden devices within a given environment. & Sensitive to environmental factors; limited feature extraction capabilities; lack of methods for low-power devices. & Prevent unauthorized device access and safeguard user privacy. \\ \hline
\multirow{1}{*}{\textbf{User Activity Inference}} & - & Supports Zigbee, BLE, and other low-power device protocols, and establishes a correlation between traffic and user activities. & Low inference accuracy and high false alarm rate in multi-device environments. & Advertising targeting; user behavior analysis; user privacy insights. \\ \hline
\multirow{3}{*}{\vbox{\vskip20pt\hbox{\hss \textbf{Malicious Traffic Analysis}}}} & \multirow{1}{*}{Attack Detection} & Variety of detection methods. Evaluated in many different datasets. & Lack of a unified evaluation standard for attack detection. & Variety of attacks on CIoT devices. \\ \cline{2-5} 
& \multirow{1}{*}{CIoT Botnet Detection} & Garnered significant academic attention and unified in research aims. High precision. Public datasets cover various types of devices. & Lots of studies based on DL may be limited by the update cycle of datasets. Detection timeliness is limited due to the time-cost of model training. & Botnet attacks raised by compromised CIoT devices. \\ \hline
\multirow{3}{*}{\vbox{\vskip15pt\hbox{\hss \textbf{Measurement}}}} & \multirow{1}{*}{Security} & Suitable for large-scale data analysis to identify device security vulnerabilities. & High data collection and processing requirements; requires legal authorization and compliance. & Evaluation of CIoT device security; vulnerability discovery and mitigation. \\ \cline{2-5} 
& \multirow{1}{*}{Privacy} & Suitable for home environments, providing insights for user privacy protection. & Involves user privacy, data collection risks, and necessitates data desensitization. & Risk assessment of data breaches on devices; protection of user privacy. \\ \hline
\end{tabular}%
}
\end{table*}

\begin{tcolorbox}[colback=black!5!white,colframe=black!300!white, breakable]
\textbf{Takeaways: }
    This section categorizes the papers based on their application goals, including device fingerprinting, user activity inference, malicious traffic analysis, and measurement, to answer RQ2. 
    \revise{We compared the advantages, disadvantages, and application scenarios for different application purposes, as shown in Table~\ref{table_application_comprision}. 
    Existing research on device fingerprinting and malicious traffic detection has made significant progress, achieving high accuracy in laboratory environments. However, the experimental environments and evaluation criteria lack unified specifications, and most studies do not include real-world scenario evaluations, making it difficult to compare model performance. In practical applications, there is still room for improvement, particularly in multi-protocol support, real-time detection, and resource-constrained deployments.}
    % Among them, device fingerprinting and malicious traffic detection have been extensively researched, achieving high accuracy in the lab environment. 
    % It is worth noting that the experimental environments and evaluation criteria of these studies do not use unified specifications, and most of them lack model evaluation in real-world scenarios, which makes it difficult to compare the performance of models horizontally. 
\end{tcolorbox}

\section{Measures Against Traffic Analysis}

\label{sec_masking}

In Section~\ref{sec_applications}, we study the feasibility of constructing device fingerprints using traffic patterns. Concurrently, a lot of research has been conducted on methods to counteract traffic fingerprinting, aiming to prevent the leakage of traffic patterns. This section categorizes the principal techniques employed in these studies, which encompass three main research directions: traffic morphing, adversarial sample generation, and differential privacy.

\subsection{Traffic Morphing}

\begin{figure}[htbp]
\centering
\includegraphics[width=\linewidth]{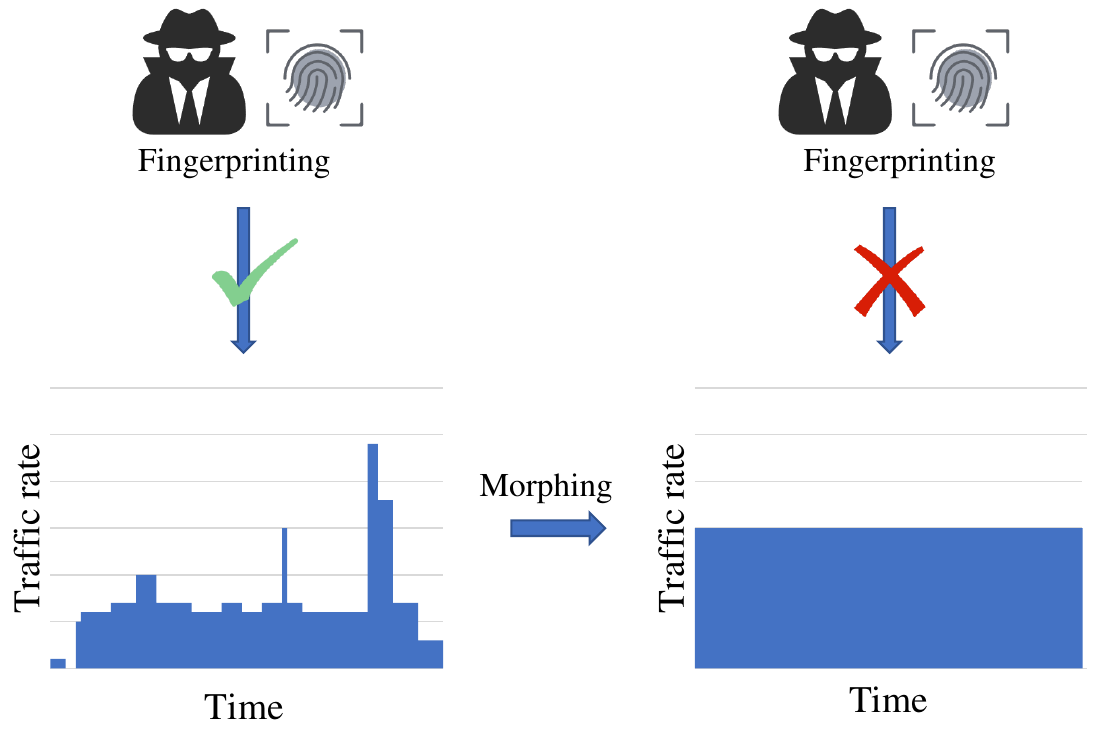}
\caption{Traffic morphing}
\label{Fig:maskingA}
\vspace{-1em}
\end{figure}

% Traffic Morphing techniques modify transmitted traffic to conceal its distinctive patterns for confusing adversaries, as shown in Figure~\ref{Fig:maskingA}. 
\revise{Traffic morphing technology changes the inherent transmission pattern of traffic by filling the original traffic with packets, which effectively prevents attackers from training the model to identify the unique traffic patterns of the device, as shown in Figure~\ref{Fig:maskingA}.}
% 流量变形技术通过向原始流量中填充数据包，以改变流量固有的传输模式，这样就能有效防止攻击者通过模型训练来识别设备独特的流量模式。
Shenoi et al.~\cite{Shenoi2023iPETPE} point out that crafting an appropriate traffic fingerprint defense algorithm may pose the following challenges: 1) defense mechanisms must not interfere with device communication; 2) reducing the network bandwidth and latency overhead caused by defense mechanisms; 3) ensuring the robustness of defense against adversarial training.

Early approaches to traffic morphing for CIoT devices primarily utilized link padding. Specifically, link padding can be categorized into Independent Link Padding (ILP) and Dependent Link Padding (DLP)~\cite{apthorpe2018keeping}. 
For ILP, traffic morphing is based on preset values, while DLP automatically adjusts according to current traffic patterns. The implementation of these two methods can utilize ML algorithms or through self-defined rules.
DLP has been used in the early days of network traffic development~\cite{shmatikov2006timing}, but it is not often used in CIoT. 
Apthorpe et al.~\cite{apthorpe2017spying} discussed different methods to prevent the inference of traffic pattern. They believe that the two most common methods, firewalls and VPNs, have certain flaws. Therefore, they employ ILP for traffic shaping, which called cover traffic. This entails filling original traffic with a fixed rate according to predetermined values to hide bursts. By implementing a control filter, the priority of device traffic always exceeds that of cover traffic, thus minimizing delay to the utmost extent. However, cover traffic consumes substantial bandwidth causing additional costs for consumers.
To reduce bandwidth consumption, Apthorpe et al.~\cite{apthorpe2018keeping} presents an improvement by introducing the concept of Random Traffic Padding (STP). This approach injects upload and download traffic with equal rate during user activity, and cyclically inserts this filler data stream into the original traffic. So that adversaries are unable to distinguish user activity from this mixture. Although STP saves bandwidth compared to ILP, it still need a significant volume of cover traffic to mask user activities effectively. For devices such as cameras, which involve audio-video streaming, STP may lead to substantial bandwidth overhead.
Further, Brahma and Sadhya~\cite{brahma2021preserving} introduced a novel defense mechanism that combines dummy packet generation with dynamic link padding. When the state of CIoT devices changes, dynamic traffic shaping introduces dummy traffic during the duration of the signature. This virtual traffic is randomly selected from the signature trace pool of other devices. This approach successfully introduces incorrect packet-level signatures, leading to a reduction in the accuracy of device identification.

% GAN
With the widespread application of ML, two defensive approaches have emerged. One of the methods is Generative Adversarial Networks (GANs)~\cite{Shenoi2023iPETPE, hou2021iotgan}.
GANs consists of two competing neural networks, namely the generator and discriminator which promote model progress through adversarial learning. 
Typically, the fingerprinting model is the discriminator, while the generator produces adversarial perturbation. The resultant generator is designed to maximally disrupt fingerprinting. 
Hou et al.~\cite{hou2021iotgan} assume a black-box scenario where no prior knowledge about the fingerprints exists. Firstly, they leveraged Knowledge of model transferability to obtain an alternative model capable of generating equivalent effects within the black-box setting, achieved through training a multi-layer fully connected neural network. In pursuit of evading device identification, the author employs the training strategy of generative models within GANs, manipulating traffic that doesn't impact the functionality of CIoT devices.
Shenoi et al.~\cite{Shenoi2023iPETPE} introduced a novel traffic morphing system named iPET, founded on adversarial perturbations. This system employs a generative DL model to generate device-specific defense perturbations. These perturbations intentionally introduce randomness between model instances and permit users to decide on a maximum bandwidth overhead. The researchers consider a model based on feature aggregates and a sequence-based device classifier. By deploying iPET, they effectively reduced fingerprinting accuracy from 96.36\% to 17\%.

%Reinforcement learning
Another is RL~\cite{tan2023youcan}. At its core, agent autonomously learns decision-making by observing feedback in the form of reward signals. The fingerprint adversarial model uses the probability of traffic originating from a certain device as a reward value, aiming to reduce the accuracy of the fingerprinting model. Notably, both GANs and RL are founded on the current environmental context of the fingerprints. These can be informed by prior knowledge (white-box testing), or construct an alternative model to fit fingerprints in black-box settings.
In research by~\cite{tan2023youcan}, traffic morphing is performed at the router entry point. The core of this approach lies in utilizing the Deep Deterministic Policy Gradient (DDPG) RL algorithm to learn effective strategies for altering CIoT traffic. The rewards needed for DDPG training are derived from the Isolation Forest algorithm, approximating the construction of the fingerprints. However, this method falls under the category of gray-box techniques, where privacy preservers possess partial knowledge of the feature dimensions used by the fingerprints. Additionally, the algorithm is evaluated based on a simplified fingerprinting model, potentially imposing limitations in practical usage.

\subsection{Adversarial Sample Generation}
\begin{figure}[htbp]
\centering
\includegraphics[width=0.7\linewidth]{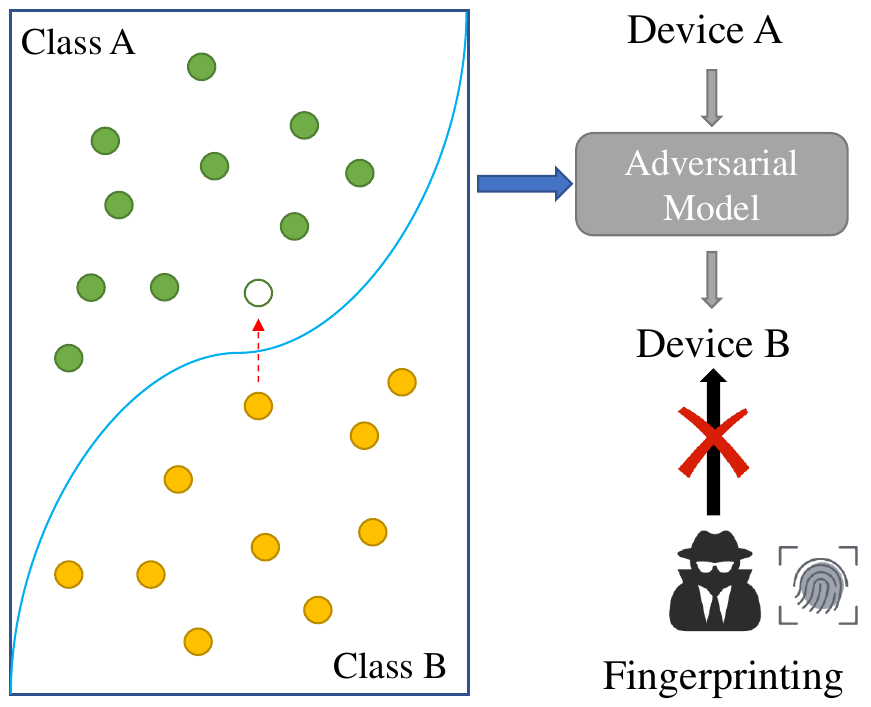}
\caption{Adversarial sample generation}
\label{Fig:maskingB}
\vspace{-1em}
\end{figure}

Due to the vulnerability of DL-based NIDS to adversarial sample attacks, a threat that has been demonstrated to be effective in various other domains, the same security concern is extended to NIDS in the CIoT ecosystem. 
% As shown in the Figure~\ref{Fig:maskingB}, introducing minor perturbations into the samples can significantly impact the detection performance of NIDS.
\revise{As shown in the Figure~\ref{Fig:maskingB}, by adding tiny and well-designed perturbations to the input data, it is possible to trick the model into producing false predictions. Adversarial samples are often very similar to the originals, and humans can't perceive these changes, but the model misclassifies them. }

Han et al. \cite{han2021evaluating} consider the gray-box and black-box attack scenarios. In gray-box attacks, the attacker possesses knowledge about the functionalities employed by the target NIDS, whereas in black-box attacks, this knowledge is entirely absent. Attackers feed pre-collected benign traffic and attack-induced malicious traffic into GAN to generate adversarial features. Subsequently, they employ a Particle Swarm Optimization (PSO) technique with predefined safe operators to automatically mutate the malicious traffic, iteratively selecting the best particles.

Further, Qiu et al.~\cite{qiu2020adversarial} explored the efficient implementation of adversarial attacks in a fully black-box scenario for NIDS. To begin, they employ model extraction techniques to replicate an adversarial example (AE) generation model, coupled with saliency maps to identify crucial features affecting fingerprinting accuracy. Ultimately, they utilize conventional AE to generate the necessary perturbations. Through attack demonstrations on the prominent NIDS tool Kitsune, the results indicate the success of perturbation.

\subsection{Differential Privacy}

\begin{figure}[htbp]
\centering
\includegraphics[width=\linewidth]{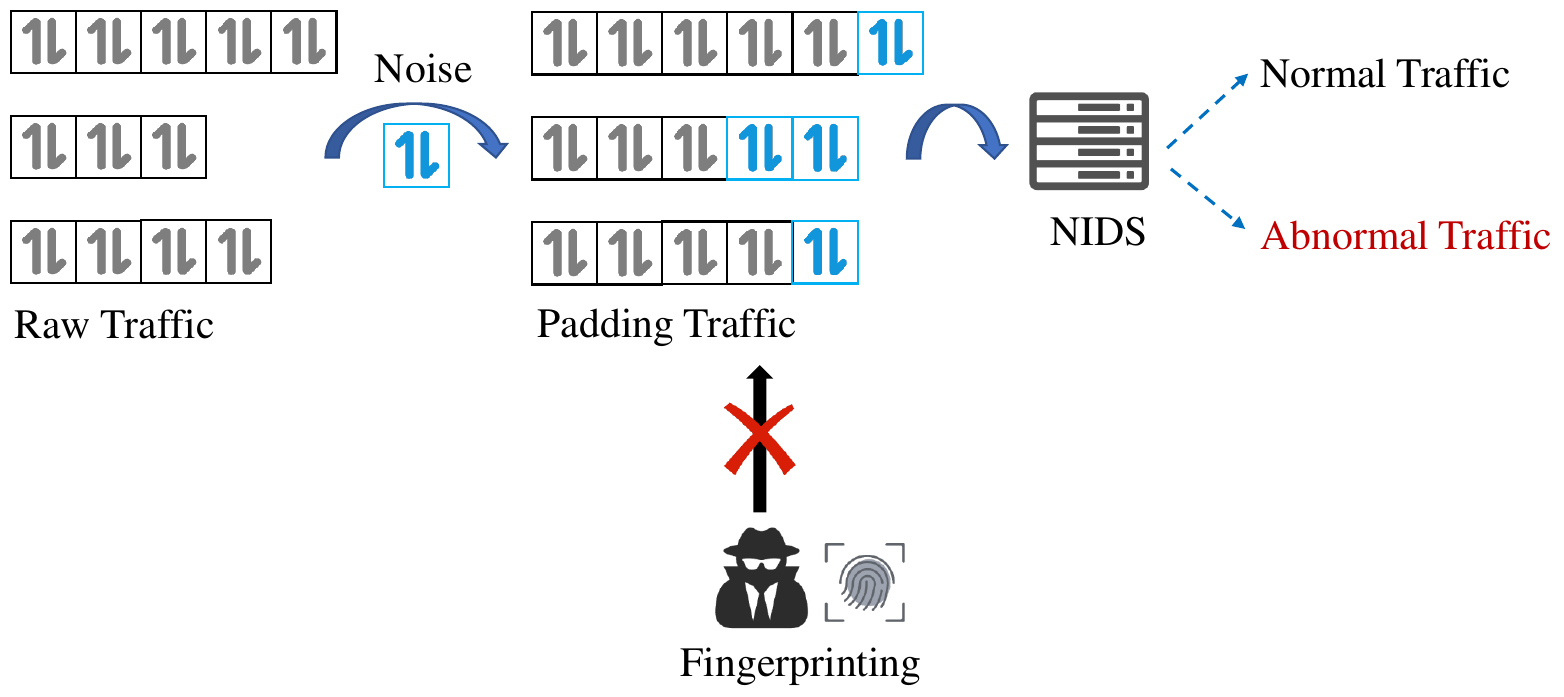}
\caption{Application of differential privacy in traffic}
\label{Fig:maskingC}
\end{figure}
To keep the user's privacy but avoid the negative impact on the NIDS's performance, researchers adopt Differential Privacy (DP) techniques to shape the CIoT traffic. \revise{As shown in Figure~\ref{Fig:maskingC}, with DP technology, CIoT traffic can be randomized or perturbed, making it impossible for external observers to easily analyze the user's real communication patterns, and at the same time, by controlling the characteristics of noise so that it conforms to a certain statistical distribution, so as not to mislead or reduce the detection accuracy of NIDS.}
% 通过DP技术，CIoT流量可以被随机化或扰动，让外部观察者无法轻易分析出用户的真实通信模式，同时通过控制噪声流量的特征，使其符合某种统计分布，从而不会误导或降低NIDS的检测准确性。

Liu et al.~\cite{liu2018epic} introduced the concept of smart communities, which direct smart home traffic toward proxy gateways before entering the Internet. By confusing traffic generated by different households, this secure and privacy-preserving multi-hop routing method can ensure non-linkability between source and destination. Notably, the gateway they used is not only a communications device but also a local computing platform. They deployed a DP mechanism to help with obfuscation.
The study by  Duan et al.~\cite{duan2022monitoring} is a notable example of simultaneously considering NIDS and user privacy protection, which also employed DP. The traffic intended for obfuscation is generated in a controlled manner without altering the individual system's spatiotemporal characteristics, thus not affecting the security monitoring system. However, from an overall perspective, the added noise can confuse attackers, preventing them from distinguishing device behaviors. Its essence lies in disrupting the robust correlation among traffic signatures, device states, and device triggers.
Similarly, Xiong et al.~\cite{xiong2022network} established a rigorous event-level DP model on discrete event packet flows and proposes an event-level $(\varepsilon _{s}, \varepsilon _{t})$-DP shaping mechanism. This mechanism leverages a discrete memoryless $max(\varepsilon _{s}, \frac{\varepsilon _{t} }{2} )$-LDP channel c to conceal packet sizes and time information, thereby mitigating traffic analysis attacks.

\vspace{3pt}\noindent\textbf{Summary: } 
This section introduces traffic morphing-related works that defend against traffic analysis. Current traffic morphing techniques often come with network performance overheads or a negative impact on NIDS's reliability. For instance, traffic padding increases network bandwidth usage, and adversarial sample generation may lower the detection performance of NIDS. Future research could consider adaptive traffic morphing systems that select the most suitable morphing strategies based on the dynamic characteristics of CIoT traffic. Meanwhile, attackers continuously develop new traffic analysis techniques to break traffic morphing. Researchers need to continually improve the robustness of traffic morphing. For example, exploring traffic features based on the protocol level or application level to enhance the effectiveness of traffic morphing. On the other hand, traffic morphing techniques based on adversarial samples push forward the improvement of robust traffic analysis AI models. Potential directions in the future include the use of GNN to capture higher-order dependencies and adversarial training for network traffic analysis.

\section{Challenges and Future Research}
\label{sec_challenges}

\revise{Despite significant contributions by researchers in the field of CIoT traffic security and privacy, it still reveals that challenges remain within this domain. Therefore, in this section, we summarize the challenges and future research directions from the perspective of CIoT traffic analysis processes, specifically focusing on four key aspects: CIoT traffic collection, CIoT traffic processing, analysis algorithm, and new applications. This analysis aims to address RQ3.}
% In this section, we answer RQ3 by summarizing the challenges and future research directions of CIoT traffic analysis in security and privacy based on the analysis process and application goals. 

\subsection{CIoT Traffic Collection}
\label{sec_challenges_trafficcollection}

Compared to PC and mobile apps, CIoT devices exhibit many special characteristics, which influence the traffic collection process, as discussed in Section~\ref{sec_systematic_newCharacteristics_trafficcollection}. In the following part, we summarize potential future research directions regarding CIoT traffic collection. 

\vspace{0pt}\setlength{\parindent}{1em}\textit{(1) More Comprehensive Datasets.} 
Firstly, compared with general network analysis like website or application fingerprinting, establishing a physical CIoT environment or testbed is more time-consuming and costly, especially considering numerous CIoT vendors, types, and models. 
Secondly, as outlined in Section~\ref{sec_systematic_dataset}, despite researchers establishing several datasets independently and even Huang et al.~\cite{huang2020iotinspector} collected traffic of thousands of devices by crowdsourcing, an up-to-date, unified, and large-scale CIoT traffic dataset available for researchers is still important. 
Considering the distinctive feature of CIoT devices, the dataset ought to span the entire CIoT lifecycle and be labeled with fine granularity, thereby fulfilling diverse future application objectives. 
% 为了解决这一问题，我们建议开发一个开源协作平台，鼓励研究人员和制造商贡献数据，确保动态更新，同时通过差异隐私等技术保护隐私。此外，利用机器学习的自动化数据标记工具可以显著减少人工工作量并提高数据集质量。
\revise{To address this, we suggest developing an open-source collaborative platform that encourages researchers and manufacturers to contribute data, ensuring dynamic updates while preserving privacy through techniques like differential privacy. Additionally, an automated data labeling tool leveraging LLM could significantly reduce manual effort and improve dataset quality.}
Lastly, as shown in Table~\ref{dataset}, almost all commonly used CIoT datasets are self-collected in laboratory settings. Therefore, how to construct an open real-world dataset that does not expose any users' privacy to support evaluations of various methods in practice is a valuable topic. 

\vspace{0pt}\setlength{\parindent}{1em}\textit{(2) Cost-minimal CIoT Traffic Collection Methods.} 
As discussed above, building a comprehensive CIoT traffic dataset is essential but remains time-consuming and costly. This complexity is due to the diverse device interaction modes, such as physical control and automation rules, as discussed in Section~\ref{sec_background_consumeriot}. Ren et al.~\cite{10.1145/3355369.3355577} to some extent automated the collection process by \textit{Monkey Application Exerciser} included in Android Studio. However, not all interaction modes can be fully simulated in an automated way, e.g., the device binding process. Therefore, automating the traffic collection process to minimize human labor is a promising future direction. 
\revise{Researchers can develop automatic mechanical operation devices to simulate diverse physical interactions. }
Meanwhile, traffic simulation generation can effectively reduce economic costs, particularly for network evaluation and the construction of datasets that contain malicious traffic~\cite{alomari2014design, lee2011traffic}. However, constructing data sets through virtual environments remains uncommon in the CIoT domain, possibly due to various modes of user interaction and device types. Consequently, simulating realistic CIoT scenarios and generating realistic traffic remains an unresolved challenge. Firmware rehosting may be a solution to this challenge. 

\vspace{0pt}\setlength{\parindent}{1em}\textit{(3) Non-IP Traffic.}
Most works focus on analyzing the TCP/IP network layer traffic and above. However, as for non-IP (Zigbee, Z-Wave, and Bluetooth) devices, existing works~\cite{8116438,8761559,8664655,pinheiro20198} often use a smart hub and collect traffic at the router, which only captures IP packets. In scenarios where the attacker is near the victim's home, significant information about the user and devices can be inferred from link-layer packets. For instance, Gu et al.~\cite{gu2020iotspy} successfully inferred user behaviours through Zigbee packets. Thus, there is a pressing need for a comprehensive CIoT traffic dataset containing various communication protocols and techniques, including Zigbee, Z-wave, Bluetooth, and 4G/5G. This will enable future works to analyze and infer information across different scenarios. 
\revise{One solution is to develop a hybrid traffic collection system that can capture non-IP traffic (e.g., Zigbee, Z-Wave) at various positions in the network. }

\vspace{0pt}\setlength{\parindent}{1em}\textit{(4) Malicious CIoT Traffic. } 
Catillo et al.~\cite{catillo2023machine} show that the malicious traffic datasets have certain limitations. 
The origins of the CIoT botnet attack are widespread, and the number is substantial, making the creation of an up-to-date dataset difficult. Meanwhile, most researchers rely on popular third-party datasets, such as BotIoT, N-BaIoT, and IoT23, to evaluate their methods. Therefore, it remains uncertain if their methods can reliably identify attack traffic from previously unidentified botnets.
\revise{To overcome this, we suggest deploying honeypots to capture real-time botnet traffic and create a dynamically updated malicious traffic dataset. }

\subsection{CIoT Traffic Processing}
\vspace{0pt}\setlength{\parindent}{1em}\textit{(1) Local Traffic.} 
In Section~\ref{sec_systematic_compare}, we analyze the differences in traffic between CIoT devices and traditional computing devices. Researchers should consider these distinct characteristics when designing and implementing traffic analysis methods. We observed that most research focuses on analyzing the communication traffic between CIoT devices and the cloud. Compared to other studies, there is a scarcity of research on local communication traffic analysis between CIoT devices and companion apps, although local communication could also show a lot of information (as Girish et al.~\cite{girish-imc23} studied). Therefore, effectively analyzing local communications is a valuable research topic. 
\revise{This will help detect potential vulnerabilities that do not rely on cloud services. We propose developing a lightweight local traffic analysis framework using edge computing to process and analyze traffic in real time. Additionally, graph neural networks could be employed to model complex interactions between devices.}
 
\vspace{0pt}\setlength{\parindent}{1em}\textit{(2) Vendor Proprietary Protocols.} 
Most works learn information about devices by standard protocols~\cite{255244, 10.1145/3419394.3423650, 9230403}. 
However, due to security and real-time communication requirements, many manufacturers opt for proprietary or private encryption protocols based on UDP. These protocols often render regular monitoring tools ineffective. Consequently, developing new techniques with protocol reverse engineering ability becomes essential for analyzing these protocols. 
\revise{For example, combining symbolic execution and dynamic analysis to decode proprietary protocols. }

\vspace{0pt}\setlength{\parindent}{1em}\textit{(3) Feature Optimization.} 
It is necessary to continuously optimize the CIoT traffic feature processing using new technologies in the field of AI. 
Frontier research on network traffic has proposed many methods of traffic representation. For instance, Xie et al.~\cite{xie2023rosetta} employed more robust TLS features. Bronzino et al.~\cite{bronzino2021traffic} presented a framework and system that evaluates the system-level costs of various traffic representation methods. The work by Zola et al.~\cite{zola2022network} employed a graph-based approach that addresses class imbalance issues and enhances the supervised node behavior classification.
Holland et al.~\cite{holland2021new} automated various aspects of traffic analysis and introduced the tool nPrint for generating unified packet representations. 
In addition to these cutting-edge approaches in network traffic, a variety of feature selection optimization methods in ML could be employed, including tree-based feature importance evaluation algorithms, recursive feature elimination (RFE), LASSO, GA, etc. 

\subsection{Analysis Algorithm}
\vspace{0pt}\setlength{\parindent}{1em}\textit{(1) Open-world Problem.} 
Most studies we reviewed evaluate their algorithms based on lab-crafted traffic datasets. However, there are numerous types and models of CIoT devices. The dataset for training the model cannot cover all CIoT devices worldwide. Thus, the model's ability to identify devices in an open world needs to be effectively verified in the future. 
\revise{A promising direction would be to focus on Transfer Learning (TL) techniques that can enable the model to recognize unknown devices based on previously observed traffic patterns. 
}

\vspace{0pt}\setlength{\parindent}{1em}\textit{(2) Challenges of Edge and Multifunctional CIoT Devices.} 
Existing research has not sufficiently addressed the potential confusion in traffic patterns caused by multifunctional and edge CIoT devices. Smart TVs, as typical edge CIoT devices, mostly feature Android-based operating systems, which could cause their traffic characteristics to be confused with the background traffic of devices such as smartphones. This overlap may obscure the intrinsic features of smart TVs. Moreover, analyzing the traffic of multifunctional devices, like smart TVs with voice assistance and sweeping robots that combine cameras, brings new challenges to existing algorithms focusing on single-function devices. For example, as smart TVs add functions like voice assistants, their traffic becomes more like that of smart speakers, deepening the confusion. 
\revise{Therefore, the algorithm should isolate traffic patterns specific to each function (e.g., voice recognition vs. video streaming) to prevent traffic from being mislabeled or misclassified. Collaborative filtering or multi-view learning could be useful to distinguish these mixed functionalities.}

\vspace{0pt}\setlength{\parindent}{1em}\textit{(3) Unified Standard Evaluation.} 
Our survey noted that the majority of studies lack comparative evaluations within a unified standard test environment (both datasets and algorithms), as outlined in \ref{sec_challenges_trafficcollection}. Unfortunately, the fragmented CIoT ecosystem results in haphazard evaluations. We recommend establishing a unified standard evaluation framework for algorithms addressing the same objective. Such a framework would facilitate more reliable comparisons across different studies, promote transparency, and accelerate the development of more effective solutions. 
% It would also help identify best practices and gaps in the current methodologies, fostering collaboration and innovation in the CIoT field.

\vspace{0pt}\setlength{\parindent}{1em}\textit{(4) Applying New AI Techniques.} 
Since 2023, the success of large language models (LLM)~\cite{kaddour2023challenges}, particularly ChatGPT, has attracted researchers to apply them in solving issues across various computer science domains. 
There is growing interest in utilizing LLM for traffic analysis applications. For instance, the recent release of Traffic LLM by Tsinghua University\footnote{Traffic LLM by Tsinghua University, visit \url{https://github.com/ZGC-LLM-Safety/TrafficLLM}}, the use of LLMs to automatically analyze HTTP banners in internet scans~\cite{sarabi2023llm}, optimizing network performance~\cite{habib2024llm}, and malicious traffic analysis~\cite{guastalla2023application, hassanin2025pllm} are notable examples. The adaptability and efficiency of LLMs can significantly enhance the processing and interpretation of complex traffic patterns.
However, the application of LLMs in the CIoT domain remains relatively limited. This may be due to the diversity of CIoT devices, which increase the difficulty of data standardization. Additionally, the resource constraints in CIoT environments pose challenges for directly applying these models. Researchers should explore more customized and lightweight models to meet the unique needs of CIoT.
\revise{Another way is using optimization techniques such as pruning or knowledge distillation to reduce resource usage. }

\vspace{0pt}\setlength{\parindent}{1em}\textit{(5) Applications in Edge Computing Scenarios.} 
Edge computing integrates big data, IoT, and AI technologies, enabling applications to operate on local servers and bringing computation closer to terminal devices~\cite{10529137, 9573429}. This approach not only reduces data transmission costs but also enhances network security by minimizing reliance on central cloud infrastructures, thereby lowering the risk of DoS attacks in global data centers. 
As data transmission increasingly shifts towards the network edge rather than central servers or data centers, edge computing can substantially decrease packet header lengths and network latency. 
This brings advantages to deploying models on devices and helps develop distributed algorithms, which further enables network management and security policies to adapt more rapidly to dynamic network conditions.

\subsection{New Applications}
CIoT is a system that involves multiple components working collaboratively. In addition to the application goals introduced in Section~\ref{sec_applications}, traffic analysis could possibly provide more insight into the CIoT ecosystem. Currently, utilizing companion applications to understand CIoT devices is becoming increasingly common. Security testing and research on the device encounter limitations due to firmware not being public~\cite{zhao2022large}. Therefore, certain studies have resorted to static analysis of companion applications to identify potential risks related to user data exposure~\cite{schmidt2023iotflow, YuhongNan2023areyou}, while others employ these applications as proxies for fuzz testing~\cite{chen2018iotfuzzer}. We believe that the traffic of CIoT companion apps is a valuable research area for assessing the CIoT ecosystem, e.g., the supply chain ecosystem. 
Furthermore, existing research indicates that CIoT traffic exposes much private information. Besides the existing goals (fingerprinting device and user behaviors), the traffic could possibly be used to fingerprint more CIoT applications that expose user privacy, such as interactions with smart voice assistants and automation rules~\cite{cobb2020risky} (e.g., IFTTT). Studies have shown that IFTTT is vulnerable to malware flows~\cite{cobb2020risky,10.1145/3243734.3243841}. Therefore, the security of third parties who have the authority to cloud API control the device also needs to be studied urgently.
Third, traffic analysis can also help identify vulnerabilities in CIoT, as some researchers demonstrated~\cite{10154338,8486369}. Given that recent studies have highlighted numerous security concerns, traffic analysis of the CIoT could provide more significant information to discover vulnerabilities. 

\revise{Last but not least, for detecting DDoS attacks on 5G networks and devices, some researchers have proposed some algorithms as is shown in Section ~\ref{sec_applications_MaliciousTrafficAnalysis}. However, till now, researchers mostly extract 5G attacks' feature from signaling changes because DDoS attack on 5G systems mainly happens on the control plane~\cite{10118854}, there is still a lot of future work to be done in the field of traffic. Meanwhile, APT (advanced persistent threats) attacks targeting CIoT devices are also worth paying attention to. Researchers have proposed different detection methods based on both DL~\cite{10609886} and DRL~\cite{10899756}. Besides the exploitation of compromised devices for launching DDoS attacks as discussed in Section~\ref{Sec:Sub-CBD}, attackers may achieve more covert and persistent threats through CIoT device infiltration. For instance, tampering with smart home sensor data like temperature thresholds or smoke alarm configurations to fabricate false environmental conditions could raise problems such as misleading smart air-condition-systems into overload operations that may result in circuit failures.}
Finally, we mentioned in Section 5 that existing studies have shown that CIoT device firmware is not updated frequently, but there is currently no work specifically studying whether the traffic model of CIoT will change with a major update, which will directly affect the timeliness of the model that use traffic patterns as testing criteria.

\begin{tcolorbox}[colback=black!5!white,colframe=black!300!white, breakable]
\textbf{Takeaways: }
    In this section, we summarize the challenges and future directions in CIoT traffic analysis through three key steps and application scenarios \revise{to answer RQ3}. Firstly, given the diverse CIoT devices, there is a need to develop a more comprehensive dataset to facilitate the establishment of a standard evaluation framework. 
    Secondly, the unique characteristics of CIoT traffic, such as local communication and proprietary protocols, necessitate a greater focus on feature optimization to capture its complexity. 
    Furthermore, with the rapid advancement of technologies like LLM and edge computing, researchers should explore customized and lightweight algorithms to address the evolving CIoT traffic in an open world. 
    Finally, attention must be given to under-researched application scenarios, such as the security of IFTTT and the impact of device updates on traffic patterns. These efforts will accelerate the development of safer and more efficient solutions in the CIoT field.
\end{tcolorbox}

\section{Conclusion}
\label{sec_conclusion}

We surveyed 310 papers on traffic analysis in the field of CIoT security and privacy from 52 conferences and journals of high reputation. We reviewed the literature according to the proposed three RQs and answered them in the takeaways. First, we summarize the process of CIoT traffic analysis in three steps and identify new characteristics of CIoT traffic, especially the complexity of traffic collection and processing. Next, we looked at the four application goals of current studies and concluded their contributions and deficiencies, and classified the measures against traffic analysis. Finally, we summarized the challenges and pointed out future directions. Compared to general traffic analysis, network architectures, communication protocols, and application scenarios of CIoT devices present new challenges in traffic analysis. We hope to inspire more researchers to analyze the security and privacy issues of CIoT from a traffic perspective.

\begin{acknowledgement}
This work is supported by the National Natural Science Foundation of China (No.62032012, 62432012, 62102198), the National Key R\&D Program of China (2022YFB3103202), and the Fundamental Research Funds for the Central Universities (079-63243152). 
\end{acknowledgement}

\footnotesize
\bibliographystyle{fcs}
\bibliography{ref}

\begin{biography}{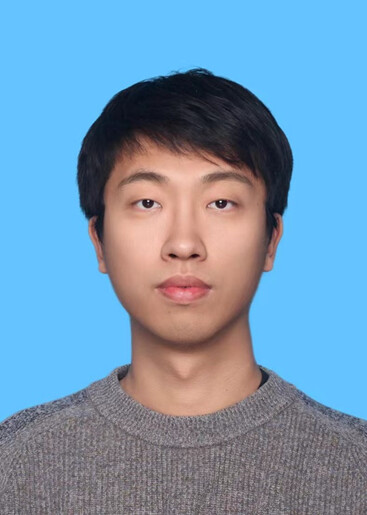}
\textbf{Yan Jia} is an Associate Professor at Nankai University in Tianjin, China. He received the B.S. and Ph.D. degree from Xidian University, Xi’an, China, in 2015 and 2020 respectively. His research interests include IoT security and privacy, vulnerability mining, Web security, usable security and privacy, network and system security, etc. He published several papers in top security conferences, such as IEEE S\&P, USENIX Security, ACM CCS and NDSS. His work helped many high-profile vendors improve their products’ security, including Amazon, Microsoft, Apple, and Google.
\end{biography}

\begin{biography}{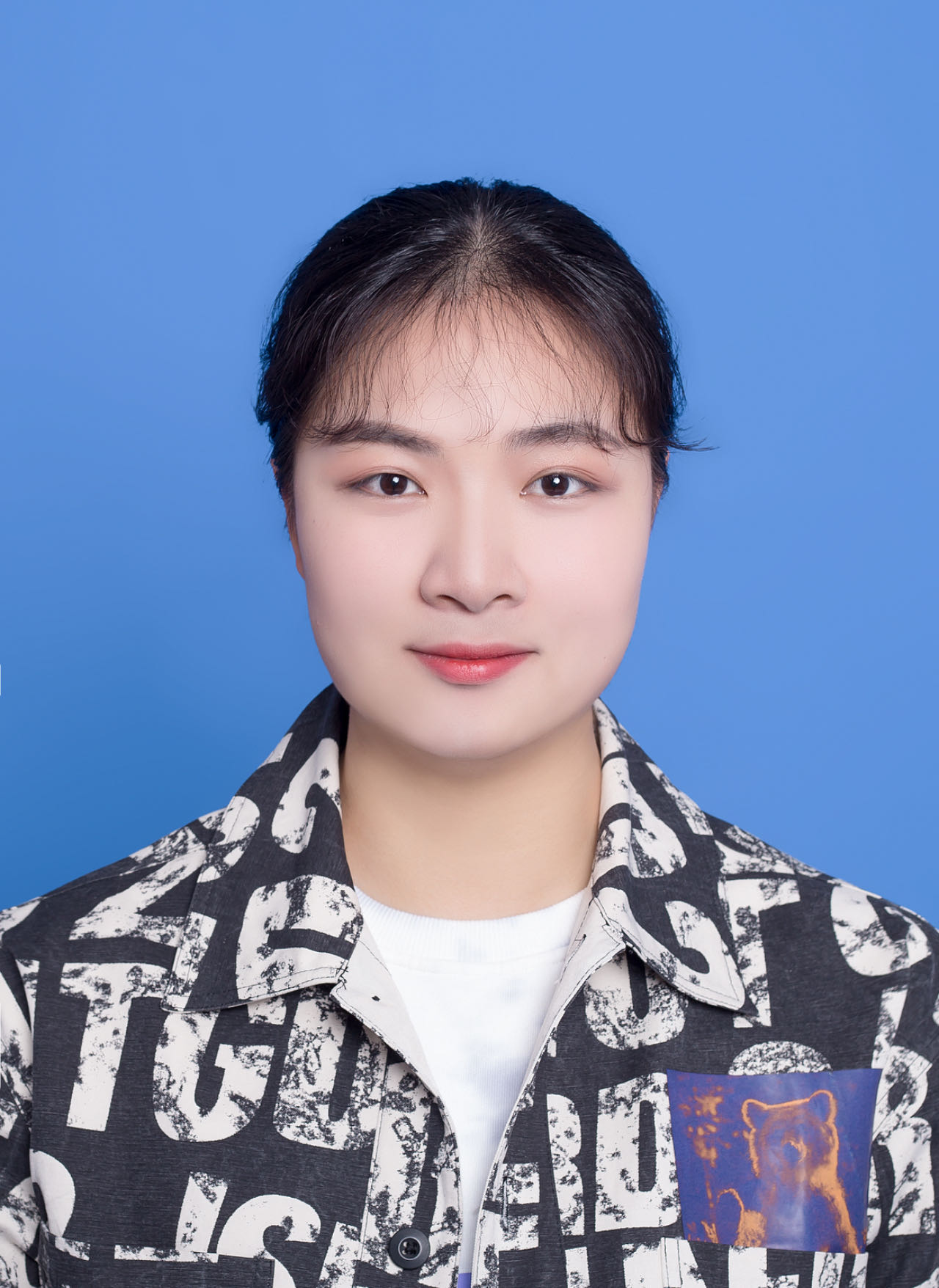}
\textbf{Yuxin Song} is now working toward the master's degree in Computer Technology at the College of Cryptology and Cyber Science, Nankai University, Tianjin, China. Her research interests mainly include IoT security, IoT privacy, and traffic analysis technology. 
\end{biography}

\begin{biography}{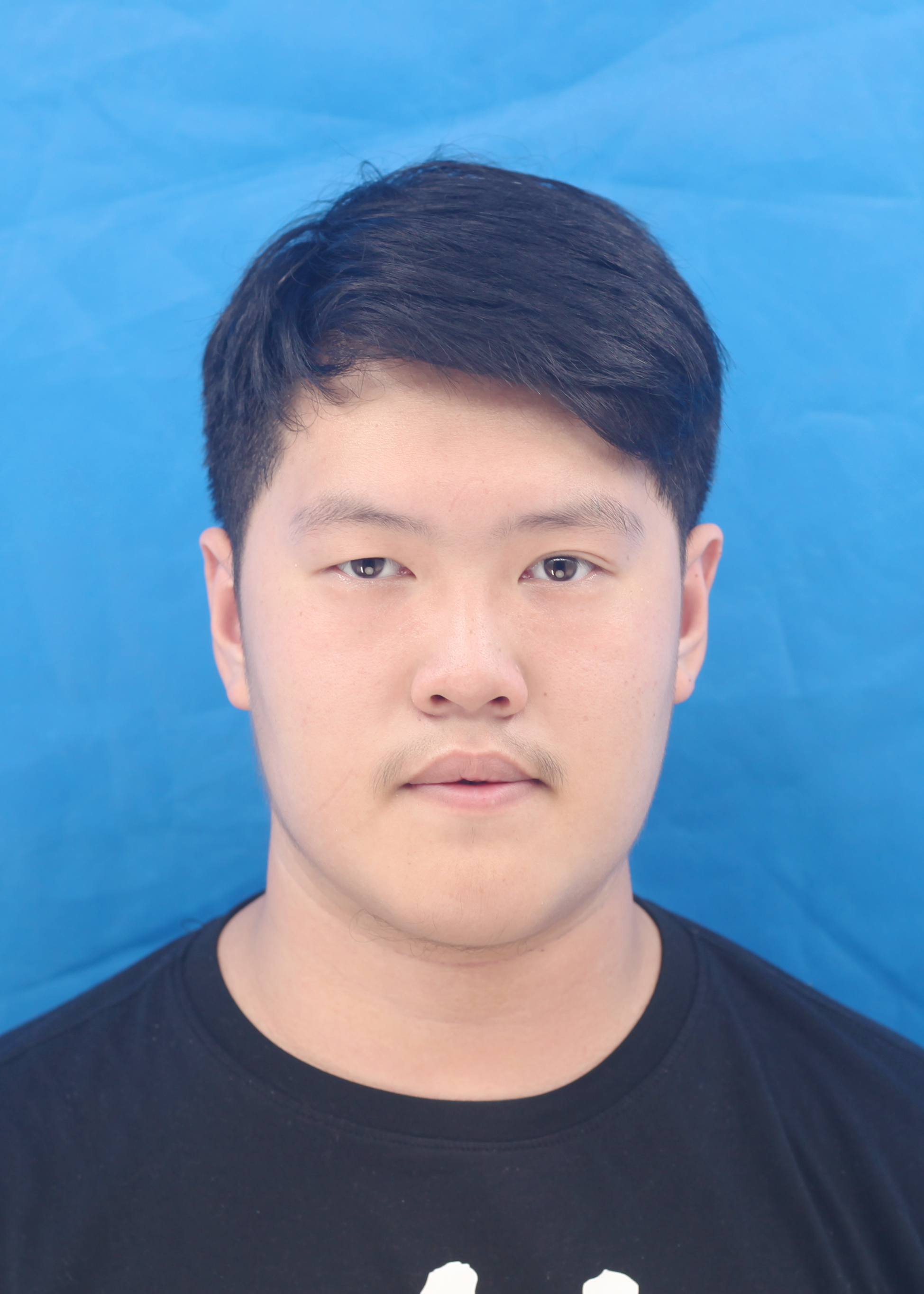}
\textbf{Zihou Liu} received the B.E. degree in Internet of Things Engineering from the Nankai University, Tianjin, China, in June 2022. He is currently pursuing the Master's degree with the College of Computer Science, Nankai University, Tianjin. His research interests include Machine Learning, Artificial Intelligence security, IoT security, and cyberspace security situational awareness.
\end{biography}

\begin{biography}{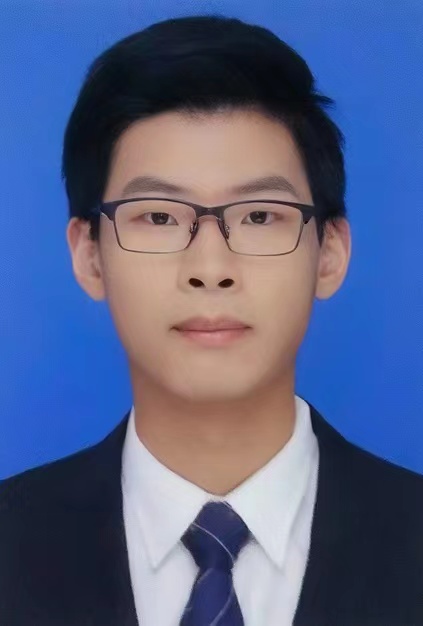}
\textbf{Qingyin Tan} received the B.E. degree in information security from the Nankai University, Tianjin, China, in June 2020. He is currently pursuing the Ph.D. degree with the College of Cryptology and Cyber Science, Nankai University, Tianjin. His main research interests include Web application security, system security, and IoT security. 
\end{biography}

\begin{biography}{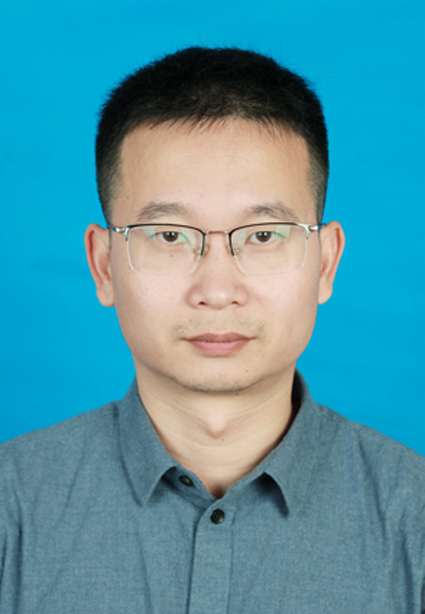}
\textbf{Yang Song}, an associate researcher at Hangzhou Dianzi University, holds a doctorate from the University of Chinese Academy of Sciences. He was formerly the Chief Scientist of Instruction Set Intelligence Technology Co., Ltd., and once served as a researcher at the Innovation Institute of Shanda Group and a senior technical expert at Alibaba Group. His main research directions include artificial intelligence, operating systems, etc.
\end{biography}

\begin{biography}{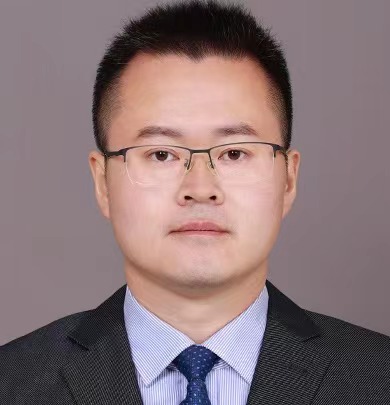}
\textbf{Yu Zhang} received the B.E. degree in computer science and the Ph.D. degree in computer system architecture from the Harbin Institute of Technology, Harbin, China, in 2004 and 2010, respectively. After graduation, he joined the College of Computer and Control Engineering, Nankai University, Tianjin, China. He is currently an Associate Professor with the College of Cyber Science, Nankai University. He has authored or coauthored more than 30 academic papers in international conferences and journals. His research interests include  machine learning, data mining, artificial intelligence security and network security, particularly cyberspace security situational awareness.
\end{biography}

\begin{biography}{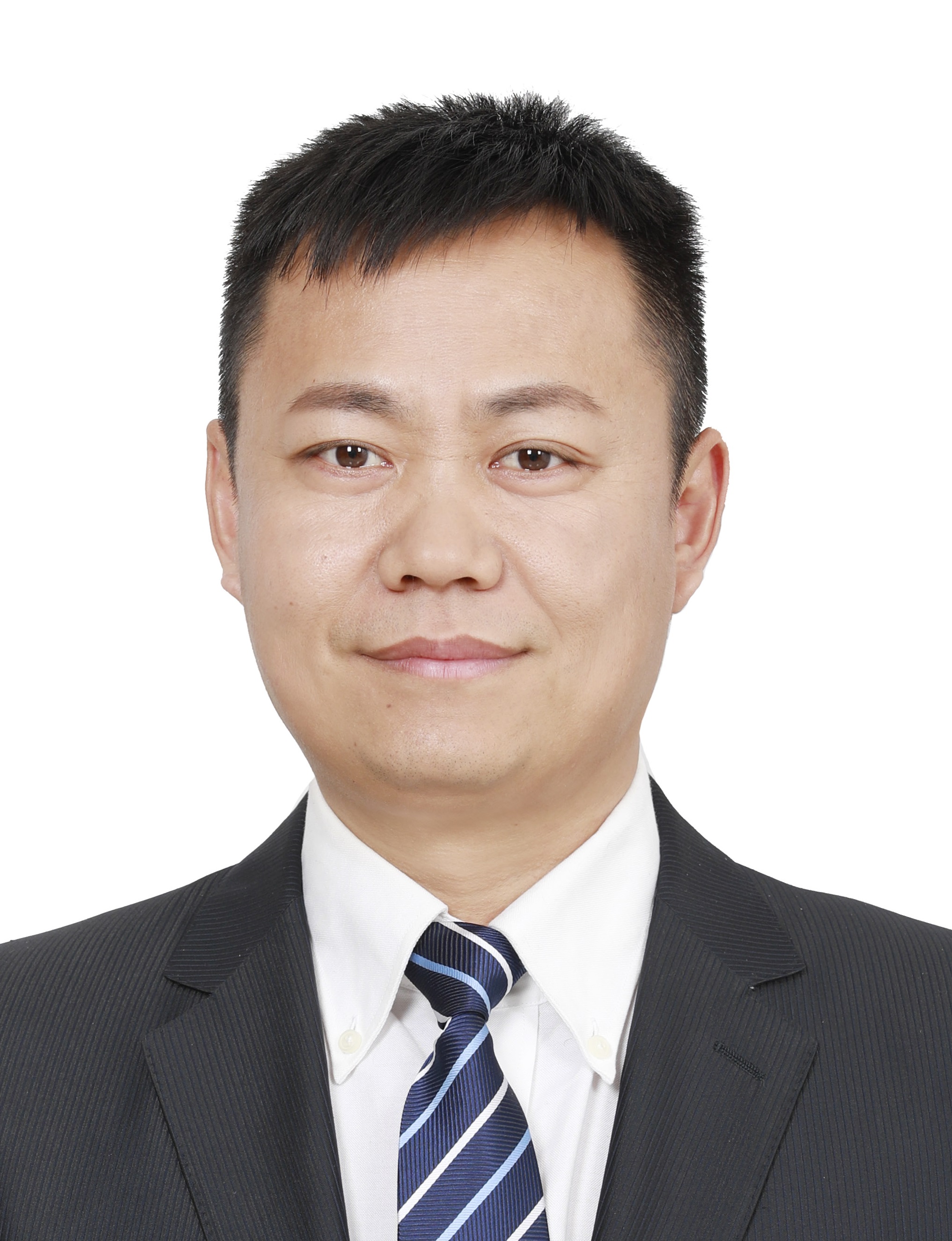}
\textbf{Zheli Liu} received the BSc and MSc degrees in computer science from Jilin University, China, in 2002 and 2005, respectively. He received the PhD degree in computer application from Jilin University in 2009. After a postdoctoral fellowship in Nankai University, he joined the College of Computer and Control Engineering of Nankai University in 2011. Currently, he works at Nankai University as a Professor. His current research interests include applied cryptography and data privacy protection.
\end{biography}

\vspace*{\fill}

\vspace*{\fill}

\end{document}